\documentclass[aps,pra,epsf,superscriptaddress,amsmath,amssymb,amsfonts,twocolumn,showpacs,nofootinbib, floHTFix]{revtex4-1}

\usepackage{ulem}
\usepackage{graphicx}
\usepackage{epsfig}
\usepackage{dcolumn}
\usepackage{bm}
\usepackage{braket}
\usepackage{amsmath}
\usepackage{mathtools}
\usepackage{graphicx,color,xcolor}
\usepackage{hyperref}
\newcommand{\abs}[1]{\left| #1 \right|} 
\usepackage{hyperref}
\usepackage{multirow}
\usepackage{afterpage}
 \usepackage{overpic} 
 \usepackage{float}
 \usepackage{physics}

\usepackage{lipsum}
\usepackage{tikz}

\begin{document}

\title{Mechanisms of localization in a finite harmonically confined optical superlattice}

\author{A. Katsaris$^{\dagger}$}
\email{andkatsa@phys.uoa.gr}
\affiliation{Department of Physics, National and Kapodistrian University of Athens, GR-15784 Athens, Greece} 
\author{I. A. Englezos$^\dagger$}
\affiliation{Zentrum für Optische Quantentechnologien, Luruper Chaussee 149, Universität Hamburg, 22761 Hamburg, Germany}
\author{C. Weitenberg}
\affiliation{Department of Physics, TU Dortmund University, 44227 Dortmund, Germany}
\author{F. K. Diakonos}
\affiliation{Department of Physics, National and Kapodistrian University of Athens, GR-15784 Athens, Greece} 
\author{P. Schmelcher}
\affiliation{Zentrum für Optische Quantentechnologien,  Luruper Chaussee 149, Universität Hamburg, 22761 Hamburg, Germany}
\affiliation{The Hamburg Centre for Ultrafast Imaging,
University of Hamburg, Luruper Chaussee 149, 22761 Hamburg, Germany}

\date{\today}

\begin{abstract}
We investigate the impact of harmonic confinement in a finite optical superlattice and reveal the different mechanisms that can lead to the emergence of localized states. The optical superlattice, with odd or even number of unit cells, can exhibit either a trivial or a non-trivial underlying topology, characterized by the corresponding Zak phase. We focus on a distinct localization mechanism in the intermediate harmonic trapping frequency regime. Specifically, the four lowest-lying eigenstates in this regime form an effective four-level system in the topologically non-trivial configuration. Larger trapping frequency values drive the system into a harmonic trap dominated regime, featuring classical pairing and localization of all states of the lower band, as in a usual optical lattice. For the lower trapping frequency regime, the fate of topological edge states is discussed. Our results are based on exact diagonalization and on a tight-binding approximation that maps the continuous to a discrete system. We address several aspects relevant to the experimental implementation of optical superlattices and provide a brief illustration of the dynamics, highlighting direct ways to observe and distinguish between the different localization mechanisms. 
\end{abstract}

\maketitle
\def\thefootnote{$\dagger$}\footnotetext{These authors contributed equally.}

\section{Introduction} 

Ultracold atomic systems offer a highly versatile platform for the study of a wide range of quantum phenomena in atomic, molecular and condensed matter systems~\cite{BlochLatticeRev,BlochNature2012}. It is noteworthy, that ever since the first experimental realization of Bose-Einstein condensation (BEC)~\cite{BECexpAnderson,BECexpBradley,BECexpDavis}, the field has undergone rapid development. Two of the cornerstones of this development are the precise control of the inter-atomic interaction strengths, via Feshbach resonances~\cite{FeshbachExpCourteille,FeshbachExpInouye,Chin2010RevModPhysFeshbach}, and the modern optical techniques capable of shaping the external confinement enabling nearly arbitrary trap geometries, including setups of low-dimensional lattices~\cite{bakr2009quantum}.
This unprecedented level of control has enabled among others the realization of Hubbard models and the observation of the superfluid-Mott insulator transition~\cite{JakschBH,GreinerExplattice}.  Moreover, topological phases of matter have been realized~\cite{weitenberg2021tailoring, L_onard_2023}, including the quantum Hall effect and topological insulators. In particular, one of the most widely studied 1D systems exhibiting non-trivial topology, the SSH model, has also been realized~\cite{DalibardReview}.

Owing to its simplicity the SSH model is an excellent starting point for understanding the topological phases of matter. Its chiral symmetry allows one to predict when topological edge states (TESs) are supported by the system, due to the strong bulk-edge correspondence of the BDI symmetry class \cite{Ryu_2010}. The robustness of the TESs against certain disorders renders them promising candidates for potential applications in quantum information processing \cite{PhysRevA.103.052409,PhysRevResearch.2.033475,PhysRevA.106.022419}. In addition, several extensions of the SSH model to higher dimensional systems with enriched topological phases \cite{PhysRevB.106.205111,PhysRevB.106.245109} or interacting systems \cite{PhysRevLett.110.260405, PhysRevA.99.053614,PhysRevB.107.054105} are available. The experimental realizations of the SSH model in optical lattices have been mainly focusing on the quantized charge transport both for fermionic \cite{Nakajima_2016} and bosonic \cite{Lohse_2015} atoms. 

The unique properties of TESs, such as those emerging in the SSH model, make them a central feature for probing topological phases. Accordingly, edge states have been experimentally observed in 2D topological systems, including a driven honeycomb lattice with a sharp wall potential \cite{Braun_2024} and a rotating trap \cite{Yao_2024}. Further observations of edge states utilize artificial dimensions stemming from internal states to define sharp edges \cite{PhysRevLett.108.133001, Mancini_2015, Stuhl_2015, Chalopin_2020, Kanungo_2022}. Taking advantage of the fact that edges occur naturally in tweezer arrays with Rydberg interactions, edge states have also been experimentally probed in those setups \cite{doi:10.1126/science.aav9105}. Focusing on the optical lattice platforms, some of the main challenges in realizing and manipulating edge states include: i) the realization of a well defined boundary which requires optical control at single site level using e.g. a quantum gas microscope \cite{Gross_2021} ii) the careful compensation for the residual harmonic confinement to achieve a flat geometry~\cite{Schneider2012BoxHubbard,Gaunt_Hadzibabic2013Box, Navon_Hadzibabic2021Box} and iii) the long periods associated with their transport dynamics, especially for larger system sizes~\cite{DalibardReview}.

Evidently, the realization of edge states has attracted significant attention in studies of topological systems. The underlying topology of a system is just one of several mechanisms through which localized states can emerge. For instance, it is well established that localization can also arise from disorder and quasi-periodicity, as in the Anderson \cite{Anderson1958} and the Aubry-Andr\'e models \cite{Aubry1980}, respectively. Moreover, especially in optical lattice setups, external potentials such as harmonic traps break the translational invariance, resulting in energetically paired and localized states in the high harmonic trapping frequency regime~\cite{Batrouni2002HOL,Hooley2004HOL,Ruuska2004HOL,Rigol2004HOL,Pezze2004HOL,Ott2004HOL,Rey2005HOL,Ali2025HOL}. In this work, we identify a distinct localization mechanism arising from finite-size effects and the presence of a harmonic trap in the underlying periodic lattice with open boundary conditions, leading to localization that is related to, yet distinct from, pure Wannier–Stark localization \cite{PhysRev.117.432, PhysRevB.36.7353, PhysRevLett.61.1639}. We conclude that, since localization can arise from different mechanisms, it is essential to identify which mechanism manifests the observed phenomenology in each system.

For these reasons, we systematically study the impact of superimposing a harmonic trap onto a continuous optical superlattice potential which can be mapped, within a tight-binding (TB) approximation at zero trapping frequency, to an SSH or extended SSH (eSSH) model~\cite{Katsaris2024}. In this way, we can examine three distinct regimes with respect to the value of trapping frequency, each one corresponding to a different localization mechanism. For low but finite harmonic trapping frequencies the TESs survive and the system behaves similarly to the usual SSH or eSSH models. In contrast, for large frequency values the harmonic trap dominates resulting in similar behavior to the case of a harmonically confined simple optical lattice. Interestingly, we find an intermediate frequency regime where intra-band avoided crossings occur resulting in the four lowest lying eigenstates to isolate and resemble an effective four-level system in the topological configuration. 
We identify this as a new form of localization in optical superlattice setups, arising from the interplay between the optical superlattice and the harmonic trap.

This work is structured as follows. In Section~\ref{sec:setup}, we present the potential landscapes and the TB approximation for comparison between the continuous and the corresponding discrete systems. In Section~\ref{sec:interplay}, we briefly discuss the interplay between the harmonic trap and the finite optical lattice. We then focus on the harmonically confined optical superlattice system and present its unique phenomenology, while providing a comparison with the optical lattice system. In Sec.~\ref{sec:4lvl}, we present an in-depth analysis of the optical superlattice parameters dependence and we address relevant experimental aspects. Furthermore, in Section~\ref{sec:fateofedges}, we discuss the fate of the topological edge states in the harmonically confined optical superlattice system. Finally, in Section~\ref{sec:SummaryAndOutlook} we summarize our results and highlight future perspectives.

\section{Harmonically Confined Optical (super)lattice Potentials and Tight-binding approximation}\label{sec:setup} 

In this section, we present the potentials we consider throughout this work and the TB approximation, which we employ for the detailed comparison between the continuous and the corresponding discrete systems. Specifically, in Sec.\ref{sec:Potentials} we introduce the optical lattice and superlattice potentials, along with the linear extension of the potential area, which is necessary for the restoration of the TESs in the finite system \cite{Katsaris2024}. Also, we discuss the addition of a harmonic trap potential in our description, in order to account for the usual experimental setups. Lastly, in Sec.\ref{sec:TB} we review the TB approximation, which helps to illuminate the phenomenology of the continuous system.

\subsection{Potential Landscapes and Eigenvalue Problem}\label{sec:Potentials}

An optical lattice can be realized experimentally by forming a standing wave, utilizing two counter propagating laser beams~\cite{BlochLatticeRev}, leading to a potential of the form:
\begin{align}\label{OL_Potential}
V_{OL}(x)&= V_0 \cos^2\left(k_r x\right) 
\end{align}
where $V_0$ is the amplitude of the standing wave, $k_{r}= 2\pi/\lambda_0$ is the single-photon recoil momentum, and $\lambda_0$ is the wavelength of the lattice. For a system of $\mathcal{N}$ sites and total length $2L$ it holds that $\lambda_0 = 2L/\mathcal{N}$. Superimposing a second standing wave with different amplitude ($V_1$) and frequency, can lead to a sequence of $M=\mathcal{N}/2$ double wells~\cite{Anderlini_2007, F_lling_2007, DalibardReview} as required to realize the SSH and eSSH models. This yields an optical superlattice potential of the form:
\begin{align}\label{SL_Potential}
V_{OS}(x)= V_0 \cos^2\left(k_r x\right) + V_1 \cos^2\left(\frac{k_r x}{2} + \phi\right) 
\end{align}
 The heights of the two barriers of the optical superlattice potential shown in Fig.~\ref{fig:SL_schematic} are:
\begin{align}
V_{low}  = V_0(1 - \frac{V_1}{4V_0})^2,\quad V_{high} &= V_0(1 + \frac{V_1}{4V_0})^2
\end{align}
The phenomenology related to the optical superlattice potential directly depends on the height of the higher barrier $V_{high}$ and the ratio of the lower to the higher barrier heights, $u = V_{low}/V_{high}$. Hence, in the following, we express our results in those terms rather than the $V_0$ and $V_1$ amplitudes. Clearly, by setting $u=1$ (or $V_1=0$) the optical superlattice potential reduces to the optical lattice potential of Eq.~\eqref{OL_Potential}. The additional phase factor $\phi$, in combination with the number of cells $M$, determines whether the system corresponds to an underlying topologically trivial or non-trivial configuration (for details, see Appendix~\ref{app:ZakPhase}), as shown on the following table:
\begin{table}[ht]
    \centering
    \renewcommand{\arraystretch}{1.5}
    \begin{tabular}{|c|c|c|}
        \hline
        ~$M$ (Number of Cells)~ & $\phi = 0$ & $\phi = \pi/2$ \\
        \hline
        Even & ~Trivial~     & ~Non-Trivial~ \\  
        \hline
        Odd & ~Non-Trivial~ & ~Trivial~ \\
        \hline
    \end{tabular}
    \caption{Topological classification of the optical superlattice system with respect to the phase factor $\phi$ and the number of cells $M$.}
    \label{Table:TopoClas}
\end{table}

\noindent In order to characterize the topology of each optical superlattice configuration, we numerically calculate the Zak phase associated with the corresponding continuous system. Specifically, we compute the discretized Berry phase, by summing the phases of overlaps between adjacent Bloch eigenstates along the Brillouin zone. 

As shown in~\cite{Katsaris2024}, hard wall boundary conditions (HWBC) strongly affect the ability of the optical superlattice system in the topological configuration to support TESs. The sharp confinement enforces vanishing wavefunction amplitudes at the lattice edges, thereby modifying the profile and energies of states localized near the edges. For a finite optical superlattice system to support TESs, the effect of HWBC must be compensated. For this purpose, a linear extension of the following form was proposed:
\begin{equation}
\label{Potential_Extension2}
V_{ext}(x)= \begin{cases}
      ~V_{SL}(-x_0) + \dfrac{dV_{SL}}{d x}\Bigr|_{x=-x_0}(x + x_0),  & x\in D_{1}\\[5pt]
      ~0,& x\in D_{2} \\[5pt]
      ~V_{SL}(+x_0) + \dfrac{dV_{SL}}{d x}\Bigr|_{x=+x_0}(x-x_0), & x\in D_{3}
      \end{cases}
\end{equation}
with $x_0$ a variable starting point for the extension of the potential, $D_1 = [-L-d,x_0]$, $D_2 = (-x_0,x_0)$, $D_3 = [x_0,L+d]$ and $d$ is the length of the extension area. Owing to the symmetry of the potential, the slopes at the symmetric points $\pm x_0$ mirror one another with opposite sign. This form of the extension ensures that both the total potential and its first derivative are continuous. Conversely, in the topologically trivial configuration the system cannot support TESs regardless of the extension, in complete agreement with the SSH and eSSH models.

In most experiments with OLs or OSs, some residual harmonic confinement is also present. Although it can be largely compensated, creating a perfectly flat geometry requires considerable and in many cases experimentally not desirable effort ~\cite{Schneider2012BoxHubbard,Gaunt_Hadzibabic2013Box,Navon_Hadzibabic2021Box}, hence it is useful to explicitly include this additional confinement into our analysis. In this work, we consider the harmonic trap to have the standard form:
\begin{align}\label{HO_Potential}
V_{HT}(x)&= \frac{1}{2} m \omega^2 x^2,~~x\in[-L-d,L+d] 
\end{align}
where $m$ is the atomic mass, and $\omega$ is the trapping frequency of the harmonic confinement. Thus, the total potential reads:
\begin{equation}
V(x) = V_{lat}(x) + V_{ext}(x) + V_{HT}(x)
\end{equation}
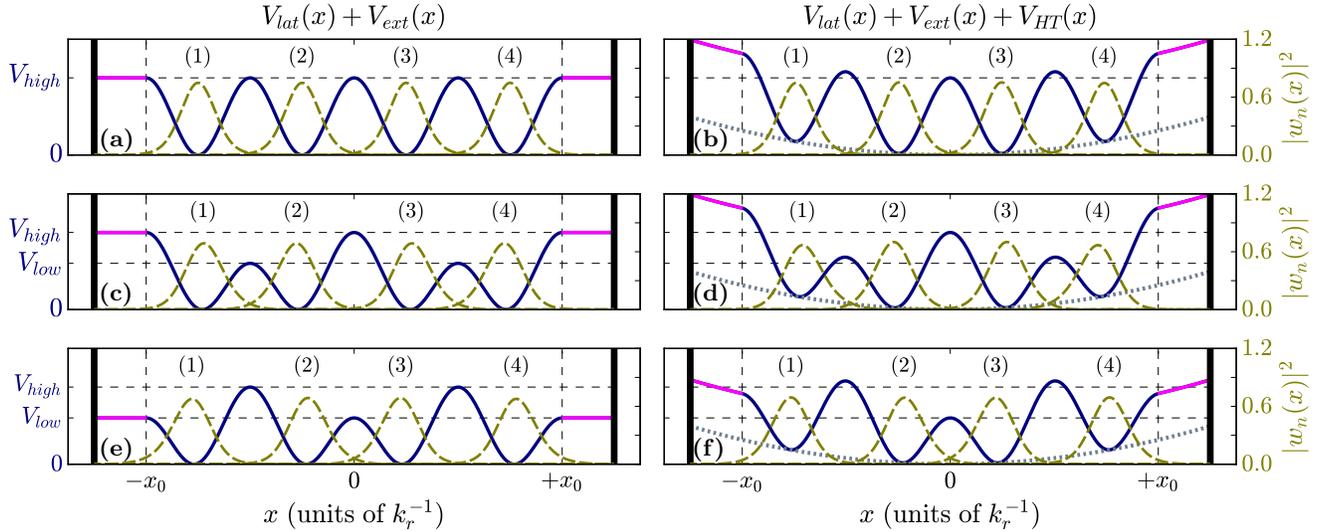
\begin{figure*}[ht]
\begin{center}
\begin{overpic}[width=\textwidth]{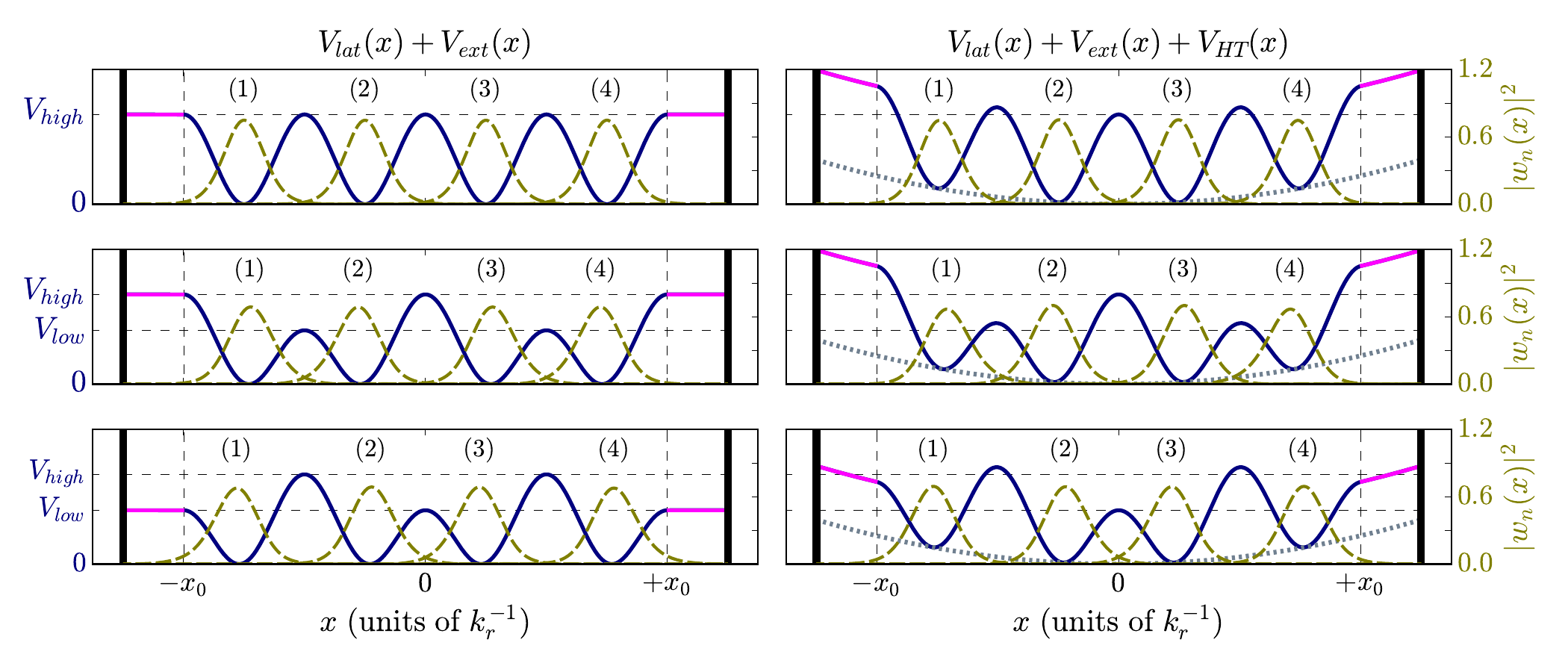}
\put(8.2, 30.5) {\textbf{(a)}}
\put(52.2, 30.5) {\textbf{(b)}}
\put(8.2, 19.) {\textbf{(c)}}
\put(52.2, 19.) {\textbf{(d)}}
\put(8.2, 7.5) {\textbf{(e)}}
\put(52.2, 7.5) {\textbf{(f)}}
\end{overpic}    
    \caption{Schematics of the extended optical lattice and superlattice potentials in the left column (a,c,e), and the same potentials with the addition of the harmonic confinement in the right column (b,d,f). Specifically, (a,b) the extended optical lattice, (c,d) and (e,f) the extended optical superlattice in the trivial and topological configuration, respectively. The solid lines represent the potentials, and the dashed lines show the corresponding Wannier functions. The vertical dashed lines highlight the starting point of the extension of the potential area. The dotted lines show the harmonic oscillator function. The right y-axis shows in olive color (medium gray in grayscale) the squared norm of the corresponding Wannier functions for each setup.}
    \label{fig:SL_schematic}
\end{center}
\end{figure*}
shown schematically in the right column of Fig.~\ref{fig:SL_schematic}. The $V_{lat}(x)$ term is either the $V_{OL}(x)$ in 
Fig.~\ref{fig:SL_schematic}(a,b) or the $V_{OS}(x)$ in Fig.~\ref{fig:SL_schematic}(c-f). In the following, to avoid repeated verbose statements, we will refer to the case where TESs may be present at $\omega=0$ as a topological configuration and as a trivial configuration otherwise. Note that the cases of zero and non-zero trapping frequency can always be connected adiabatically, by tuning the trapping frequency sufficiently slowly.

In order to obtain the complete spectrum of energy eigenvalues and eigenstates for each of the aforementioned systems, we solve numerically the stationary Schr\"odinger equation
\begin{equation}
\hat{H}\ket{\Psi_n} = E_n\ket{\Psi_n},~\text{with}~ \label{Schrodinger}~\hat{H} = 
-\frac{\hslash^2}{2m} \frac{\partial^2}{\partial x^2} 
 + \hat{V}
\end{equation}
via exact diagonalization (ED), using a nine-point finite difference stencil to approximate the second derivative. Finally, for computational convenience, we recast the Schr\"odinger equation in a dimensionless form by expressing the length in terms of $k_r^{-1}$ and the energy in terms of the recoil energy $E_r = \hbar^2 k^2_{r}/2m$, resulting in the harmonic trapping frequency given in terms of $E_r/\hbar$, where $\hbar=h/2\pi$ with $h$ denoting Planck’s constant.

\subsection{Tight-Binding Approximation}\label{sec:TB}

In a deep enough lattice, the system can be truncated to the first, or first few, energy bands~\cite{WannierReview,tweezerArrays2024}, known as the TB approximation. For a system with $\mathcal{N}=2M$ minima (sites), where $M$ is the number of unit cells, and restricted to the lowest band, the TB Hamiltonian is:
\begin{equation}\label{TB_Hamilt}
    \hat{\mathcal{H}}_{TB} = \sum_{i=1}^{\mathcal{N}}\sum_{j=1}^{\mathcal{N}} h_{i,j} \hat{\alpha}^\dagger_{i}\hat{\alpha}_{j},
\end{equation}
where $\hat{\alpha}^\dagger_{i}$ ($ \hat{\alpha}_{i}$) are the creation (annihilation) operators, creating (annihilating) either a bosonic or a fermionic particle in the lowest band at $i$-th site, and the matrix elements
\begin{equation}\label{TB_elements}
    h_{i,j} = \int w^*_i(x) \Big[ \frac{\hbar^2}{2m}\frac{\partial^2}{\partial x^2} - V(x) \Big] w_j(x) dx
\end{equation}
are defined in terms of the Wannier functions $w_i(x)$ localized at the $i$-th site. The potential is real and mirror-symmetric, so the resulting matrix exhibits reflection symmetry both along the main diagonal, $h_{i,j} = h_{j,i}$ and along the anti-diagonal $h_{i,j} = h_{N+1-j,\,N+1-i}$. In the following, we define the Wannier functions as
\begin{equation}\label{eq:wannier}
    w_i(x)=\bra{x}\ket{\chi_i}
\end{equation}
where $\ket{\chi_i}$ are the eigenstates of the position operator restricted to the lowest band ($\hat{\mathcal{X}}_{band}$):
\begin{equation}
\hat{\mathcal{X}}_{band} =  \sum_{n=1}^{\mathcal{N}}  \sum_{m=1}^{\mathcal{N}} \ket{\Psi_n}\bra{\Psi_n}\hat{x}\ket{\Psi_m}\bra{\Psi_m},
\end{equation}
and $\ket{\Psi_{n}}$, $\ket{\Psi_{m}}$ are the single particle eigenstates, defined in Eq.~\eqref{Schrodinger} and computed by ED.
In 1D these are the uniquely defined maximally-localized Wannier functions, even when the construction is generalized to include higher bands~\cite{WannierStates}. We show illustrative examples of the Wannier functions that corresponds to our systems in Fig.~\ref{fig:SL_schematic}.

\begin{figure}[h]
\begin{center}
\begin{overpic}[scale = 1]{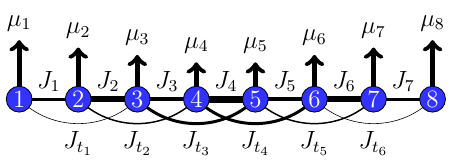}
\end{overpic}
\end{center}
\caption{Schematic of the discrete model that is produced by the TB approximation of a continuous optical lattice or superlattice system with $\mathcal{N}=8$ minima (sites). Each site corresponds to a Wannier function, and the connecting lines show the hopping amplitudes expressed by the $J_i$'s and $J_{t_i}$'s parameters. The upward arrows denote the on-site potential $\mu_i$ at site $i$. The widths of the lines and the heights of the arrows respectively reflect the mirror symmetry of the lattice about its central site.}
\label{Fig:Disc_Schematic}
\end{figure}

In the case of the harmonically trapped optical superlattice potential the TB Hamiltonian may be modeled by an extended SSH model with site-dependent hopping amplitudes and on-site potential terms:
\begin{equation}
\hat{\mathcal{H}}_\text{TB} = \hat{\mathcal{H}}_{\mu} + \hat{\mathcal{H}}_\text{NN} + \hat{\mathcal{H}}_\text{NNN},
\label{TB_eq}
\end{equation}
with 
\begin{align}
\nonumber
\hat{\mathcal{H}}_{\mu} & =  \sum_{i=1}^{\mathcal{N}} \mu_i
\hat{a}^\dagger_{i}\hat{a}_{i} \\
\hat{\mathcal{H}}_\text{NN} & =  \sum_{i=1}^{\mathcal{N}-1} J_i
\hat{a}^\dagger_{i}\hat{a}_{i+1} + h.c \label{TB_eq_Terms}\\
\nonumber
\hat{\mathcal{H}}_\text{NNN} & =  \sum_{i=1}^{\mathcal{N}-2} J_{t_i}
\hat{a}^\dagger_{i}\hat{a}_{i+2} + h.c
\end{align}
where $\mu_i$ denote the on-site potentials, $J_i$ the nearest-neighbor hopping amplitudes, and $J_{t_i}$ the next-nearest-neighbor hopping amplitudes. A schematic of this discrete model is shown in Fig.~\ref{Fig:Disc_Schematic}.

\begin{figure}[h]
\begin{center}
\begin{overpic}[scale = 0.5]{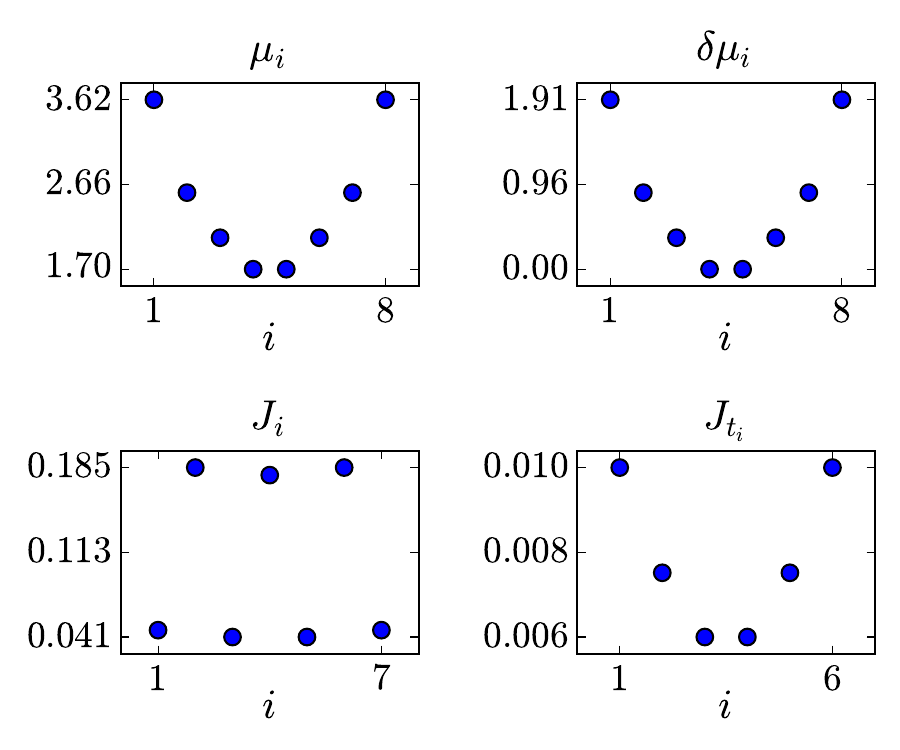}
\put(18,76.5) {\textbf{(a)}}
\put(68,76.5) {\textbf{(b)}}
\put(18,36) {\textbf{(c)}}
\put(68,36) {\textbf{(d)}}
\end{overpic}
\end{center}
\caption{Profiles of the (a) on-site, (b) on-site relative to the central site, (c) nearest neighbor hopping and (d) next-to-nearest neighbor hopping amplitudes of the TB approximation for an optical superlattice system with the addition of a harmonic trap. The extended optical superlattice system has $M=4$ ($\mathcal{N}= 8$) cells (minima/sites), $V_{high} = 5.0E_r$, $u = 0.6$, $x_0 = L$, $d=0.36\pi/k_r$, $\omega = 0.25E_r/\hbar$. All amplitudes
are in units of $E_r$.}
\label{TB_Hoppings_Example}
\end{figure}
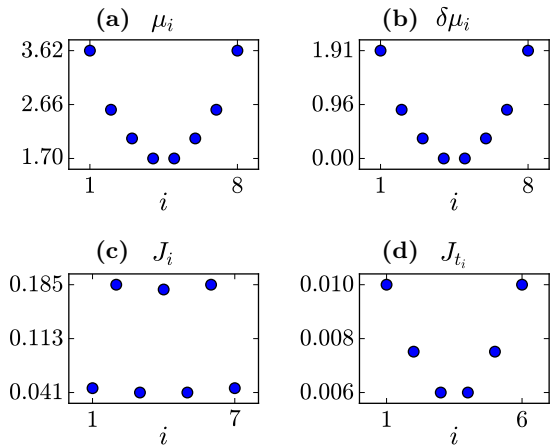

As shown in Fig.~\ref{TB_Hoppings_Example}(a) the effect of the harmonic trap potential is reflected in the on-site potentials of the discrete system, since they follow exactly the parabolic profile enforced by it. In Fig. \ref{TB_Hoppings_Example}(b) we use as point of reference the on-site amplitude of the middle sites ($\mu_{M,M+1}$), and plot the difference $\delta\mu_i = \mu_i - \mu_{M,M+1}$, since in the TB approximation we can always subtract a constant on-site amplitude, and only the difference is relevant for the phenomenology of the system. Also, from Fig.~\ref{TB_Hoppings_Example}(c) we see that the nearest-neighbor hoppings are site-dependent, though the effect is weak, and they retain the alternating pattern dictated by the optical superlattice potential. The next-to-nearest neighbor hopping terms have relatively small values as we see in Fig.~\ref{TB_Hoppings_Example}(d), however we include them in our analysis for clarity. The even higher order hopping terms are negligible and we omit them. Finally, all the terms have mirror symmetry, owing to the parity symmetry of the corresponding continuous system.

The above description applies only if a well-defined band structure exists, with states assigned to each band, making possible the restriction to the lowest band. Adding the harmonic trap to the system shifts all energies upward, ultimately leading to avoided crossings between states from lower and higher bands at some finite trapping frequency $\omega_{crit}$. Beyond this $\omega_{crit}$, crossings with states not included in the TB model occur, and the very notion of bands becomes ill-defined. In our analysis, we use the TB approximation to describe and explain the phenomenology of our systems in the low to mid trapping frequency regime, where in each case we are sufficiently below the corresponding $\omega_{crit}$ value.

\section{Interplay of the harmonic trap and the finite optical (super)lattices}\label{sec:interplay} 

In order to demonstrate the interplay between the harmonic trap potential and the optical lattice or superlattice potentials, we first address the behavior of the energy spectrum and the lowest-lying eigenstates, for fixed lattice parameters ($V_{high}$, $u$ and $M$), and varying trapping frequency ($\omega$). In Sec. \ref{sec:OLvsOmega} we address the case of the optical lattice potential ($u=1$) without extension of the potential landscape, which already has been discussed in detail in several previous works~\cite{Batrouni2002HOL,Hooley2004HOL,Ruuska2004HOL,Rigol2004HOL,Pezze2004HOL,Ott2004HOL,Rey2005HOL,Ali2025HOL}. We begin with this simpler case in order to illustrate the two extreme regimes of a perturbatively weak or a dominant harmonic confinement. Then, in Sec. \ref{sec:OSvsOmega} we discuss the optical superlattice potential with the linear extension of the potential landscape and emphasize in detail the similarities to and differences from the optical lattice potential. We investigate the direct effects of the configuration (trivial or topological) and the number of cells (even or odd) of the optical superlattice, as both are shown to play a decisive role in the resulting phenomenology.  

\begin{figure*}[ht]
\begin{center}
\begin{overpic}[width=\textwidth]{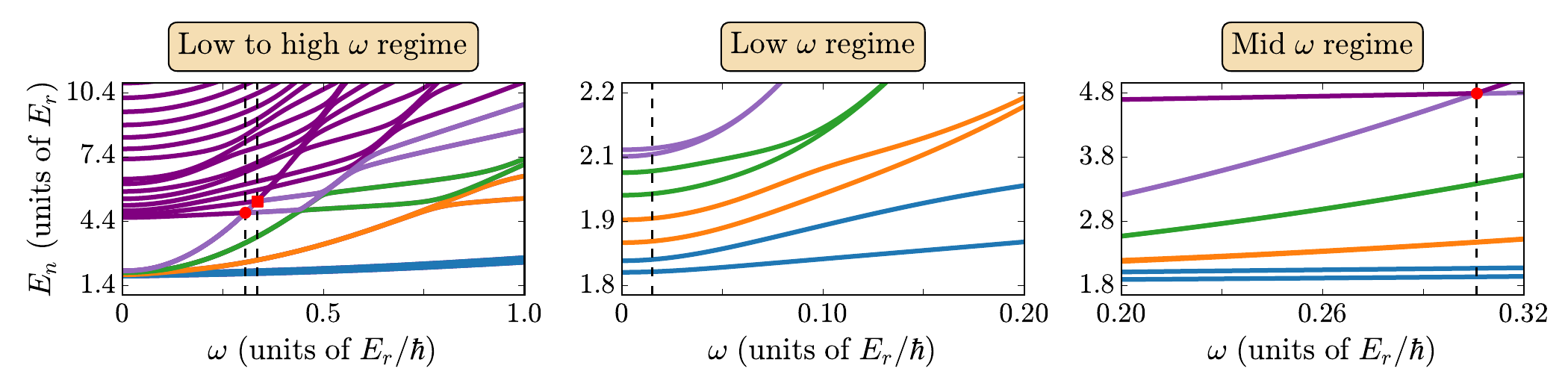}
\put(5.25,21.5) {\textbf{(a)}}
\put(40.5,21.5) {\textbf{(b)}}
\put(72.5,21.5) {\textbf{(c)}}
\end{overpic}
\begin{overpic}[width=\textwidth]{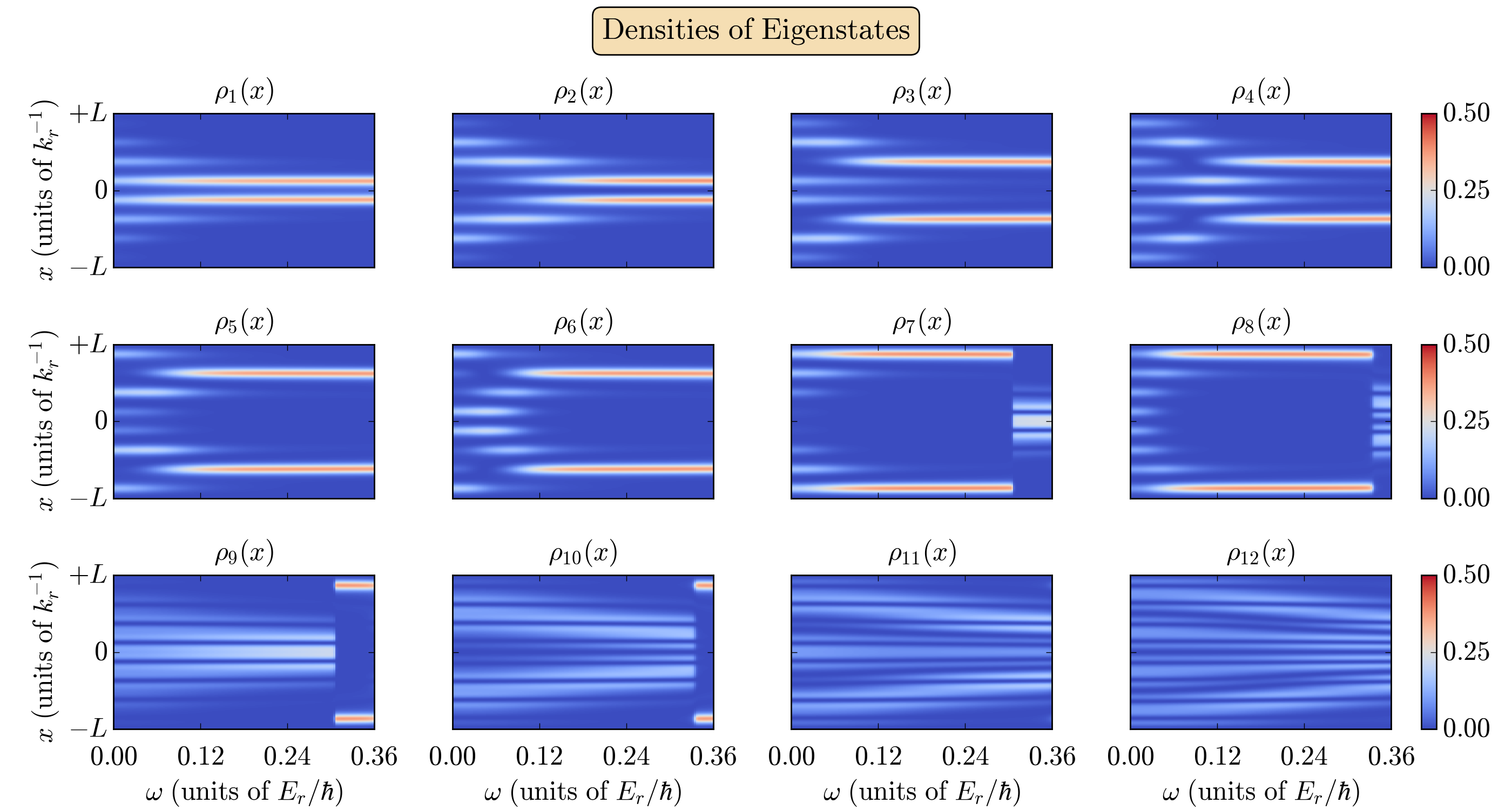}
\put(34,51) {\textbf{(d)}}
\end{overpic}
\end{center}
\caption{(a-c) Energy spectrum versus the trapping frequency $\omega$ of an optical lattice potential with the addition of a harmonic trap. The optical lattice has $\mathcal{N}=8$ minima (sites) and $V_{0}=5E_r$. (a) Energy spectrum from low to high $\omega$ regime, (b) and (c) the low and mid regimes, respectively. In (a) and (c) the vertical dashed lines indicate the values of $\omega$ where crossing with the upper band occurs, and in (b) the approximate value of $\omega$ where the effect of the harmonic trap starts to become apparent in the system. (d) Spatial profile of the densities of the first twelve (12) eigenstates of the system for low values of $\omega$. The first eight (8) are from the first band, and the next four (4) correspond to the first states of the second band.}
\label{fig:OL_Spectra_States_vs_omega}
\end{figure*}

\subsection{Optical Lattice and Harmonic Trap}\label{sec:OLvsOmega}

We begin our analysis with a relatively shallow optical lattice potential of $V_0 = 5E_r$ and $\mathcal{N}=8$ minima (sites), without considering an extension ($d=0$). In this system we add the harmonic trap and vary its trapping frequency ($\omega$), which leads to the energy spectrum and the densities of eigenstates $\rho_n(x) = \abs{\psi_n(x)}^2$, shown in Fig.~\ref{fig:OL_Spectra_States_vs_omega}. 

For sufficiently weak $\omega$ (see Fig.~\ref{fig:OL_Spectra_States_vs_omega}(b)), the energy spectrum is largely unaffected by the harmonic trap, maintaining a clear band structure, being only slightly shifted upwards due to the influence of the harmonic trap. Namely, for $\omega\lessapprox 0.015 E_r/\hbar$ which is highlighted by the vertical dashed line in Fig.~\ref{fig:OL_Spectra_States_vs_omega}(b), the corresponding eigenstates are delocalized across the whole lattice, as can be seen from the corresponding densities in Fig.~\ref{fig:OL_Spectra_States_vs_omega}(d). As $\omega$ increases, Fig.~\ref{fig:OL_Spectra_States_vs_omega}(b) shows that the energies begin to pair and become degenerate, starting from the highest eigenstates of the first band and progressively moving to the lower ones. We note that the first two eigenstates, the ground state and the first excited state, are never fully paired. However the energy gap separating them from the rest of the spectrum becomes sufficiently large, hence they practically become isolated and effectively behave as a pair. This pairing and degeneracy becomes also evident in the spatial profile of eigenstates in Fig.~\ref{fig:OL_Spectra_States_vs_omega}(d), which also become localized in pairs of parity symmetric sites with increasing $\omega$. Clearly, for the specific system we show for an approximate range of $\omega$ from $0.16$ to $0.28E_r/\hbar$ all the states in the first band are paired and localized.

We observe that further increasing $\omega$ causes the paired energies to start shifting upwards parabolically, until eventually they perform consecutive avoided crossings with the states of the excited bands. The first and second avoided crossings are indicated in Fig.~\ref{fig:OL_Spectra_States_vs_omega}(a) and (c) with markers and vertical dashed lines, respectively. Notably, the involved states exchange their character, as shown e.g. at the right edge of Fig.~\ref{fig:OL_Spectra_States_vs_omega}(d) for the $\rho_{7}(x)$ to $\rho_{10}(x)$ densities. The localized states observed before entering the avoided-crossing regime in $\omega$ can be termed ``quasi-classically" localized, as they resemble classical particles confined to the parity symmetric local minima of the potential. Each pair of states forms a nearly degenerate doublet that is energetically well separated from the rest of the spectrum, effectively behaving as an isolated two-level system localized on parity symmetric sites. These ``quasi-classically" localized states can be superimposed to enable quantum transport dynamics. In particular, if the system is initialized with population on one site of a paired set, we observe coherent tunneling dynamics between the two paired sites, even though they are spatially separated. However, the period of the transport is very long, due to the near degeneracy of the parity symmetric states, as we show in Sec.~\ref{dynamics}, in the context of the optical superlattice potential.

\begin{figure}[!ht]
\begin{center}
\begin{overpic}[width=0.45\textwidth]{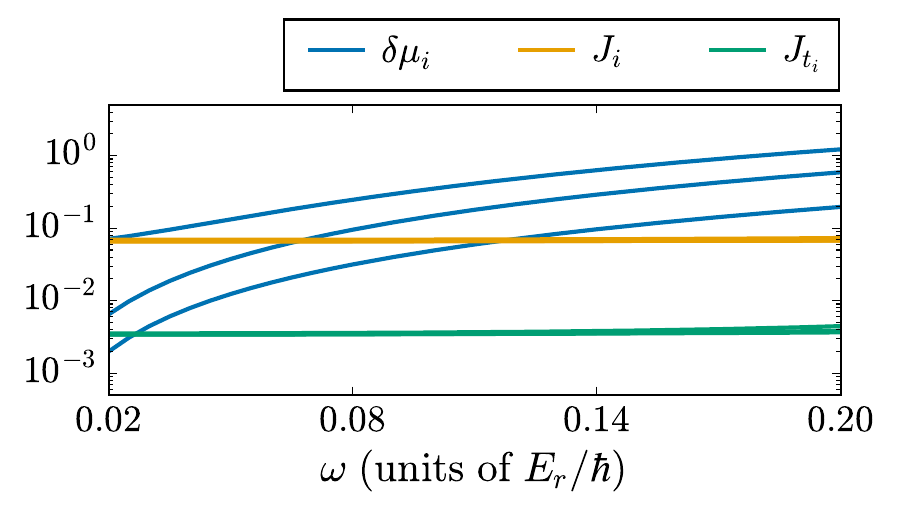}
\end{overpic}
\caption{The on-site and hopping amplitudes versus the trapping frequency $\omega$ of the TB approximation for an optical lattice system with the addition of a harmonic trap. The optical superlattice system has $\mathcal{N}=8$ minima (sites) and $V_0 = 5.0E_r$ and has no extension. The index $i$ identifies each on-site and hopping term in the Hamiltonian, according to the definition in Eq.~\eqref{TB_eq_Terms}. All amplitudes are in units of $E_r$.}
\label{fig:TBApprox_vs_omega_OL}
\end{center}
\end{figure}

A particular feature of our finite system in the intermediate $\omega$ regime, is that the eigenstates localize on specific sites, with the higher energy states of each band localizing on the outermost sites and progressively moving to the lowest energy states corresponding to the inner most sites. This phenomenology can also be predicted by the behavior of the on-site and hopping amplitudes of the corresponding TB approximation for increasing $\omega$. Specifically, as we can see from Fig.~\ref{fig:OL_Spectra_States_vs_omega} and Fig.~\ref{fig:TBApprox_vs_omega_OL} the trapping frequency regime where the states are paired and localized at parity-symmetric sites, occurs when the on-site potentials $\delta\mu_i$ that correspond to these sites, exceed the hopping amplitudes $J_i$.

The connection between on-site potential and localization has been studied in depth in the context of Anderson localization~\cite{Anderson1958} in random lattices or the Aubry-Andr\'e~\cite{Aubry1980} model in quasi-periodic systems, with important experimental realizations for one-dimensional systems~\cite{Roati2008, Lahini2008,Lahini2009}. In our case, we argue that the existence of localized states stems from the superposition of the periodic potential (optical lattice) and the dominant confining potential (harmonic trap).

\begin{figure}[ht]
\begin{center}
\begin{overpic}[width=0.45\textwidth]{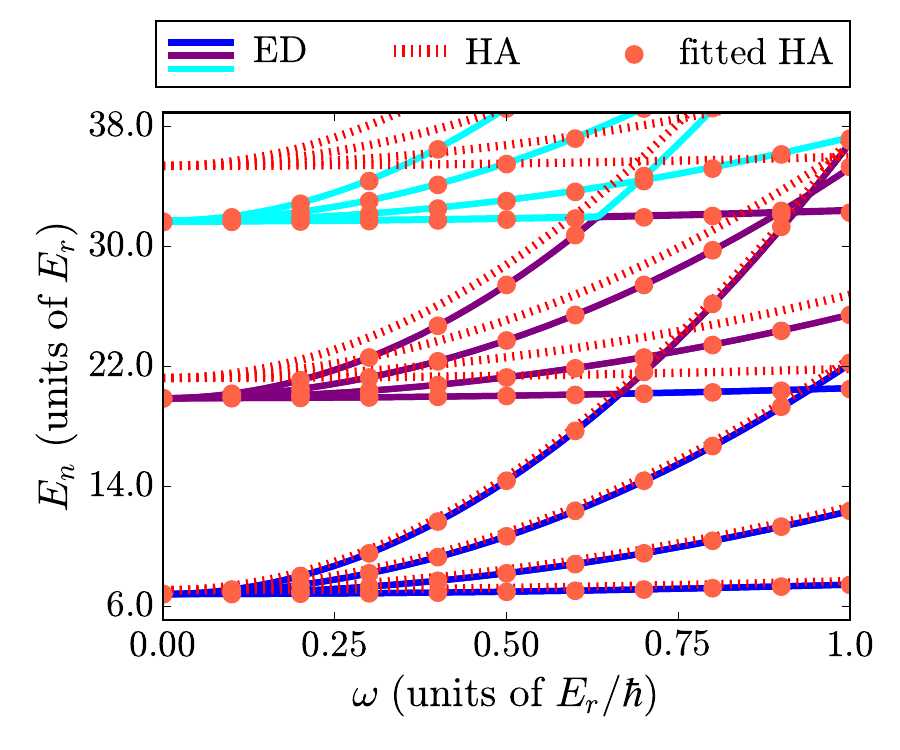}
\end{overpic}
\caption{The energy spectrum of a very deep optical lattice ($V_0 = 50 E_r$ and $M = 8$) versus the trapping frequency $\omega$. Solid and dotted lines represent the results from the ED and the harmonic approximation (HA), respectively. Scatter circle markers shows the results from the fitted harmonic approximation, using as $\epsilon_n$ at $\omega = 0$, the values from the ED.}
\label{fig:Harmonic_Approx}
\end{center}
\end{figure}

In order to describe the general behavior of the spectrum of the continuous system with respect to $\omega$, we can treat the values of the harmonic trap potential $V_{HT}(x)$ at the location of the minima of the optical lattice 
\begin{equation}
x_{j}=\pm \dfrac{\pi}{2k_r}(2j+1),~\text{for}~~j=0,~1,~\dots,~M=\frac{\mathcal{N}}{2},
\end{equation}
as effective energy shifts, resulting in
\begin{equation}
    \epsilon_{shift}^{j} = \frac{\pi^2\omega^2}{16}(2j+1)^2 .\label{energyShift}
\end{equation}
Also, using a harmonic approximation for the optical lattice around its minima, results in the standard harmonic oscillator levels of the form
\begin{equation}
\epsilon_{n^\star} = 2 \sqrt{V_0}(n^\star+\frac{1}{2})
\end{equation}
where $2\sqrt{V_0}$ is the effective harmonic frequency and $n^\star = 0,~1,~2,\dots$ is the band index. The approximated total energy levels of the system are: 
\begin{align}
\nonumber
E_{n^\star}^{j}(\omega) &= \epsilon_{n^\star} + \epsilon_{shift}^{j}(\omega)\\[2pt] &= \sqrt{V_0}(2n^\star+1) + \frac{\pi^2\omega^2}{16}(2j+1)^2
\end{align}
Eq. (17) describes both the parabolic dependence on the trapping frequency and the direct connection between the $m^{th}$ energy level of each band and the states localized at the positive and negative $x_m$ minima. As shown in Fig.~\ref{fig:Harmonic_Approx} even for a deep enough optical lattice, the harmonic approximation fails to describe the higher bands directly. Nevertheless, replacing $\epsilon_n$ with the numerically calculated value (from the ED) lowest energies of each band at zero trapping frequency $\epsilon_{n^\star} = E_{n^{\star}\mathcal{N}}(\omega=0)$ and using $\epsilon_{shift}^m$ to capture the $\omega$-dependence (termed ``fitted harmonic approximation" (fHA)), leads to near perfect agreement with the exact results from the ED. Note, however, that this description does not hold for trapping frequency values where crossings with the upper bands have occurred, as there is no direct correspondence between the energy levels and specific sites.

\subsection{Optical Superlattice and Harmonic Trap}\label{sec:OSvsOmega}

We move on to the case of the optical superlattice potential and compare it with the optical lattice potential. For the considered optical superlattice systems the extension of the potential landscape starts at $x_0 = L$ and possess the optimal extension length $d_{opt}$ enabling the appearance of TESs. See Appendix~\ref{app:Optimal_d} for details. Even though $d_{opt}$ is defined in the topological configuration, we use the same value for the trivial configuration too, since in that case the HWBC (confinement) effect fades away for smaller values and the behavior remains the same for larger values of $d$.

\subsubsection{Trivial Configuration}

In the trivial configuration without the harmonic trap ($\omega = 0$), the spectrum of the optical superlattice system features a two-sub-band structure. Due to its underlying double-well cell geometry, it hosts $M=\mathcal{N}/2$ states in each sub-band, separated by a clear intra-band gap. All eigenstates are delocalized over multiple sites as the system does not support any localized (or topological edge) states. 

\begin{figure*}[ht]
\begin{center}
\begin{overpic}[width=0.49\textwidth]{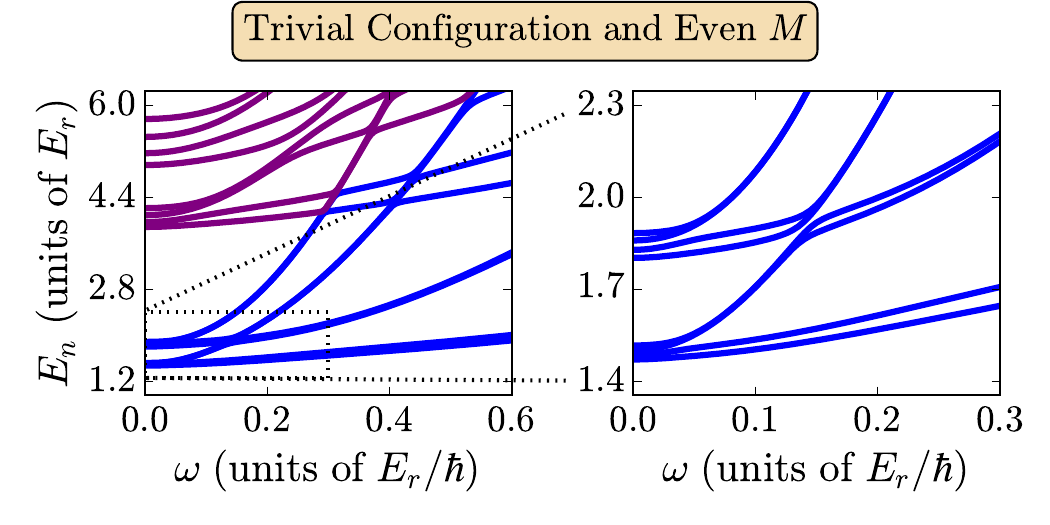}
\put(11.5,46) {\textbf{(a)}}
\end{overpic}
\begin{overpic}[width=0.49\textwidth]{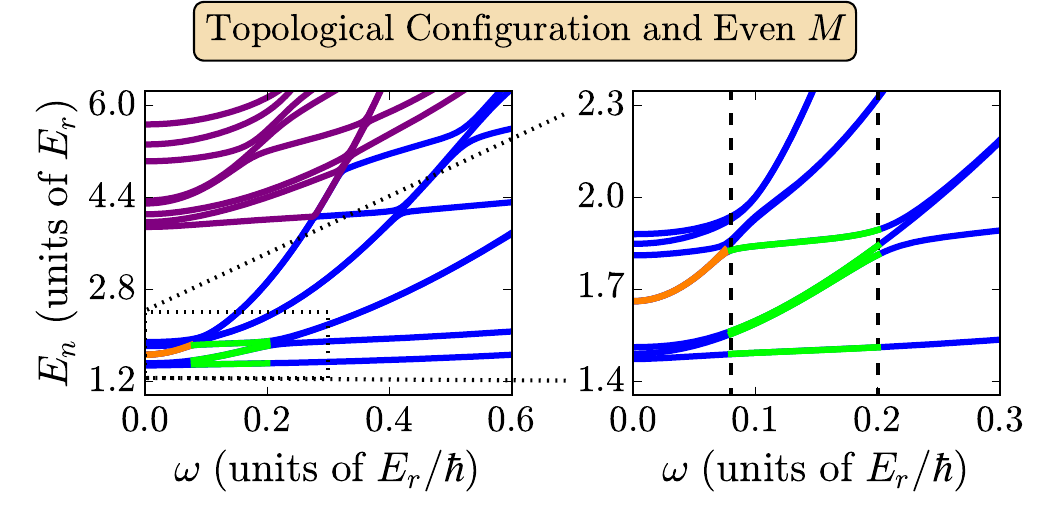}
\put(8.5,46) {\textbf{(b)}}
\end{overpic}
\begin{overpic}[width=0.49\textwidth]{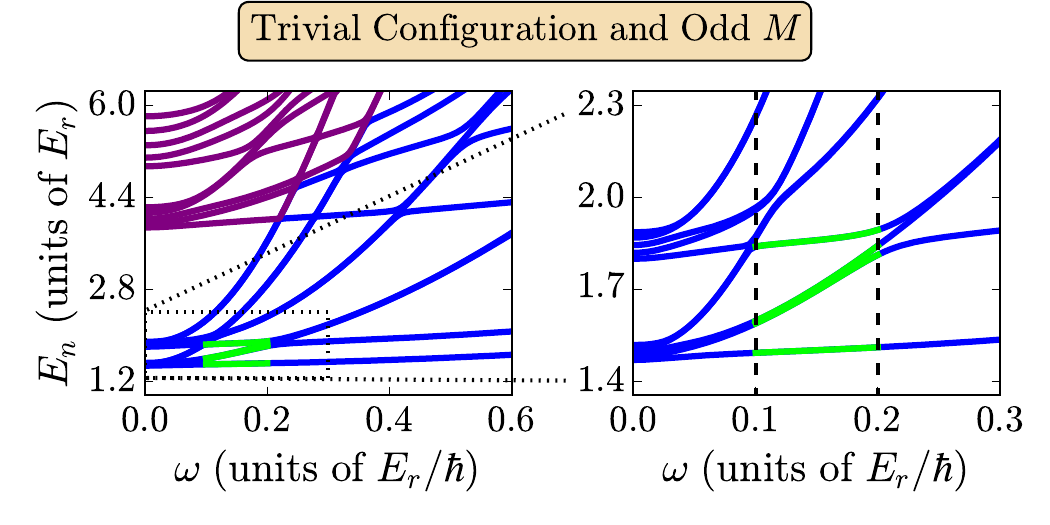}
\put(12.5,46) {\textbf{(c)}}
\end{overpic}
\begin{overpic}[width=0.49\textwidth]{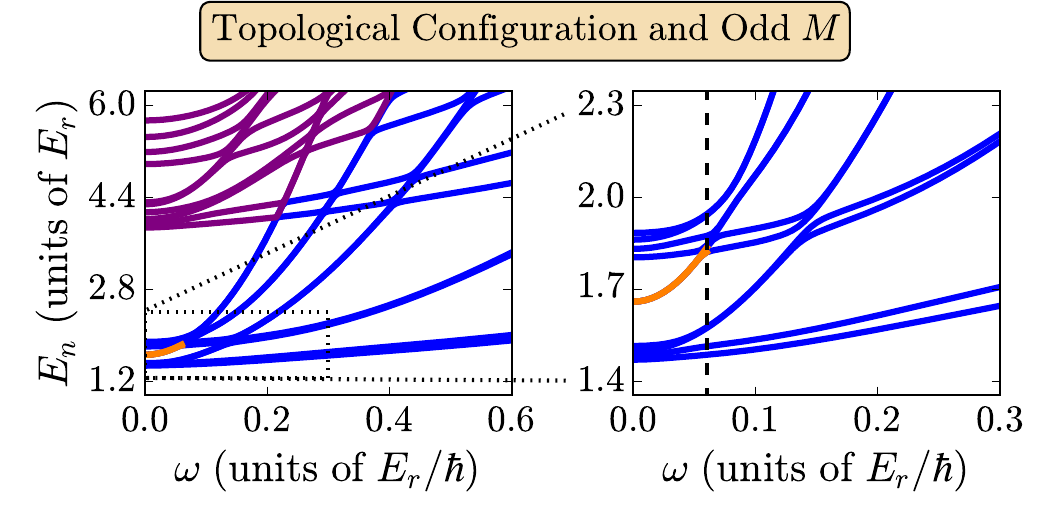}
\put(8.5,46) {\textbf{(d)}}
\end{overpic}
\begin{overpic}[width=0.98\textwidth]{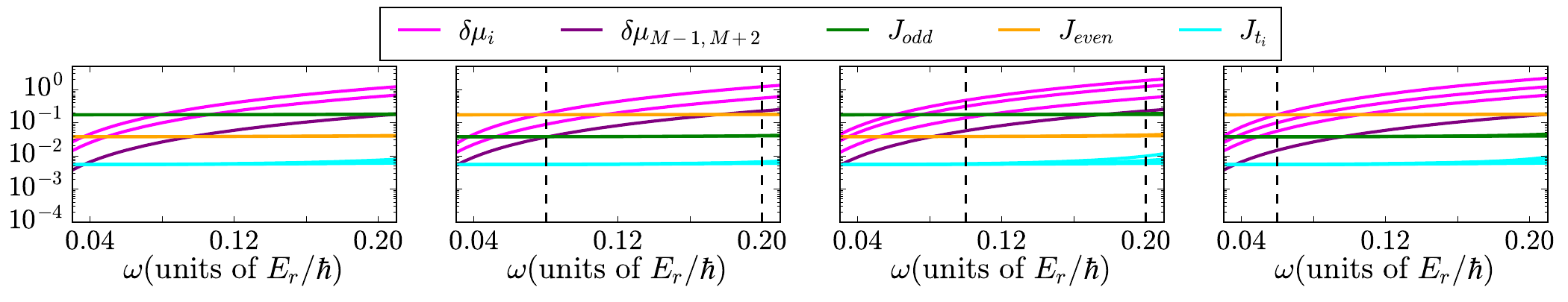}
\put(21.2,6.5) {\textbf{(e)}}
\put(45.2,6.5) {\textbf{(f)}}
\put(69.2,6.5) {\textbf{(g)}}
\put(94.8,6.5) {\textbf{(h)}}
\end{overpic}
\caption{(a-d) Energy spectra and (e-h) the on-site and hopping amplitudes of the TB approximation versus the trapping frequency $\omega$ of an extended optical superlattice potential with the addition of a harmonic trap. The optical superlattice system has $V_{high} = 5.0E_r$, $u=0.6$ and (a,b,e,f) $M=4~(\mathcal{N}=8)$ and (c,d,g,h) $M=5~(\mathcal{N}=10)$ cells (sites/minima), corresponding to even and odd number of cells, respectively. Also, it is in (a,c,e,g) the trivial and (b,d,f,h) the topological configuration. The extension starts at $x_0=L$ and has the optimal length (a,b) $d=0.36\pi/k_r$ and (c,d) $d=0.38\pi/k_r$. (a-d) The dotted rectangles highlight the low to mid $\omega$ regime we zoom in. (b-d) The orange (medium gray in grayscale) lines correspond to the energies of the topological edge states, while the light-green (light-gray in grayscale) lines correspond to the energies of the effective four-level system. (b,c,f,g) The vertical dashed lines highlight the region of $\omega$ for which the first 4 states corresponds to the minimal discrete system of $M=2$ in the topological configuration, shown in Appendix~\ref{app:MinimalSystem}. (d,h) The vertical dashed line shows the value of $\omega$ up to which the edge states survive. (e-h) The hoppings $J_{odd}$ and $J_{even}$ correspond to the $J_i$ with odd and even indices, respectively. All amplitudes of the TB approximation are in units of $E_r$.}
\label{fig:OS_Spectra_TBApprox_vs_omega}
\end{center}
\end{figure*}

\begin{figure*}[ht]
    \begin{overpic}[width=0.98\linewidth]{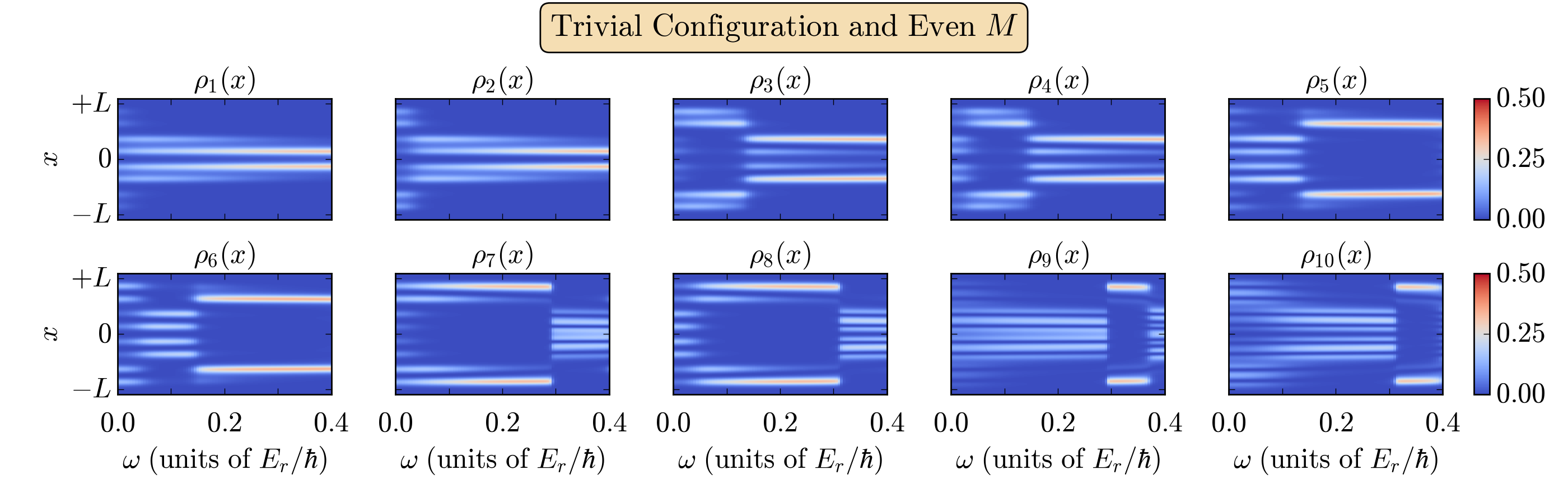}
    \put(29.5,29) {\textbf (a)}
    \end{overpic}
    \\[10pt]
    \begin{overpic}[width=0.98\linewidth]{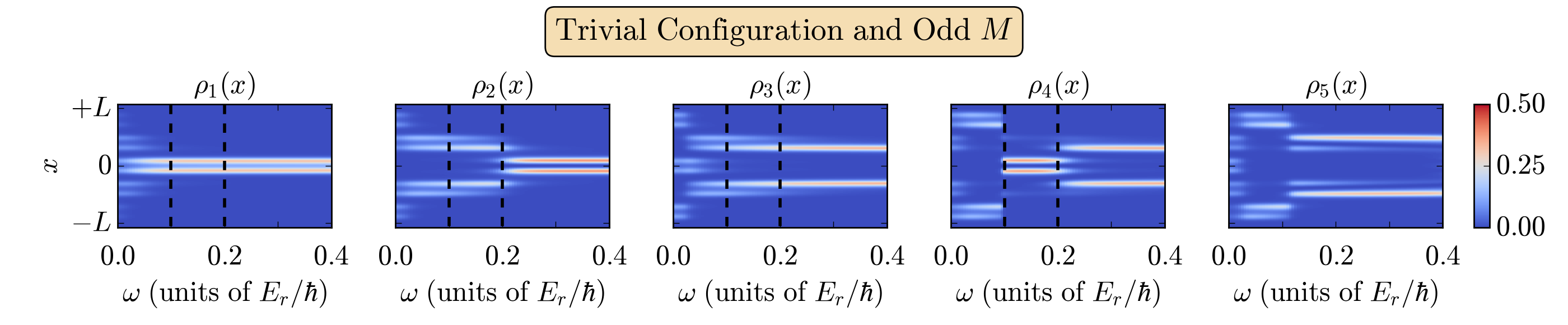}
    \put(29.5,18) {\textbf (b)}
    \end{overpic}
    \\[10pt]
    \begin{overpic}[width=0.98\linewidth]{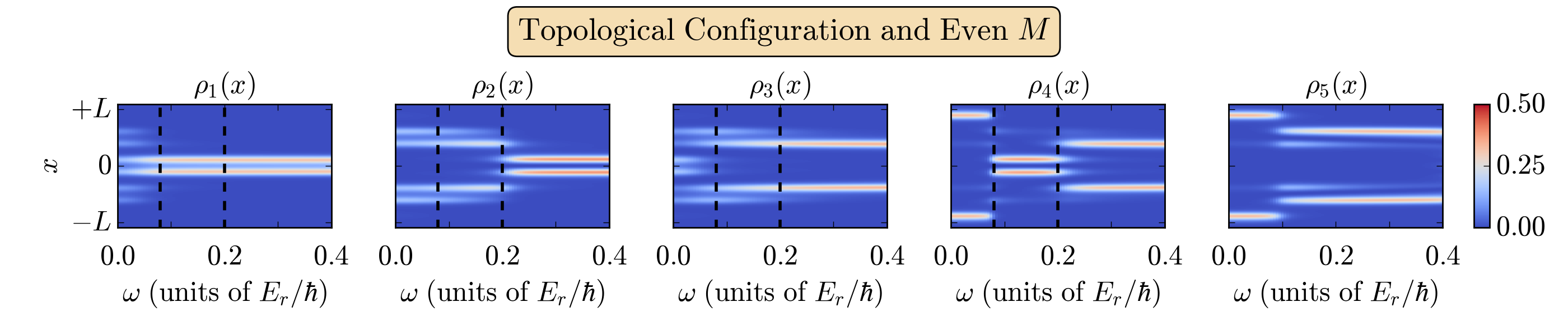}
    \put(27.5,18) {\textbf (c)}
    \end{overpic}
    \begin{overpic}[width=0.9\linewidth]{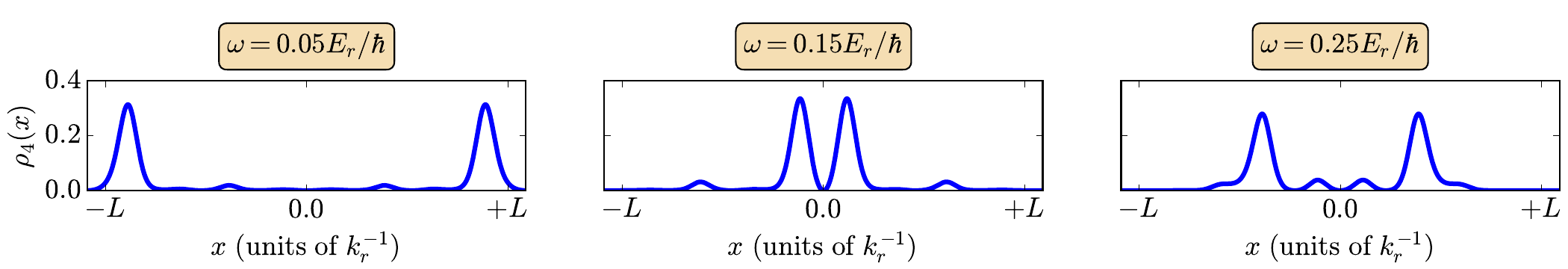}
    \put(7,13.75) {\textbf (c.1)}
    \put(39.75,13.75) {\textbf (c.2)}
    \put(72.5,13.75) {\textbf (c.3)}
    \end{overpic}
    \caption{(a-c) Spatial profiles of the densities of an extended optical superlattice system versus the trapping frequency $\omega$. The optical superlattice system has $V_{high} = 5.0E_r$, $u=0.6$ and (a,c) $M=4$ and (b) $M=5$ cells, corresponding to even and odd $M$, respectively. Also, it is in (a,b) the trivial and (c) the topological configuration. The extension starts at $x_0=L$ and has the optimal length (a,c) $d=0.36\pi/k_r$ and (b) $d=0.38\pi/k_r$. (b,c) The vertical dashed lines highlight the region of $\omega$ for which the first 4 states corresponds to an effective four-level system in the topological configuration. (c.1–c.3) Spatial profiles of the $\psi_4(x)$ density for distinct values of the trapping frequency $\omega$, each corresponding to a different trapping frequency regime, resulting in localization around different minima of the optical superlattice.}
    \label{fig:OS_States_vs_omega}
\end{figure*}

As shown in Fig.~\ref{fig:OS_Spectra_TBApprox_vs_omega}(a) for an optical superlattice with even $M$, increasing the trapping frequency causes the highest-lying states of the first band to merge into degenerate pairs, while the intra-band gap becomes filled by the highest-lying states of the lower sub-band. Similar to the case of the optical lattice potential this pairing leads to the ``quasi-classical" localization of the eigenstates, first only for the outer sites (corresponding to the highest-lying states of the first band) and eventually for all sites as $\omega$ increases more, as shown especially at the right edge of the graphs in Fig.~\ref{fig:OS_States_vs_omega}(a). Further increasing $\omega$, results in a parabolic increase of the energies of the first band until they cross with the states of the second band, exchanging their character and shifting the localized character to the higher lying states. 

On the other hand, as seen in Fig.~\ref{fig:OS_Spectra_TBApprox_vs_omega}(c), an important difference arises when the optical superlattice system in the trivial configuration has odd $M$. In this case, each sub-band hosts an odd number of states. Simultaneously, the states begin to form degenerate pairs, starting from higher and moving to lower energy levels.
This means that at each sub-band the lowest energy state will become isolated from the other states. The influence of this effect becomes readily evident after consecutive avoided crossings, leading to the point where the $\psi_4(x)$ state is isolated and remains so for a wide range of trapping frequency $\omega$. In particular, beyond this point and until they encounter an avoided crossing with the higher-energy states, the first four states remain energetically isolated from the rest and span mostly the four central sites of the optical superlattice: $M-1,M,M+1,M+2$. Those states resemble an effective four-level system, which corresponds to the minimal discrete system of $M=2$ in the topological configuration (see Appendix~\ref{app:MinimalSystem}). The spatial profiles of the densities of the first four eigenstates of the system depicted in Fig.~\ref{fig:OS_States_vs_omega}(b), confirm that they are indeed localized mostly in the four central sites. Specifically, $\psi_1(x)$ and $\psi_4(x)$ are localized at the middle sites ($M$ and $M+1$) and $\psi_2(x)$ and $\psi_3(x)$ at the corresponding edge (boundary) sites ($M-1$ and $M+2$). Nevertheless, there is some residual occupation to the nearest neighboring sites ($M-2$ and $M+3$). This is to be expected, since the effective four-level system is not completely isolated, as it would be in the presence of a sharp confining potential at its edges, such as in the hard-wall boundary scenario (discussed in Appendix~\ref{app:MinimalSystem}). This unique phenomenology persist up to a certain value of the trapping frequency, beyond which the harmonic trap term becomes dominant, leading once again to all first band states to be paired and localized. 

The above results can be explained by the behavior of the corresponding discrete system in the low to mid trapping frequency regime. As we can see in Fig.~\ref{fig:OS_Spectra_TBApprox_vs_omega}(e) and (g), there are no substantial differences for the amplitudes of the TB approximation of the optical superlattice system for the trivial configuration, whether $M$ is even or odd. The hopping amplitudes in both cases satisfy $J_{even}<J_{odd}$, since in the trivial configuration the first and last hopping amplitudes ($J_{1,\mathcal{N}-1}$) must be strong. So regardless of even or odd $M$, when we increase $\omega$ (starting from zero) the localization of the states at the outermost sites starts to appear when $\delta\mu_{1,\mathcal{N}}>J_{1,\mathcal{N}-1}$. This phenomenology is similar to the optical lattice system. However, further increasing $\omega$ results in distinct phenomenologies for the two cases of even or odd number of cells. The difference originates from the hopping amplitudes connecting the four central sites $J_{M}$ and $J_{M\pm1}$. Clearly, for even $M$, $J_{M}=J_{even}$ (weak) and $J_{M\pm1}=J_{odd}$ (strong), while for odd $M$, $J_{M}=J_{odd}$ (strong) and $J_{M\pm1}=J_{even}$ (weak). This essentially means that in the case of odd $M$, we have a topological configuration of weak-strong-weak hopping amplitudes for the central site. The on-site energies of the two corresponding edge (boundary) sites of the effective four-level system compete with the strong central hopping. Hence, the system resists the tendency for localization for a larger range of the trapping frequency, compared to the case of even $M$. On the other hand, for even $M$ the corresponding effective four-level system has a trivial configuration of strong-weak-strong hopping amplitudes and the on-site energies compete with the weak central hopping. That explains why for odd $M$ the intermediate regime of the trapping frequency where the unique phenomenology of the effective four-level system appears, coincides with the region where $J_{even}<\delta\mu_{M-1,M+2}<J_{odd}$. While, for the case of even $M$, increasing $\omega$ leads directly to the phase of the system where all the states are localized, characterized by $\delta\mu_i>J_i~ \forall i$. 

We reach the conclusion that there are only minor quantitative differences between the optical superlattice system in the trivial configuration with even $M$ and the optical lattice system. This is a strong indication of the fundamental similarities in the underlying mechanism responsible for the observed phenomenology. Namely, the key factor for the existence of localized states in both systems is the combination of the periodic potential (optical (super)lattice) and the harmonic trap. As long as $\omega$ remains below a finite value, the behavior of the first band of the continuous system can be mapped to a weakly connected one-dimensional discrete lattice with well separated on-site energies at mirror symmetric sites. For larger values of the trapping frequency the behavior of both the optical lattice and optical superlattice systems is dominated by inter-band crossings requiring the use of extended multi-band TB models. In contrast, the odd $M$ system features an intermediate trapping frequency region where the effective four-level system in the topological configuration appears. In Sec.~\ref{sec:4lvl} we explore thoroughly this unique phenomenology.

\subsubsection{Topological Configuration}
\label{subsec:TopoCon}

In the absence of the harmonic trap in the topological configuration the spectrum of the optical superlattice system, also features a two-sub-band structure, due to the underlying double-well cell geometry of the optical superlattice. However, in this case, when the system is properly extended, each sub-band hosts $\mathcal{N}/2-1$ states, while the remaining two states reside in the middle of the intra-band gap. The eigenstates of the sub-bands are delocalized over multiple sites (bulk states). The two states in the middle of the intra-band gap are localized at the outermost sites corresponding to the TESs of the SSH or eSSH models. 

As seen in Fig.~\ref{fig:OS_Spectra_TBApprox_vs_omega}(b) and (d) for relatively small but finite values of the trapping frequency ($\omega \lesssim 0.02 E_r/\hbar$), the energies of the two edge states remain in the middle of the sub-band gap for both parities.  Beyond a certain value of the trapping frequency the edge states begin to spread towards the inner part of the lattice, as the two highest-lying states of the first upper sub-band pair and become ``quasi-classically" localized at the outer-most sites. Also, in Fig.~\ref{fig:OS_States_vs_omega}(c), for the two edge states ($\psi_4(x)$ and $\psi_5(x)$) of the optical superlattice system with even $M$, we see that they remain localized at the edges, until the avoided crossing with the state of the upper sub-band occur. The edge states of the optical superlattice system with odd $M$ have the same behavior for small trapping frequency values. In Sec.~\ref{dynamics} we present the transport dynamics that can occur between the edge states of topological origin, which are mostly unaffected by the presence of the trap for small trapping frequency values. This result is particularly relevant for experimental considerations, as it indicates that the edge states can be readily observed even if the residual harmonic trap has not been completely compensated~\cite{Schneider2012BoxHubbard}.

Regarding the intermediate trapping frequency regime, in the topological configuration we observe the same phenomenology as in the trivial one, but with the roles of even and odd $M$ reversed. That is, if $M$ is even, the number of states in each sub-band ($M-1$) is odd and vice versa. This leads us directly to the conclusion that in the topological configuration, the even $M$ situation generates the mechanism for the effective four-level system to be present in the intermediate trapping frequency regime. In contrast, for odd $M$, after the suppression of TESs via avoided crossings with the upper sub-band state, we get the same behavior as for the optical superlattice with even $M$ in the trivial configuration. Firstly, these observations are supported by the energy spectra in Fig.~\ref{fig:OS_Spectra_TBApprox_vs_omega}(b) and (d). It is readily seen that, to the right of the first dashed lines in both graphs, the spectral behavior is exactly the same as in Fig.~\ref{fig:OS_Spectra_TBApprox_vs_omega}(c) and (d), respectively. Secondly, in Fig.~\ref{fig:OS_States_vs_omega}(c) is shown that, for the intermediate trapping frequency values the first four states of the optical superlattice system with even $M$ are localized mostly in the four central sites. The region is highlighted by the vertical dashed lines in the graph.

Lastly, we support our arguments once again with the results from the TB approximation. We start from Fig.~\ref{fig:OS_Spectra_TBApprox_vs_omega}(f), for the optical superlattice system with even $M$ in the topological configuration. We observe that the intermediate trapping frequency regime, where the effective four-level system can be considered, coincides with the region where the inner most on-site potentials ($\delta\mu_{M-1,M+2}$) are between the values of the $J_i$'s. In this case the inner most on-site potentials ($\delta\mu_{M-1,M+2}$) compete with the stronger even hopping amplitude $J_{M}$. On the other hand, in the case of odd $M$, the central hopping is the weak one, so the system passes right away to the ``quasi-classical" localization regime. 

Upon further increase of $\omega$, the optical superlattice system starts to behave similarly to the optical lattice and superlattice in the trivial configuration systems, since the harmonic trap term becomes dominant. 

\subsubsection{Overview of the Phenomenology of the Optical Superlattice System}

As the optical superlattice system can support four distinct cases with respect to its configuration and number of cells, we close this subsection with an overview of the available scenarios and the corresponding phenomenologies. We start from the case of a trivial configuration and even $M$. This is the optical superlattice system with the most similarities to the optical lattice system, since neither edge states of topological origin nor an effective four-level system can be supported. The localization of the states stems only from the superposition of the optical superlattice with the progressively dominant harmonic trap.
We proceed with the topological configuration with odd $M$, where with the proper extension, TESs are supported up to a certain value of the trapping frequency. Beyond that value, the system again supports only the ``quasi-classically" localized states as the optical lattice system. We continue to the optical superlattice system in the trivial configuration with odd $M$, which is one of the two cases where the effective four-level system can be supported at intermediate values of the trapping frequency. In line with the other cases, for larger $\omega$ the ``quasi-classically" localized states are supported.

Finally, we reach to the case with the richest phenomenology, namely the optical superlattice system in the topological configuration with even $M$. In this system, we can observe TESs for small trapping frequency values and proper extension of the potential landscape, and for intermediate trapping frequency values the effective four-level system. Also, as with all the other cases, for large trapping frequency we find the ``quasi-classically" localized states. 
Especially, in the case of  $M=4$ the $\psi_4(x)$ state can be a TES, or a ``bulk" state of the effective four-level system localized at the central sites, or a state localized at the middle sites, in each scenario depending on the value of the trapping frequency (see Fig.~\ref{fig:OS_States_vs_omega}(c.1-c.3)). Armed with these observations, we focus in Sec.~\ref{sec:4lvl} on the intermediate region of the optical superlattice system and provide strong evidence for the universality of the emergence of the effective four-level system.

\section{Emergence of the effective four-level system in the harmonically confined optical superlattice system}\label{sec:4lvl}

In this section, we thoroughly study the regime in which the harmonic trap is neither weak enough to be treated as a simple perturbation nor strong enough to dominate the system's behavior. In Sec.~\ref{subsec:details} we address in detail how varying the optical superlattice parameters ($V_{high}$, $u$, $M$) interplays with the trapping frequency to shape the phenomenology discussed in Sec.~\ref{subsec:TopoCon}. In this way, we illustrate the universality of the emergence of the effective four-level system in an optical superlattice system with a harmonic trap. Then, in Sec.~\ref{subsec:ExpAsp} we address two important aspects related to the experimental implementations of the system. Namely, we show why the extension of the potential landscape is not necessary for the observation of the effective four-level system. Also, we show the dependence of the observed phenomenology on the relative position of the harmonic trap with respect to the center of the optical superlattice. We close with a brief discussion on the dynamics of the system. Bellow we set the phase factor to $\phi = \pi/2$, which corresponds to the cases of the optical superlattice system that we are interested in, i.e. both the trivial and topological configuration for odd and even $M$, respectively.

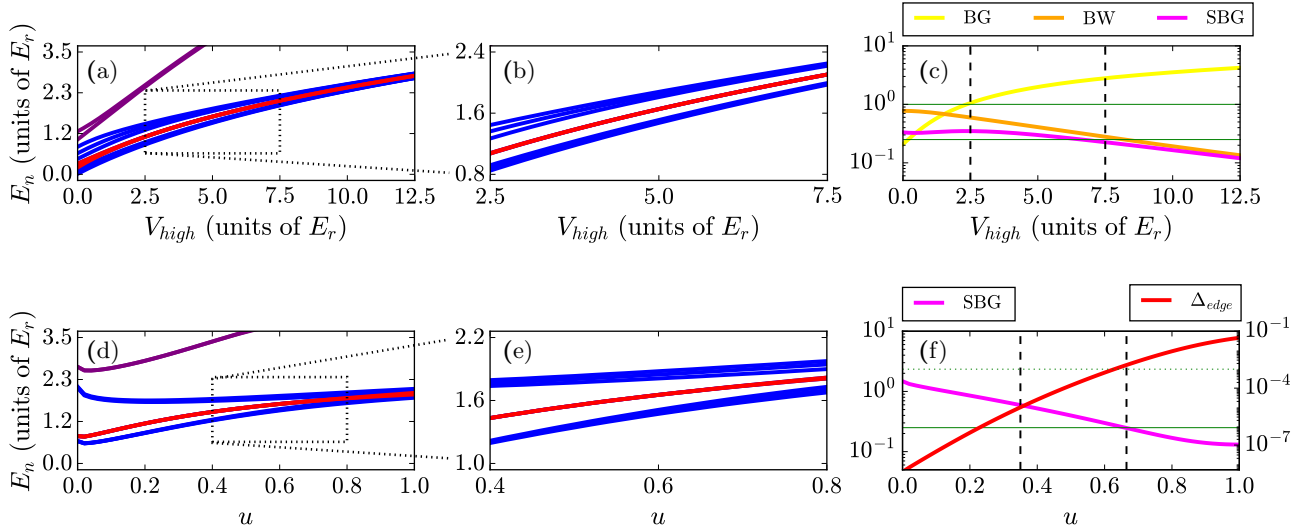
\begin{figure*}[ht]
    \centering
    \begin{overpic}[width=0.98\linewidth]{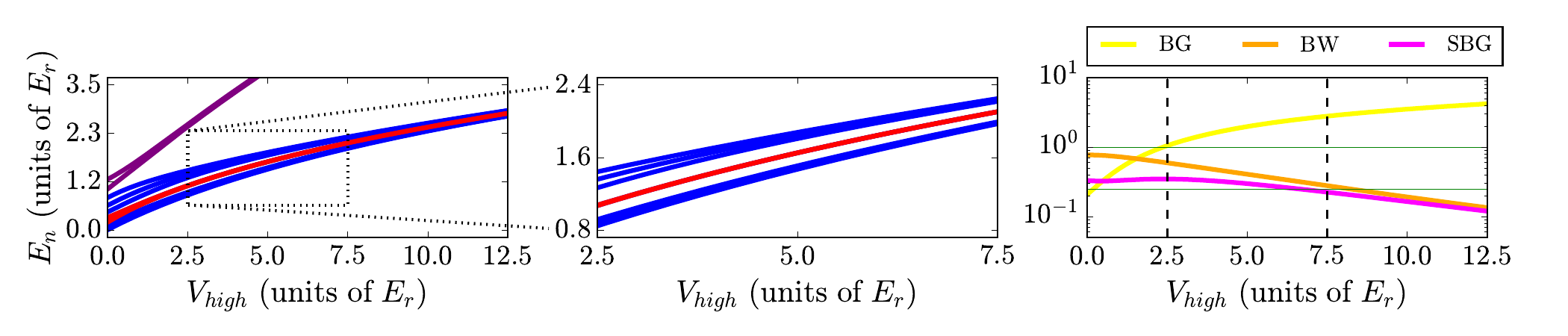}
    \put(7.5,14) {\textbf (a)}
    \put(39,14) {\textbf (b)}
    \put(70.5,14) {\textbf (c)}
    \end{overpic}
    \begin{overpic}[width=0.98\linewidth]{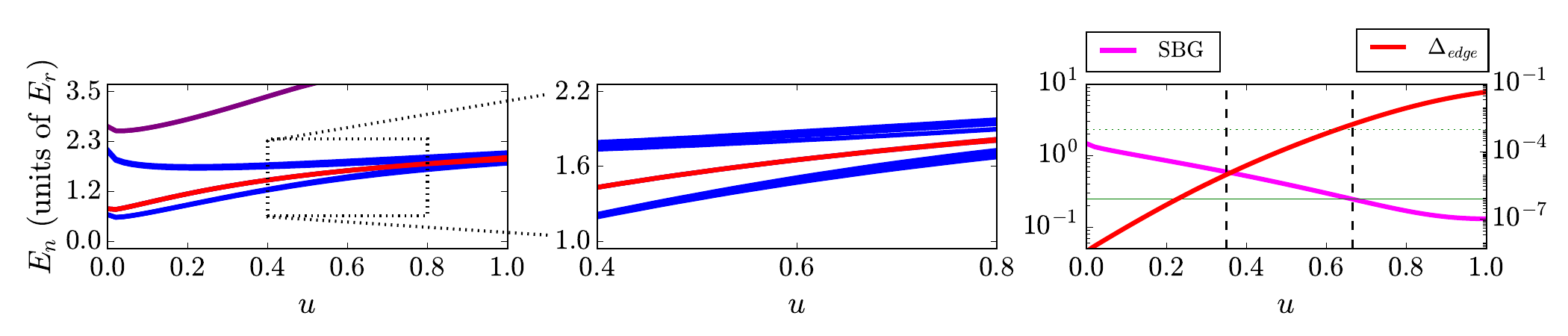}
    \put(7.5,14) {\textbf (d)}
    \put(39,14) {\textbf (e)}
    \put(70.5,14) {\textbf (f)}
    \end{overpic}
    \caption{(a) Energy spectrum versus the height of the higher barrier $V_{high}$ of an extended optical superlattice potential without the harmonic trap.  
    The dotted square highlight the $V_{high}$ regime that is zoomed in and shown in (b). (c) The band gap (BG) between the first and second bands,
    bandwidth (BW) and sub-band gap (SBG) of the first band versus $V_{high}$. (a-c) The optical superlattice system has $M=4$, $u=0.6$. (d,e) The same as (a,b), but versus the ratio of the heights of the barriers $u = V_{low}/V_{high}$. (f) The SBG and the edge state energy splitting $\Delta_{edge}$ versus $u$. (d-f) The optical superlattice system has $M=4$, $V_{high} = 5.0 E_r$. In (c) the horizontal lines shows the values $0.25E_r$ and $1E_r$, which are the boundaries we set for the BG, BW and SBG values. 
    In (c) and (f) the y-axis are in units of $E_r$ and the vertical dashed lines show (c) the $V_{high}$ and (f) the $u$ regimes we expect the effective four-level system to be observed. (a-f)
    The starting-point of the extension is $x_0 =L$ and the length is constant $d = 0.5\pi/k_r$.}
    \label{fig:OS_SpectraGaps_vs_Vhigh_and_u}
\end{figure*}

\subsection{Dependence of the Phenomenology on Optical Superlattice Parameters (\texorpdfstring{$\vb*{V_{high}},\vb*{u}, \vb*{M}$}{Vhigh, u, M})}\label{subsec:details}

We begin by presenting important characteristics of the energy spectrum of the optical superlattice system, without the harmonic trap, that depend on the parameters $V_{high}$ and $u$. First, it is well-established that $V_{high}$ controls the energy scale of the system. Most importantly though, increasing $V_{high}$ causes the bandwidths to decrease and the band gaps to increase. So there exists a trade-off between the isolation of bands, which is necessary to realize a discrete system, and the required resolution to observe the intra-band structure.
As shown in Fig.~\ref{fig:OS_SpectraGaps_vs_Vhigh_and_u}(a-c), in order for the four-level system to be more clearly observed we choose the values of $V_{high}$ in the range $2.5$ to $7.5E_r$. The lower value emerges due to a lower limit of one energy unit ($1E_r$) for the band gap between the first and second bands. While, the higher value is set due to the lower limit of $E_r/4$ for the bandwidth and sub-band gap of the first band. Of course, different choices, depending e.g. on the experimental conditions, can be made.

Let us next consider the dependence on the ratio between the lower to higher barrier heights $u$. In Fig.~\ref{fig:OS_SpectraGaps_vs_Vhigh_and_u}(d,e) we see that increasing $u$ leads to an overall energy shift upwards and at the same time to a decrease in the bandwidth. However, the ratio $u$ is mostly related to the sub-band gaps (SBG), so in Fig.~\ref{fig:OS_SpectraGaps_vs_Vhigh_and_u}(f) we show only its effect on the SBG of the first band (the $\Delta_{edge}$ quantity will be discussed in Sec.~\ref{fateofedge}). As we can see, using the same lower limit of $E_r/4$, the highest acceptable ratio is roughly $u\approx0.67$. In practice these limits have to be adjusted to the resolution of the experimental setup. In our theoretical analysis when the harmonic trap is added to the system, these values account for the emergence of a well-isolated effective four-level system in the intermediate trapping frequency regime. 

\begin{figure*}[ht]
    \centering
    \begin{overpic}[width=0.98\linewidth]{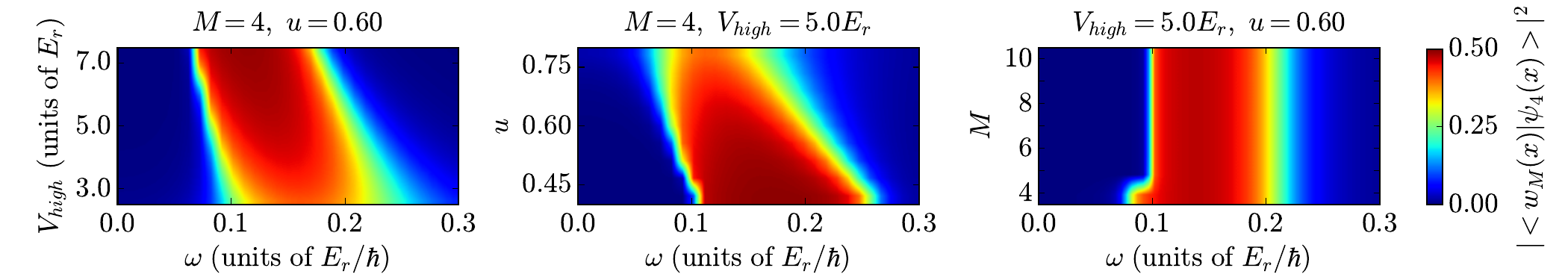}
    \put(7.4,16.) {\textbf (a)}
    \put(35,16.) {\textbf (b)}
    \put(63.6,16.) {\textbf (c)}
    \end{overpic}
    \\[10pt]
    \begin{overpic}[width=0.98\linewidth]{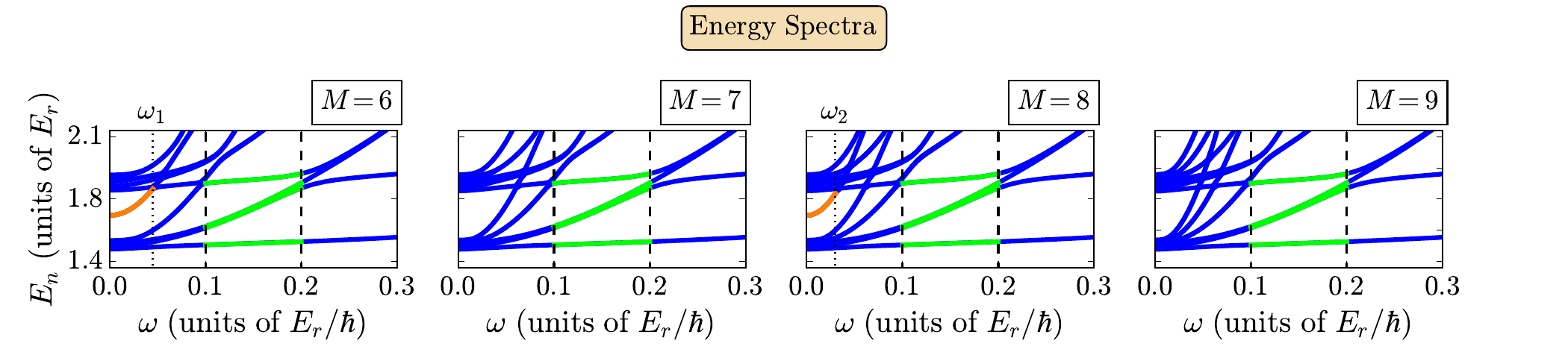}
    \put(38,21) {\textbf (d)}
    \end{overpic}
    \caption{(a-c) The squared absolute value of the overlap between the Wannier function $w_{M}(x)$ corresponding to the central left site and the 4-th eigenstate of the extended optical superlattice system with the addition of the harmonic trap. (d) Energy spectra versus the trapping frequency $\omega$ for the extended optical superlattice system under harmonic confinement. The extended optical superlattice has (a) $M=4$, $u=0.6$, (b) $M=4$, $V_{high}=5.0E_r$ and (c,d) $V_{high} = 5.0E_r$ and $u=0.6$. In (a-d) the
    starting-point of the extension is $x_0 =L$ and the extension length is (a,b) $d=0.5\pi/k_r$ and (c,d) $d=d_{opt}$ with respect to the number of cells $M$. The extended optical superlattice has phase $\phi = \pi/2$ for all values of $M$, enabling it to support the effective four-level system. $w_{M}(x)$ is calculated at $\omega = 0$. (d) For $M = 6, 8$, the vertical dotted lines highlight the trapping frequency values $\omega_1$, and $\omega_2$, beyond which the TESs cease to exist and cross into the upper band. The vertical dashed lines highlight the region of $\omega$ for which the first 4 states resemble the effective four-level system in the topological configuration.}
    \label{fig:Cverlaps_WannierM_Psi4_and_Spectra}
\end{figure*}

In Fig.~\ref{fig:Cverlaps_WannierM_Psi4_and_Spectra}(a-c) we present color plots of $\abs{\bra{w_M(x)}\ket{\psi_4(x)}}^2$, which is the squared absolute value of the overlap between the Wannier function $w_M(x)$ and the 4-th eigenstate $\psi_4(x)$ of the extended optical superlattice system with the addition of the harmonic trap. 
Essentially $w_M(x)$ refers to the central left site of the optical superlattice. Then, we use this overlap as an indicating measure for the emergence of the effective four-level system. In the absence of the harmonic trap, there is no localization mechanism that confines the $\psi_4(x)$ state solely to the central sites of the system. We consider the effective four-level system to emerge and mimic the minimal system of four sites in the topological configuration (shown in Appendix~\ref{app:MinimalSystem}), when $|\bra{w_M(x)}\ket{\psi_4(x)}|^2$ take its maximal values. Of course, due to the parity symmetry of the system, it holds that $|\bra{w_M(x)}\ket{\psi_4(x)}|^2=|\bra{w_{M+1}(x)}\ket{\psi(x)}|^2$, so without loss of generality, we show in our graphs only the $|\bra{w_M(x)}\ket{\psi_4(x)}|^2$.
We mention that in our results we use $w_M(x)$ calculated at $\omega=0$. Although, the Wannier functions in principle depend on $\omega$, they are practically unaffected in the low to intermediate trapping frequency regime, i.e. as long as the band structure persists. 

As we can see in Fig~\ref{fig:Cverlaps_WannierM_Psi4_and_Spectra}(a), for an extended optical superlattice system of $M=4$, $u=0.6$ and $V_{high}\in[2.5,7,5]$, increasing $V_{high}$ narrows down and shifts to slightly smaller values of $\omega$ the region where the effective four-level system can be observed. At the same time, in Fig.~\ref{fig:Cverlaps_WannierM_Psi4_and_Spectra}(b) we observe the inverse behavior for increasing $u$ for the same number of cells ($M=4$) and $V_{high}=5.0E_r$. This is related to and expected from the behavior of the BW and SBG of the first band of the system with respect to $V_{high}$ and $u$. For both systems in Fig.~\ref{fig:Cverlaps_WannierM_Psi4_and_Spectra}(a,b) we consider the extension at the starting-point $x_0=L$ and a constant length $d = 0.5\pi/k_r$. 

Finally, in Fig.~\ref{fig:Cverlaps_WannierM_Psi4_and_Spectra}(c) we observe that for the case of an extended optical superlattice system with $V_{high}=5.0E_r$ and $u=0.6$, the squared absolute overlap $\abs{\bra{w_M(x)}\ket{\psi_4(x)}}^2$ remains essentially constant for $M=5,6,..,10$. Specifically, as the number of cells $M$ is increased ($\phi = \pi/2$), the system supports the effective four-level system in the same trapping frequency regime for all $M>4$. 
This is further highlighted by the energy spectra shown in Fig.~\ref{fig:Cverlaps_WannierM_Psi4_and_Spectra}(d), where the same spectral behavior is observed for the first four states of the system for all $M>4$ once the intermediate trapping frequency regime is reached, as indicated by the first dashed line in the graphs. In the case of $M=4$, the only difference is that $\psi_4(x)$ is the first state to participate in the avoided crossing with the upper band. This leads to the effective four-level system emerging at a lower trapping frequency threshold in the $M=4$ system, as can be seen by the comparison of Fig.~\ref{fig:Cverlaps_WannierM_Psi4_and_Spectra}(d) with Fig.~\ref{fig:OS_Spectra_TBApprox_vs_omega}(b). 
We would also like to mention that, depending on the system size, multiple states can successively take on the role of the energetically isolated state, each remaining unpaired and separated by finite energy gaps from its neighboring states. However, we find that the $\psi_4(x)$ state tends to be the one that remains energetically isolated for the largest range of trapping frequencies and exhibits the higher degree of localization at the inner-most wells, regardless of the system size as shown in Fig.~\ref{fig:Cverlaps_WannierM_Psi4_and_Spectra}(d).
This appears to be due to the smaller energy shift experienced by the sites closer to the origin from the harmonic trap as shown in Eq.~\eqref{energyShift}.

We conclude that the underlying mechanism related to the emergence of the effective four-level system is unaffected by the system size and also is robust with respect to variations of the values of the $V_{high}$ and $u$ parameters. Lastly, we mention that the specific characteristics of the effective system are strongly related to the profiles of the interplaying potentials. If the harmonic trap is changed to a different confining potential, effective systems with different number of states could emerge. Similarly, changing the optical superlattice to an optical lattice with three or four distinct barrier heights could lead to a different type of effective model.

\subsection{Relevant extension of the superlattice setup}
\label{subsec:ExpAsp}

In the following, we present two important aspects relevant to the superlattice setup in general and also for experimental implementation. We also further explore the effective four-level system. In Sec.~\ref{subsubsec:noext} we examine the case where no extension is considered to the system. In this way, no additional tuning regarding the extension of the boundaries is required. In Sec.~\ref{subsubsec:relativepos} we address the presence of a spatial offset between the centers of the optical superlattice and the harmonic trap. Lastly, in Sec.~\ref{subsubsec:4lvldynamics} we briefly illustrate the difference in the time evolution of an initial state $\Psi(x,t=0) = w_4(x)$ for vanishing and finite trapping frequency values. We consider this as a direct way to observe the emergence of the effective four-level system via the dynamics of the optical superlattice system.

\subsubsection{Harmonically confined optical superlattice system without extension}
\label{subsubsec:noext}

The steep confinement imposed by the HWBC on the optical superlattice leads to a substantial discrepancy in the TESs characteristics of the system, namely between what is expected from the corresponding SSH or eSSH models and the results of the ED. To address this, the linear extension of the potential domain was introduced to the optical superlattice system in \cite{Katsaris2024}. Specifically, modifying the HWBC in this way is necessary for the TESs to be observed in the optical superlattice, since otherwise the boundary conditions force the states of the system to sharply vanish at its edges. On the other hand side, the phenomenology of the effective four-level system is related to the central sites of the system where the effect of the boundary conditions is minimal. Therefore, even though the linear extension was previously considered essential, we explore in the following whether it is also necessary for the observation of the effective four-level system. Of course, we bear in mind that the finite system boundaries can always play a decisive role for the observed phenomenology. 
\begin{figure}
\begin{center}
\begin{overpic}[width=0.48\textwidth]{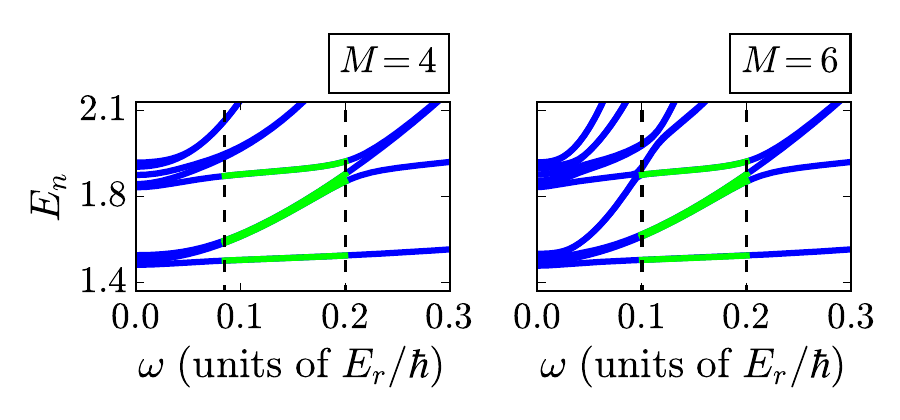}
\put(24.5,38.6) {\textbf{(a.1)}}
\put(69.3,38.6) {\textbf{(a.2)}}
\end{overpic}
\begin{overpic}[width=0.48\textwidth]{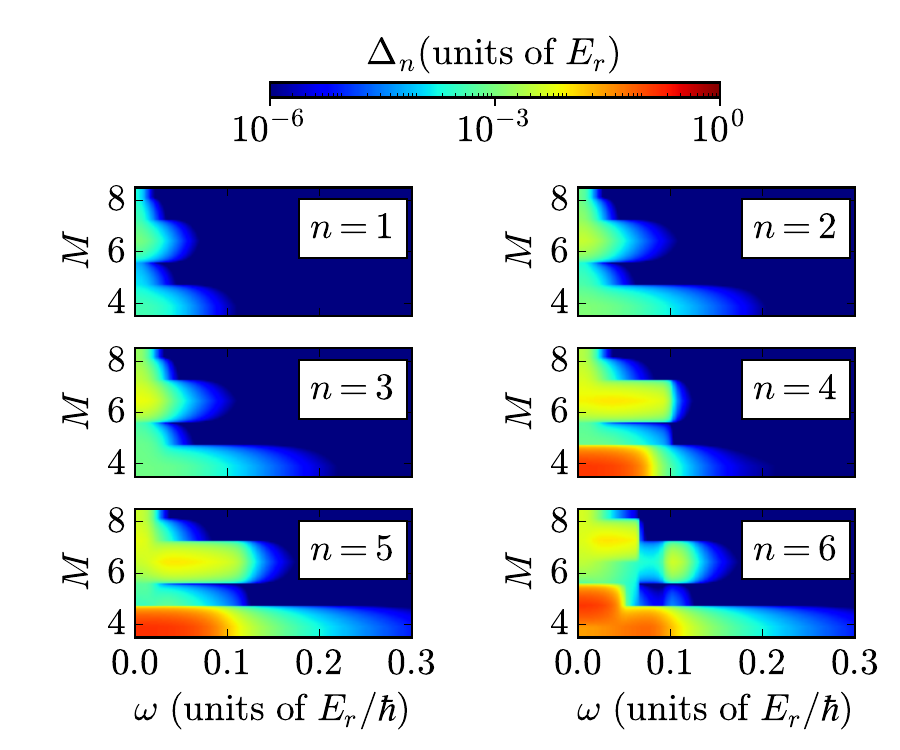}
\put(30.5,76.4) {\textbf{(b)}}
\end{overpic}
\\[10pt]
\begin{overpic}[width=0.48\textwidth]{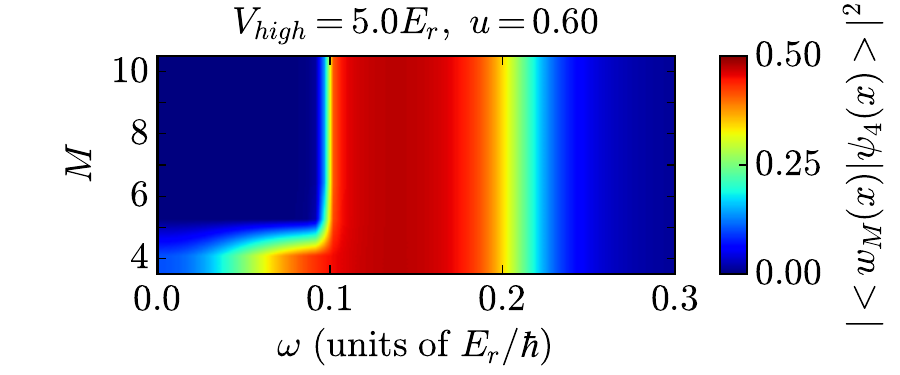}
\put(17.8,38.3) {\textbf{(c)}}
\end{overpic}
\caption{(a.1,a.2) Energy spectra versus the trapping frequency $\omega$ for the optical superlattice system without extension under harmonic confinement with $V_{high} = 5.0E_r$, $u=0.6$ and (a.1) $M=4$, (a.2) $M=6$. The vertical dashed lines indicate the trapping frequency regime where the effective four-level system in the topological configuration appears in the system. (b) The energy-level differences $\Delta_n$ between the first six eigenvalues of the optimally extended optical superlattice and the optical superlattice without extension systems versus the trapping frequency $\omega$. (c) The squared absolute value of the overlap between the Wannier function $w_{M}(x)$ corresponding to the central left site/minima and the 4-th eigenstate of the optical superlattice system without extension under harmonic confinement versus trapping frequency $\omega$.}
\label{fig:NoExtResults}
\end{center}
\end{figure}

In Fig.~\ref{fig:NoExtResults}(a.1) and (a.2) we present the energy spectrum of a harmonically confined optical superlattice system without extension in the topological configuration for $M=4$ and $M=6$ number of cells, respectively. We see that for very small trapping frequency values, the topological edge states of the system can not be observed, as expected. However, as we increase the trapping frequency the spectral behavior starts to resemble what we have discussed above. Specifically, this can become evident by the direct comparison of Fig.~\ref{fig:NoExtResults}(a.1) and (a.2) with Fig.~\ref{fig:OS_Spectra_TBApprox_vs_omega}(b) and Fig.~\ref{fig:Cverlaps_WannierM_Psi4_and_Spectra}(d) ($M=6$), respectively.

More concretely, in Fig.~\ref{fig:NoExtResults}(b) we show the differences between the first six eigenvalues of the optical superlattice system with and without optimal extension, plotted as a function of $M$ and the trapping frequency $\omega$. Specifically, for each pair $(\omega,M)$ we solve the corresponding eigenvalue problem both for the optimal extension length $d=d_{opt}$ and for no extension $d=0$, and denote the corresponding energy spectra as $E_{n,d_{opt}}(\omega,M)$ and $E_{n,0}(\omega,M)$, respectively. We then define the energy-level difference: 
\begin{equation}
\Delta_n(\omega, M) = \abs{E_{n,d_{opt}}(\omega,M)-E_{n,0}(\omega,M)}
\end{equation}
where $n$ is the energy level index. In each color map in Fig.~\ref{fig:NoExtResults}(b) $\Delta_n(\omega, M)$ is represented by the color scale, illustrating how the discrepancy between the two spectra varies with $M$ and $\omega$. As shown, increasing $\omega$ leads to an almost identical spectral behavior, as highlighted by the overall reduction of $\Delta_n$ across all energy levels. In particular, for the specific choice of parameters $V_{high} = 5.0E_r$ and $u=0.6$, the differences are reduced to values $10^{-6}-10^{-3}E_r$ for trapping frequency values of $\omega\approx 0.1E_r/\hbar$. In this analysis, significant discrepancies are observed for $M=4$ and particularly for the eigenvalues of the $\psi_4(x)$ and $\psi_5(x)$ eigenstates. This is to be expected, since in that case, these are the edge states of the system, so the deviations express exactly the impact of the extension to the system. Additionally, we observe that as we increase the number of cells $M$, the differences of the eigenvalues do not follow a specific pattern. This stems from the fact that the optimal extension is based on the restoration of the topological edge states (Appendix~\ref{app:Optimal_d}), implying that the other energy levels are not expected to behave uniformly when the effect of extension is considered.

In order to ensure the emergence of the effective four-level system in the optical superlattice system without extension, except the spectral similarities, we also calculate once again the overlap of the $M$-th Wannier function $w_M(x)$ and the $\psi_4(x)$ eigenstate.
Particularly, in Fig.~\ref{fig:NoExtResults}(c) we see that the only distinction with the previous results (Fig.~\ref{fig:Cverlaps_WannierM_Psi4_and_Spectra}(c)), appears once again for the case of $M=4$. Essentially, we find that the system enters the phase supporting the effective four-level structure already at 
$\omega \approx 0.05\,E_r/\hbar$, which is roughly half the value required in the case of optimal extension. At $\omega=0$, without the extension, the energies of the topological edge states are shifted upwards and in essence merged with the upper band. The $\psi_5(x)$ and $\psi_6(x)$ eigenstates pair, while the $\psi_4(x)$ eigenstate isolates energetically once the harmonic trap is applied. This leads to an enlargement of the trapping frequency range for which the system supports the effective four-level system.

We reach to the conclusion that regarding the emergence of the effective four-level system, the addition of an extension to the optical superlattice system is not necessary. The boundaries of the system are not affecting this phenomenology, except for the $M=4$ case. Especially, when we are not interested in the topological edge states that may be supported for small but finite values of $\omega$, it is better to not consider an extension, due to the enlarged trapping frequency regime that makes easier the observation of the effective four-level system.

\subsubsection{Harmonically confined optical superlattice system with center displacement}
\label{subsubsec:relativepos}

\begin{figure}[!h]
\begin{center}
\begin{overpic}[width=0.48\textwidth]{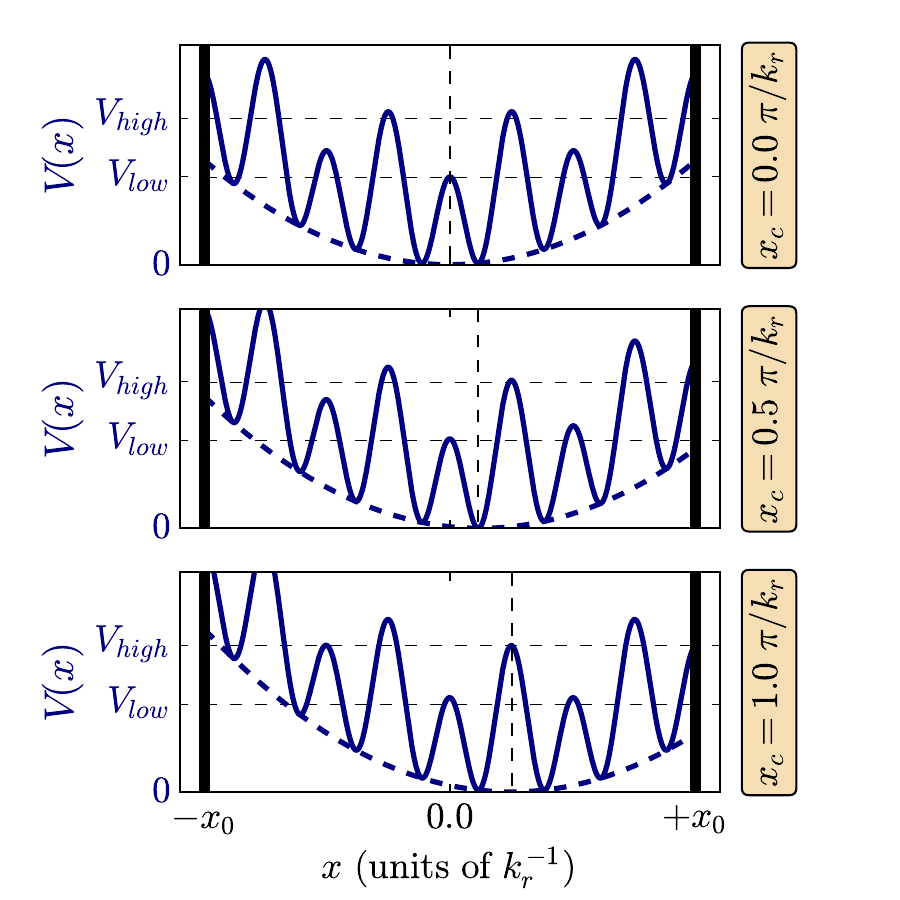}
\put(24,72.5) {\textbf (a)}
\put(24,43) {\textbf (b)}
\put(24,14) {\textbf (c)}
\end{overpic}
\caption{Schematics of the optical superlattice system without extension, superimposed with a harmonic trap, where their centers are displaced by a distance of (a) $x_c=0.0\pi/k_r$, (b) $x_c=0.5\pi/k_r$ and (c) $x_c = 1.0\pi/k_r$. $\pi/k_r$ corresponds to the distance of two consecutive local maxima. $V(x)$ are in units of $E_r$.}
\label{fig:Schematics_HarmonicDis}
\end{center}
\end{figure}

So far in our analysis we have considered the centers of the two superimposed potential landscapes $V_{lat}(x)$ and $V_{HT}(x)$ to be perfectly aligned. Such an alignment 
requires an experimental control over the trap center position and the 
lattice position well in the sub-wavelength regime, i.e. an 
interferometric stability. This can be realized by referencing both to 
an imaging lens \cite{Su_2025}. In order to account for the decentering aspect we introduce a center displacement on the potential term resembling the harmonic trap of the form:
\begin{equation}
\tilde V_{HT}(x) = \frac{1}{2}m\omega^2(x-x_c)^2 
\end{equation}
where $x_c$ is the parameter controlling the position of the trap's center. Fig.~\ref{fig:Schematics_HarmonicDis} shows schematics of the harmonically confined optical superlattice system when $x_c$ is (a) zero, (b) $\pi/(2k_r)$ and (c) $\pi/k_r$. We use $\pi/k_r$ since it is the corresponding distance of of two consecutive local maxima of the optical superlattice. 

\begin{figure}
\begin{center}
\begin{overpic}[width=0.48\textwidth]{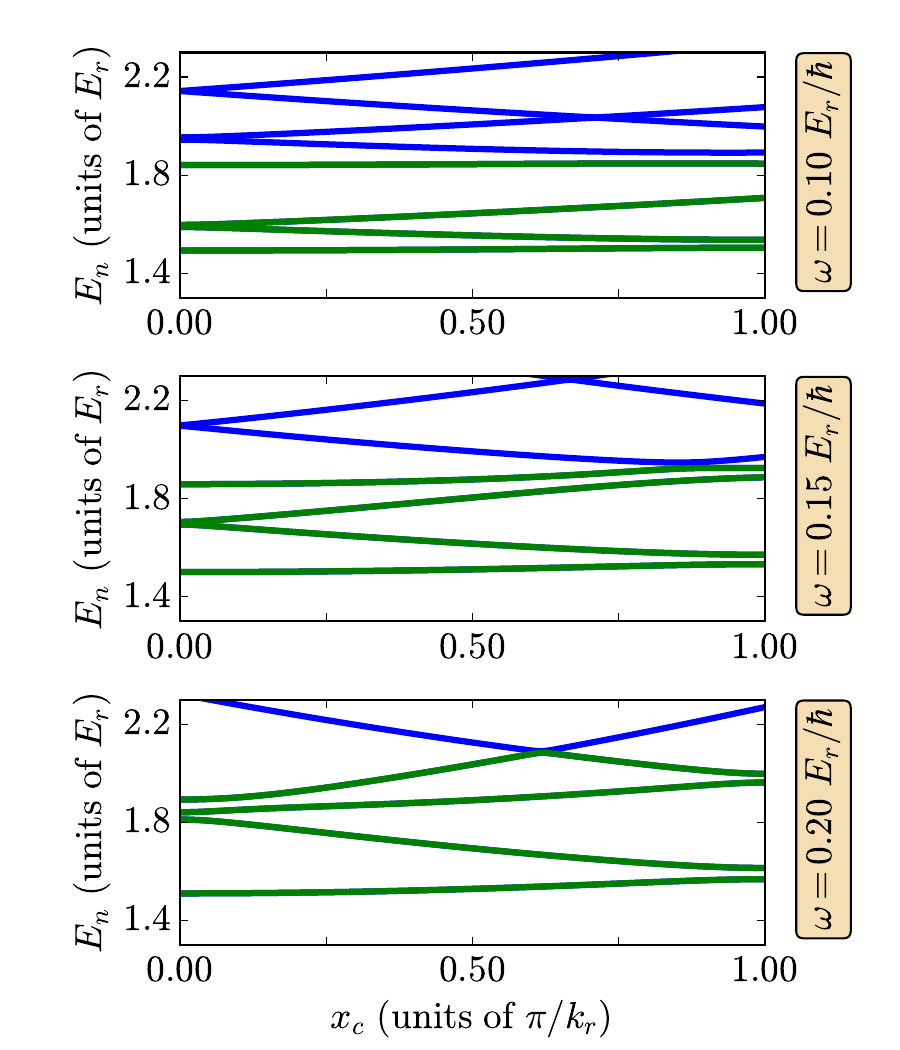}
\put(76,96.7) {\textbf (a)}
\put(76,65.7) {\textbf (b)}
\put(76,35.2) {\textbf (c)}
\end{overpic}
\begin{overpic}[width=0.48\textwidth]{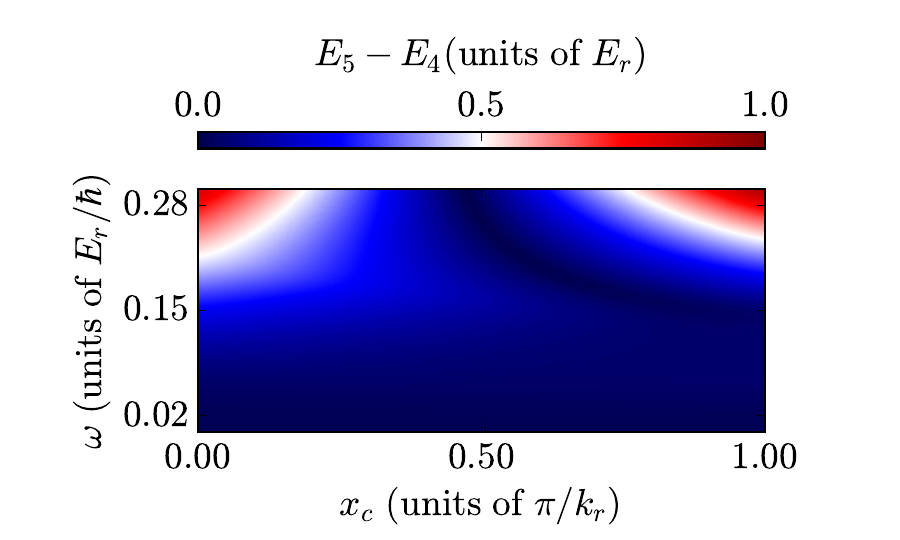}
\put(27,53) {\text (d)}
\end{overpic}
\caption{(a-c) Energy spectra versus the harmonic trap center displacement $x_c$ for the optical superlattice system without extension under harmonic confinement with $V_{high} = 5.0E_r$, $u=0.6$, $M=4$ and (a) $\omega = 0.10 E_r/\hbar$, (b) $\omega = 0.15 E_r/\hbar$ and (c) $\omega = 0.20 E_r/\hbar$. (d) The energy gap between the fifth and fourth eigenvalues system versus the trapping frequency $\omega$ and $x_c$.}
\label{fig:Spectra_DE54_HarmonicDis}
\end{center}
\end{figure}

At this point in our analysis we have established two 
ways to determine whether the four lowest-lying eigenstates represent an effective isolated four-level system in the topological configuration: by inspecting the energy levels and the spatial profile of the eigenstates. 
We follow this procedure, starting from Fig.~\ref{fig:Spectra_DE54_HarmonicDis}(a-c) where we show the energy spectra for different values of the trapping frequency $\omega$ as a function of $x_c$. 
For trapping frequency values $\omega \gtrsim 0.125 E_r/\hbar$ and for small values of $x_c$ ($\lessapprox\pi/(2k_r)$), the first four eigenvalues remain isolated from the remainder of the spectrum, forming their own band. Upon further increasing $x_c$ for constant $\omega$, we observe that the two lowest lying eigenvalues begin to pair while the energies of the other two eigenstates start to increase and eventually cross the energies of the higher-lying eigenstates. 

This behavior is best described in Fig.~\ref{fig:Spectra_DE54_HarmonicDis}(d), where we present the energy gap between the fifth and fourth eigenvalues as a function of the trapping frequency $\omega$ and the position of harmonic trap's center $x_c$. 
Evidently, when $\omega$ is sufficiently large, the system supports the effective four-level system in the regime of small $x_c$. Specifically, we establish a criterion of $E_5-E_4>0.1 E_r$, meaning that the gap is at least one tenth of the energy scale unit. Above this threshold, the first four eigenvalues are well-separated from the rest. For our specific choice of parameters ($V_{high}=5.0E_r$, $u=0.6$ and $M=4$) this approximately occurs for $\omega>0.1 E_r/\hbar$ and $x_c<0.5\pi/k_r$. 

Interestingly, there is a particular set of values for $\omega$ and $x_c$ that lead to an avoided crossing of the $E_4$ and $E_5$ eigenvalues. 
These pairs can be considered  transition points, indicating a change in the global minimum of the system. 
In particular, when $\omega$ and $x_c$ are combined such that either a large $\omega$ pairs with a small $x_c$, or a smaller $\omega$ with a larger $x_c$, the system’s effective center shifts from the $M$-th to the $(M+1)$-th site of the optical superlattice. 
In this parameter regime, the system does not support the effective four-level system, since the ``new'' central sites does not correspond to the minimal system of $M=2$ in the topological configuration. Hence, 
when this transition has occurred the system is expected to exhibit the same phenomenology as the optical lattice or superlattice with $\phi=0$ upon further increasing $\omega$.
\begin{figure}[!h]
\begin{center}
\begin{overpic}[width=0.48\textwidth]{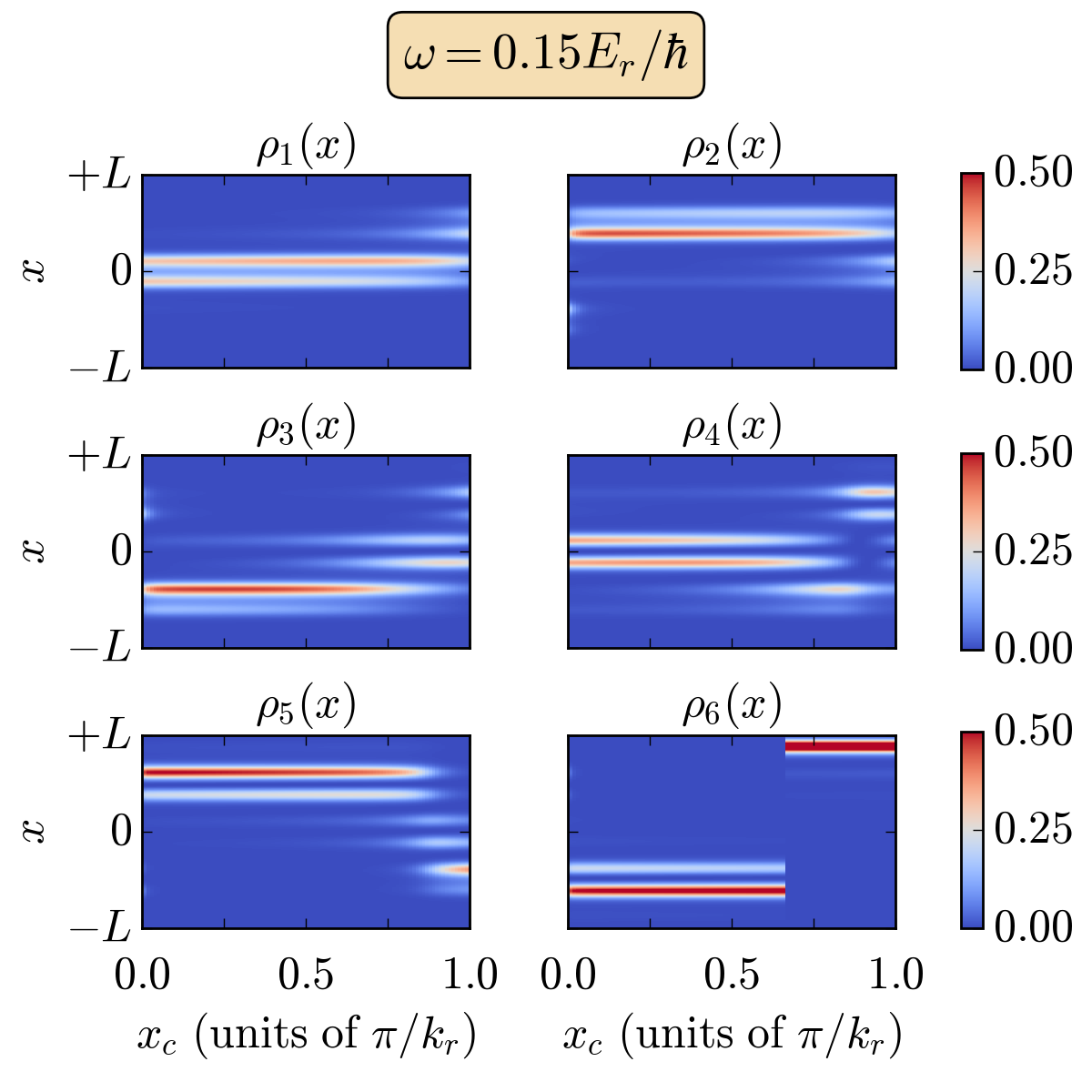}
\end{overpic}
\caption{Spatial profiles of the densities of an optical superlattice system 
without extension versus the harmonic trap center displacement $x_c$. The optical superlattice system has $V_{high} = 5.0E_r$, $u=0.6$ and $M=4$.}
\label{fig:States_HarmonicDis}
\end{center}
\end{figure}

In Fig.~\ref{fig:States_HarmonicDis} we show spatial profiles of the densities of the harmonically confined optical superlattice system 
as a function of $x_c$ and constant $\omega$. Once again, the first four eigenstates are mostly localized at the four central sites of the optical superlattice system $(M-1,M,M+1,M+2)$, before the central transition occurs at $x_c\approx0.65\pi/k_r$. 
In detail, we see that $\psi_1(x)$ and $\psi_4(x)$ are the only ones localized at the two central sites, which is a strong indication that the system again supports the effective four-level system in the topological configuration. 
However, an important feature has been added to the system due to the displacement of the harmonic trap's center: 
The mirror (parity) symmetry of the system is explicitly broken. 
This impacts the profile of $\rho_2(x)$ and $\rho_3(x)$, and $\rho_5(x)$ and $\rho_6(x)$ where the states are localized approximately at one site, as opposed to the previously observed paired occupation of lattice sites.
Most importantly, once the global minimum has been effectively shifted, the localization of $\rho_1(x)$ and $\rho_4(x)$ at the central sites begins to decrease and eventually vanishes, since the ``new" central sites no longer resemble an
effective four-level system in the topological configuration. 

\subsubsection{Effective four-level dynamics of the optical superlattice system}
\label{subsubsec:4lvldynamics}

We complete our analysis of the effective four-level system by presenting illustrative examples of the dynamics in the harmonically confined optical superlattice system. We use optical superlattice parameters $V_{high} =5.0E_r$, $u=0.6$ and $M=4$ cells. 
Three cases are considered: (i) perfect alignment with the harmonic trap and added optimal extension, (ii) perfect alignment without an extension, and (iii) misalignment without an extension. In all cases, we use as initial condition the Wannier function $w_4(x)$, calculated at zero trapping frequency, which is the equivalent to a single-site excitation of the discrete system. The key idea is that, when the system supports the effective four-level system in the topological configuration, loading one of the two central sites of the system leads to a time evolution confined to these sites, with no access to the outer sites. On the other hand, when the system has a small trapping frequency and does not support this feature, the time-evolved state spreads across the entire lattice over time. We choose the representative values for the trapping frequency regime at $\omega = 0.01~E_r/\hbar$ where no effective four-level system has been formed, and $\omega = 0.15~E_r/\hbar$ corresponding to the ``central" value of the four-level system regime.
\begin{figure}[!h]
\begin{center}
\begin{overpic}[width=0.48\textwidth]{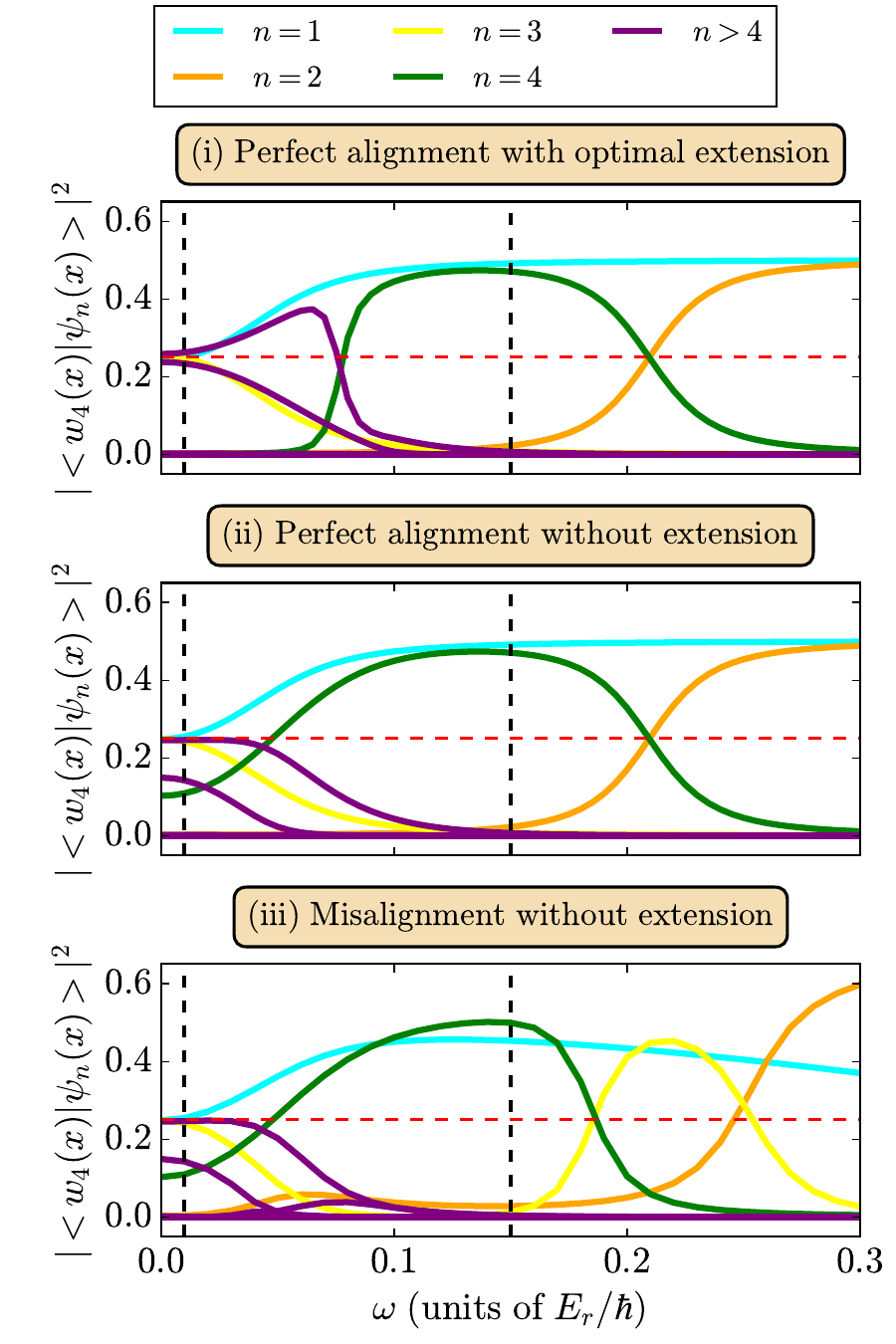}
\end{overpic}
\caption{The squared absolute value of the overlaps between the Wannier function $w_{4}(x)$ corresponding to the central left site and the first twelve eigenstates of the optical superlattice system under harmonic confinement vs the trapping frequency $\omega$. The optical superlattice system has $V_{high}=5.0 E_r$, $u=0.6$ and $M=4$ and $\phi = \pi/2$. In (i) we use as starting-point of the extension $x_0=L$ and the optimal extension length $d=d_{opt}=0.36\pi/k_r$. In (ii) and (iii) there is no extension. In (i) and (ii) the center of the harmonic trap is perfectly aligned with the center of the optical superlattice system. In (iii) there is a center displacement of $x_c = 0.25\pi/k_r$. In all graphs, the vertical dashed lines indicate the representative trapping frequency values that determine whether the system supports an effective four-level structure. The horizontal dashed line marks the 
$1/4$ squared absolute value overlap threshold.}
\label{fig:4lvlDynamics_WannierPsi_n}
\end{center}
\end{figure}
\begin{figure}
\begin{center}
\begin{overpic}[width=0.48\textwidth]{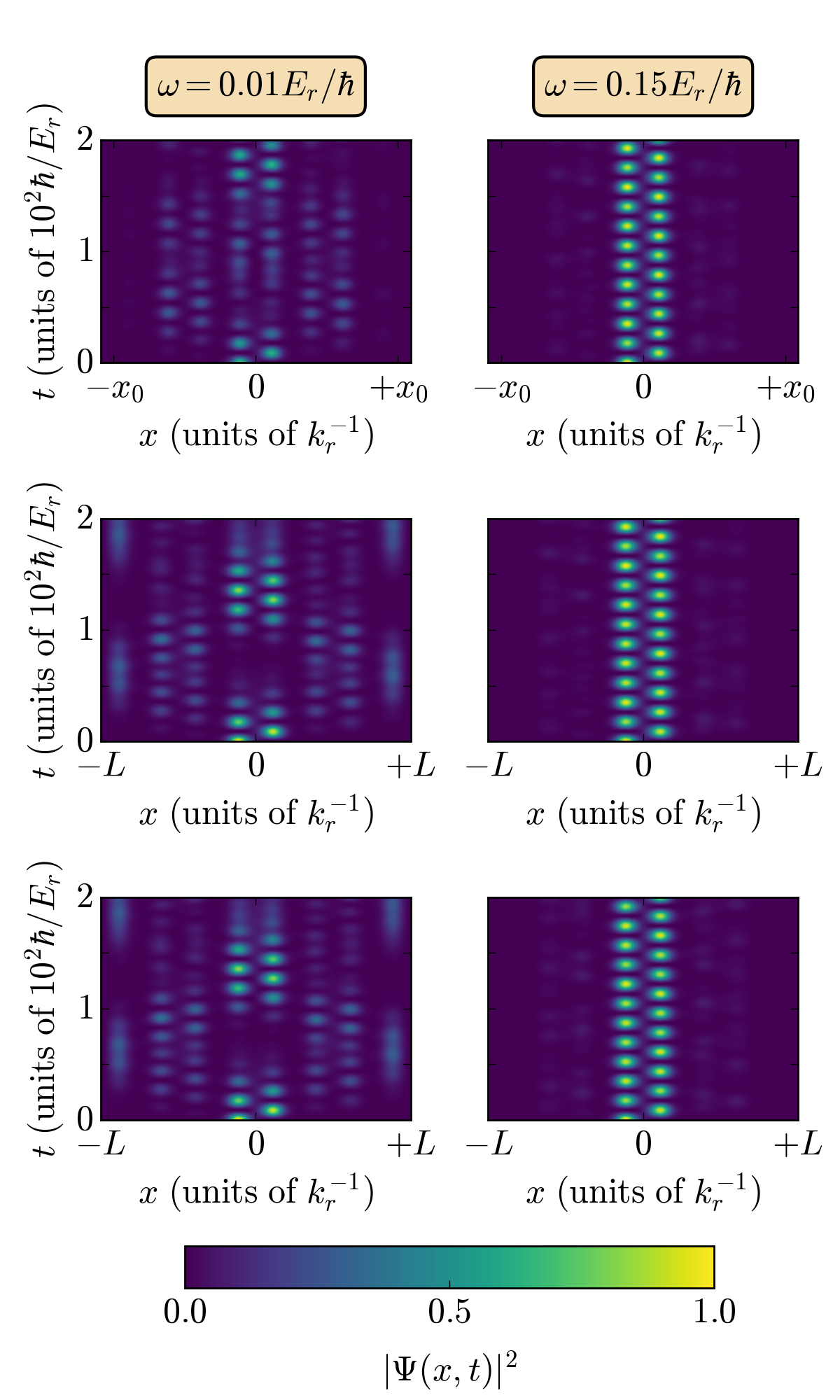}
\put(4,93.5) {\textbf{(a)}}
\put(31.7,93.5) {\textbf{(b)}}
\end{overpic}
\begin{overpic}[width=0.48\textwidth]{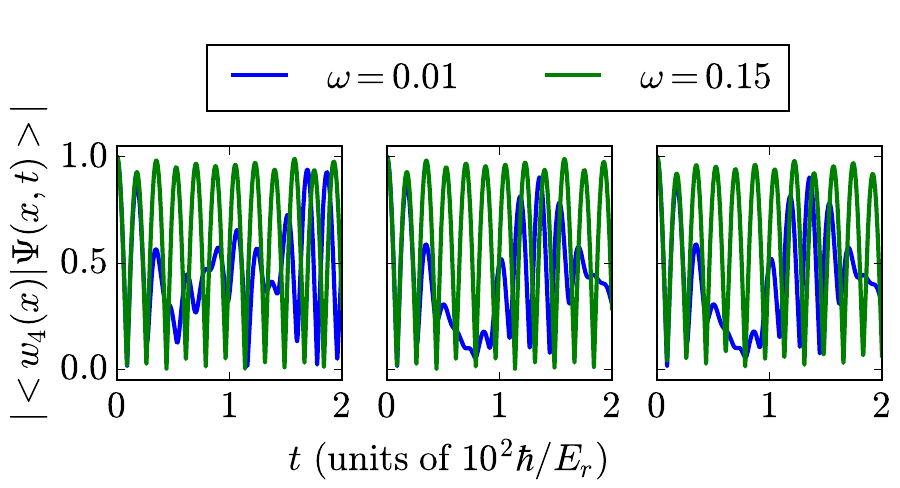}
\put(14,44) {\textbf{(c)}}
\end{overpic}
\caption{(a,b) Time evolution of an initial state 
$\Psi(x,t=0) = w_4(x)$ in an optical superlattice system for three configurations: optimal extension without center displacement (first row), no-extension without center displacement (second row) and no-extension with center displacement of $x_c = 0.25\pi/k_r$ (third row). The optical superlattice system has $V_{high}=5.0 E_r$, $u = 0.6$ and $M=4$. The systems of the left column (a)
have $\omega = 0.01 E_r/\hbar$ and of the right column (b) $\omega = 0.15 Er/\hbar$. (c) The overlap of $\Psi(x,t)$ with the Wannier function $w_4(x)$ corresponding to the left central site of the system. $w_4(x)$ is calculated for $\omega = 0$.}
\label{fig:4lvlDynamics_TimeEvol_Overlap}
\end{center}
\end{figure}

In Fig.~\ref{fig:4lvlDynamics_WannierPsi_n} we show the squared absolute value of the overlaps between $w_4(x)$ and the first twelve eigenstates of the system for the three cases discussed, plotted as a function of the trapping frequency.
For low trapping frequency values, this quantity is $1/4$ for at least three states, as indicated by the horizontal dashed line, regardless of the alignment and extension. 
This indicates that a substantial fraction of the probability density of each state (approximately $1/2$ due to mirror symmetry) is localized on the two central sites. However, the remaining $1/2$ extends across the entire lattice. We can see from Fig.~\ref{fig:4lvlDynamics_TimeEvol_Overlap}(a) that this results in a time evolution where the state have occasional occupation on the central sites for certain periods, while they spread across the whole lattice at other times. 
As we increase $\omega$, we reach the point where the effective four-level system emerge. 
This is indicated by the fact that for a certain regime only the $\psi_1(x)$ and $\psi_4(x)$ eigenstates have maximal overlaps with $w_4(x)$. 
In Fig.~\ref{fig:4lvlDynamics_TimeEvol_Overlap}(b) we see that this manifests as relatively high frequency transport dynamics between the two central sites during the time evolution, for all three cases, owing to the larger energy difference of the involved states ($E_4-E_1$).
The difference in the duration of the localization process at the two central sites between the two representative trapping frequency values is readily seen in Fig.~\ref{fig:4lvlDynamics_TimeEvol_Overlap}(c), where the time evolution of the overlap with $w_4(x)$ is shown. 
As shown in Fig.~\ref{fig:4lvlDynamics_WannierPsi_n}, increasing trapping frequency beyond a certain value ($\omega\approx0.25 E_r/\hbar$), results in only the $\psi_1(x)$ and $\psi_2(x)$ eigenstates being localized at the two central sites, signifying the dominance of the harmonic trap potential in the system.
Note, however, that in this regime the frequency of the transport dynamics between the two central sites is much slower due to the near-degeneracy of the two states.
 
To conclude, we presented the dependence of the effective four-level system on the optical superlattice parameters ($V_{high}$, $u$, and $M$), the boundary conditions, and the relative alignment of the two superimposed potentials. We demonstrated the universality of this phenomenon, which emerges under the following conditions: system extension is not required, small displacements are allowed as long as the overall minimum is preserved and the trapping frequency is sufficiently large, and the dynamics provide a direct means to observe it.

\section{Edge states of the harmonically confined optical superlattice system.}
\label{sec:fateofedges}

In this section we examine the harmonically confined optical superlattice system in the topological configuration and the low trapping frequency regime. 
Specifically, in Sec.~\ref{fateofedge} we are interested in the fate of the TESs when the harmonic trap is applied to the system. 
We consider this analysis to be crucial, in order to address possible experimental realizations. 
In essence, we present the behavior of the TESs varying optical superlattice parameters $V_{high},~u,~M$ and the trapping frequency $\omega$ of the harmonic trap.
In all cases, the system is extended by the optimal length ($d_{opt}$), since the TESs are highly sensitive to the boundary conditions. In Sec.~\ref{dynamics} we present the main difference in transport dynamics when the system's edge states have a topological origin versus when they are ``quasi-classically" localized due to the harmonic trap dominance.  

\subsection{Fate of the Topological Edge States}
\label{fateofedge}

The TESs of an extended optical superlattice system without the harmonic trap ($\omega = 0$) have been thoroughly studied in \cite{Katsaris2024}. 
The importance of the extension has been discussed here in Secs.~\ref{sec:setup} and ~\ref{sec:interplay}, while details on the optimal length are provided in Appendix~\ref{app:Optimal_d}. 
We now focus on the dependence of TESs on the parameters of the optical superlattice.
\begin{figure*}[!ht]
    \centering
  \begin{overpic}[width=0.98\linewidth]{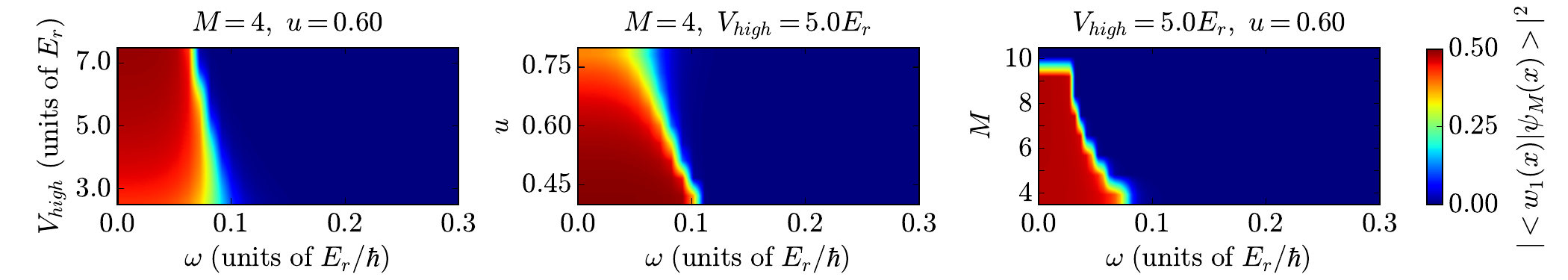}
    \put(7.4,16.) {\textbf (a)}
    \put(35,16.) {\textbf (b)}
    \put(63.6,16.) {\textbf (c)}
    \end{overpic}
\caption{(a-c) The squared absolute value of the overlap between the Wannier function $w_{1}(x)$ corresponding to the left edge site and the M-th eigenstate of an extended optical superlattice system with the addition of the harmonic trap. The extended optical superlattice system has (a) $M=4$, $u=0.6$, (b) $M=4$, $V_{high}=5.0E_r$ and (c) $V_{high} = 5.0E_r$ and $u=0.6$. In (a-c) the starting-point of the extension is $x_0 =L$ and the extension length is (a,b) $d=0.5\pi/k_r$ and (c) $d=d_{opt}$ with respect to $M$. 
In (c) the extended optical superlattice has for odd values of $M$ phase $\phi = 0$ and for even $M$ phase $\phi = \pi/2$, enabling it to support TESs. The $w_{1}(x)$ is calculated at $\omega = 0$.}
 \label{fig:Cverlaps_Wannier1_PsiM}
\end{figure*}
For a fixed number of cells (i.e. number of sites), the behavior of TESs is primarily governed by the ratio $u$. Specifically, as can be seen by the TESs energy splitting $\Delta_{edge}$ in Fig.~\ref{fig:OS_SpectraGaps_vs_Vhigh_and_u}(f), the splitting increases with increasing $u$. It is apparent, that $\Delta_{edge}$ is related to the persistence of the TESs once the harmonic trap is switched on. Obviously, the next-to-nearest hopping terms decrease for higher values of $u$, since the lower and higher barriers have similar values, making the tunneling of a particle through two consecutive cells harder. Thus, a small $u$ enhances the robustness of the TESs against the on-site potentials related to the harmonic trap potential, but leads to a corresponding eSSH with next-to-nearest neighbor hopping terms. Furthermore, the small $u$ regime corresponds to the phase where the optical superlattice system is deep in the region in which the two topological phases are well-distinguished, as quantified by the corresponding Zak phase shown in Appendix \ref{app:ZakPhase}.

In Fig.~\ref{fig:Cverlaps_Wannier1_PsiM}(a) and (b) we
show the squared absolute value of the overlap of the Wannier function $w_1(x)$, corresponding to the left edge site of the system, and of $\psi_M(x)$, which is the lowest energy TES. 
This overlap directly expresses how robust the TESs are in the presence of the harmonic trap. 
As shown, increasing $V_{high}$ or decreasing $u$ (while keeping the other parameter fixed), has the same effect on the TESs. In particular, the trapping frequency regime narrows and shifts, reducing the range over which the TESs are supported by the system.
Since we have used the same optical superlattice parameters $V_{high}=5.0E_r,$ $u=0.6$ and $M=4$, we can directly compare the behavior of $|<w_1(x)|\psi_M(x)>|^2$ and $|<w_M(x)|\psi_4(x)>|
^2$ with respect to changes of $V_{high}$ and $u$. Hence, from Fig.~\ref{fig:Cverlaps_Wannier1_PsiM} and Fig.~\ref{fig:Cverlaps_WannierM_Psi4_and_Spectra}, we observe that these two overlaps exhibit exactly the same behavior, but in different trapping frequency regimes. Each curve effectively mirrors the other: when one overlap reaches its minimum, the other attains its maximum. This is expected, since for $M=4$ two localization mechanisms compete, resulting in the $\psi_4(x)$ eigenstate being localized either at the edges or at the center of the system. This means that, when preparing the system, one should keep in mind which localization mechanism is being targeted for preparation and observation.

We now address the dependence on the system size.
The spectral behavior of the system can be extracted from Fig.~\ref{fig:OS_Spectra_TBApprox_vs_omega}(b) for $M=4$, and from Fig.~\ref{fig:Cverlaps_WannierM_Psi4_and_Spectra} for $M=6$ and $M=8$.
We observe that upon increasing the system size, the TESs survive over a smaller range of $\omega$. This is clearly connected to the fact that the influence of harmonic trap increases quadratically with the distance from the center of the system, thus, the larger the system size, the stronger the trap's influence on the edges.
This effect is also evident from Fig.~\ref{fig:Cverlaps_Wannier1_PsiM}(c). 
As we increase $M$, the overlap stays maximal over a shorter trapping frequency regime. Eventually, the overlap drops to zero as a result of the complete de-hybridization of the edge states. In essence, when the system is sufficiently large, the edge states become fully localized at individual edges, unlike the usual edge states that exhibit support on both edges.

Importantly, the critical value of the trapping frequency, $\omega_c$, at which the TESs cease to exist coincides with the value at which the harmonic trap becomes dominant at the edge sites of the system. This means that while localization at the edge sites persists, beyond a certain trapping frequency value it arises from a different mechanism, whose classification as topological or not would require a generalized topological index, which lies beyond the scope of the present work. We regard this as an open question for future investigation. In the following subsection, we present in detail how the different edge localization mechanisms affect the system's dynamics.

\subsection{Transport Dynamics of Edge States}\label{dynamics}

In order to study the transport dynamics of the edge states in the harmonically confined optical superlattice system, we consider the set of its parameters to be $V_{high} =5.0E_r$, $u=0.6$, $M=4$ and $\phi = \pi/2$. Additionally, since the TESs are significantly influenced by the boundary conditions, we only employ the extended version of our system. Explicitly, we have the extension with starting point at $x_0=L$ and optimal length $d_{opt} = 0.36 \pi/k_r$. The initial condition we use is $\Psi(x,t=0) = w_1(x)$, which is the Wannier function localized at the left edge site. 

\begin{figure}[!ht]
\begin{center}
\begin{overpic}[width=0.48\textwidth]{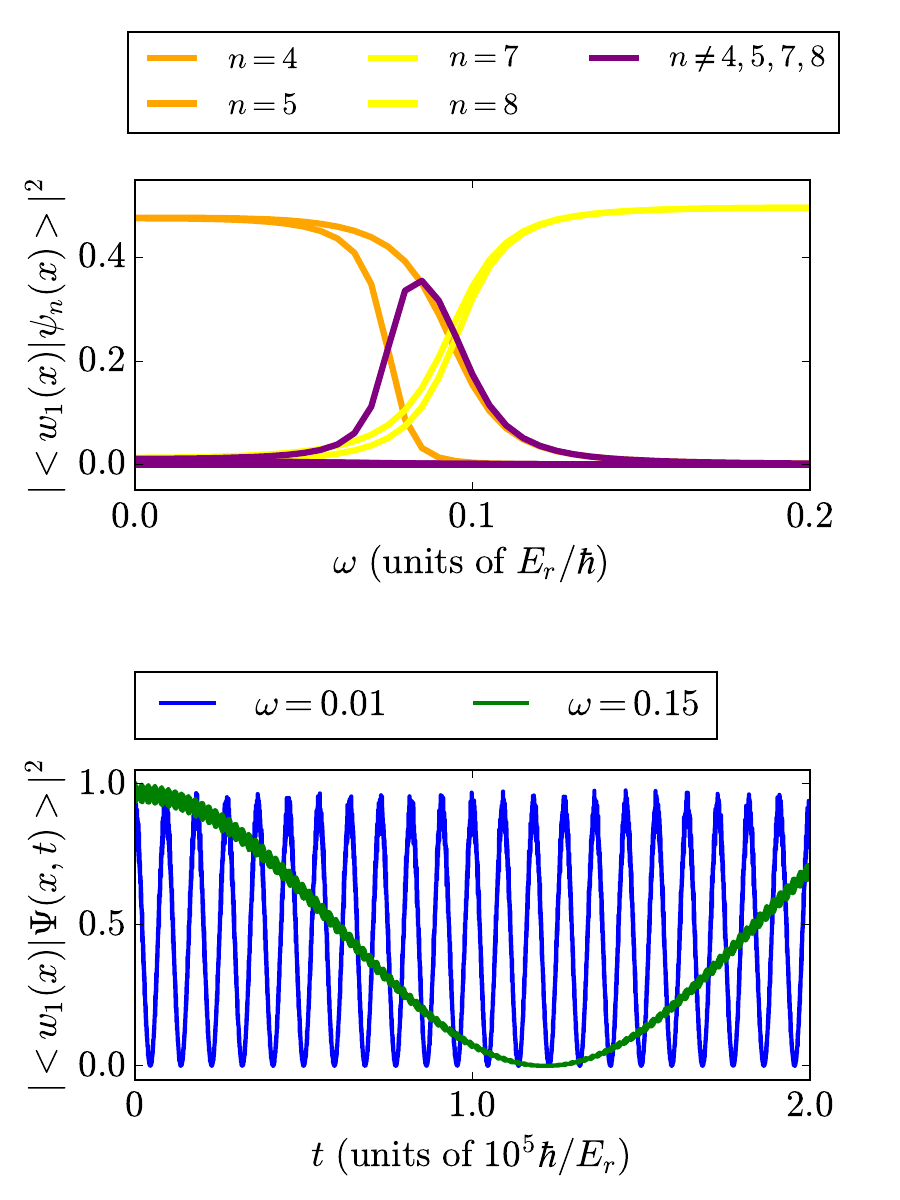}
\put(3.5,92.5) {\textbf{(a)}}
\put(3.5,40.5) {\textbf{(b)}}
\end{overpic}
\caption{(a) The squared absolute value of the overlaps between the Wannier function $w_{1}(x)$ corresponding to the left edge site and the first twelve eigenstates of the optical superlattice system under harmonic confinement vs the trapping frequency $\omega$. The optical superlattice system obeys $V_{high}=5.0 E_r$, $u=0.6$ and $M=4$ and $\phi = \pi/2$. We use as starting-point of the extension $x_0=L$ and the optimal extension $d=d_{opt}=0.36\pi/k_r$. (b) The overlap of $\Psi(x,t)$ with the Wannier function $w_1(x)$ when the latter is considered as the initial condition. $w_1(x)$ is calculated at $\omega = 0$.}
\label{fig:Dynamics}
\end{center}
\end{figure}

In Fig.~\ref{fig:Dynamics}(a) we present the squared absolute value of the overlap between $w_1(x)$ and $\psi_n(x)$ (for $n\in[1,12]$) as a function of $\omega$. We see that for small trapping frequency values, before the avoided crossings of the TESs with the upper sub-band occur, the TESs, namely $\psi_4(x)$ and $\psi_5(x)$, are the only edge localized states. Beyond a certain trapping frequency value, the role of the edge states is taken over by the two highest-lying, $\psi_7(x)$ and $\psi_8(x)$ energy eigenstates of the upper sub-band of the first band, due to the harmonic trap dominance at the edges. Based on that,
when we perform the time evolution process of the system, we once again consider two representative trapping frequency values. Specifically, we set $\omega = 0.01E_r/\hbar$ and $\omega = 0.15E_r/\hbar$ for small and intermediate trapping frequency regimes, respectively. Evidently, there is a trapping frequency regime in which the smooth transition from one localization mechanism to the other takes place. In this trapping frequency regime, numerous states display significant occupation both at the edges and across the bulk, preventing any state from being characterized as an edge state. The smooth transition must be taken into account because it causes occupation leakage into the bulk of the system, impacting the observation of edge states and the transition from one localization mechanism to the other.

In Fig.~\ref{fig:Dynamics}(b) we present the squared absolute value of the overlap between $w_1(x)$ and $\Psi(x,t)$, which serves as an indicator of how long the state remains at the edges of the system. Due to mirror symmetry, the overlap with $w_8(x)$ (other edge) is identical, so it is not shown. More precisely, starting from a state initially localized entirely at one edge, the time-evolved states exhibits transport behavior from that edge to the opposite edge without any significant leakage to the bulk, since in both cases under study the system support edge states.
As we can see, the two cases differ greatly in their transport frequencies. 
The spatial overlap and the energy gap of the edge states are the two main factors that affect this transport frequency. 
In both cases, the spatial profile of the edge states are nearly identical, resulting in similar overlap values. 
However, there is a substantial difference in the energy gaps, which are proportional to the transport frequencies. 
The TESs, due to the finite-size boundaries effect, have a small but finite energy gap that depends on the ratio $u$, as shown in Fig.~\ref{fig:OS_SpectraGaps_vs_Vhigh_and_u}(f) and discussed in the previous subsection. On the other hand, the ``quasi-classically" localized edge states are almost degenerate, causing their energy gap value to be near zero. This is precisely why the transport frequency of the TESs is nearly twenty-six times higher than that of the ``quasi-classically" localized states. 
The fundamental difference in the underlying localization mechanisms and their impact on transport dynamics should be carefully considered when designing experimental setups. 
This also implies that the two mechanisms can be distinguished and potentially exploited for different quantum processes.

\section{Summary and Perspectives}\label{sec:SummaryAndOutlook} 

We have identified and explored in detail the three distinct localization mechanisms present in the finite, harmonically confined optical superlattice. 
We use results from ED and the corresponding TB approximation. First, we discussed the similarities and differences with the harmonically confined optical lattice, a system that has been studied in the literature ~\cite{Batrouni2002HOL,Hooley2004HOL,Ruuska2004HOL,Rigol2004HOL,Pezze2004HOL,Ott2004HOL,Rey2005HOL,Ali2025HOL}. The  “quasi-classical” localization mechanism responsible for the localized states observed in both the optical lattice and superlattice originates from the progressively dominant influence of the harmonic trap at high trapping frequency values.
Then, we revealed a novel localization mechanism arising from the interplay between the optical superlattice and the harmonic trap in an intermediate trapping regime between the perturbative and trap-dominated limits.
We performed an in-depth analysis on the universality of the emergence of the effective four-level system with respect to the parameters of the optical superlattice and addressed relevant properties, important for the flexibility of a corresponding experimental realization. We also studied the fate of the TESs of the optical superlattice system, which arise from the system's topological configuration through the bulk-edge correspondence and the topology of the band structure. Lastly, we illustrated examples of the systems dynamics in order to highlight the impact of localization on the time evolution of an initial state localized at a single site. We consider this as a direct probe of the different localization mechanism, since they are shown to be characterized by clearly distinct frequencies.

Evidently, from our analysis, the effective four-level system emerges as the most promising candidate for novel applications and observation protocols. An important perspective we aspire to explore in the future is to first study adiabatic variations of the trapping frequency and then develop time-dependent protocols in which this frequency explicitly varies with time. The primary goal of implementing these protocols is to observe different localization mechanisms through the time evolution of $\psi_4(x)$ in the $M=4$ system in the topological configuration. We believe that, since the same state will be localized at different sites over time, this will provide access to interesting state transfer protocols relevant for controlled transport. Additionally, we aim to incorporate interactions into our system to investigate whether and in what regime of interactions this phenomenology persists in many-body particle systems. 
Finally, it is of interest to investigate how different periodic and/or confining potentials affect the effective system formed by the lowest-lying eigenstates, before the confinement becomes dominant. We find that this approach fully leverages the novel localization mechanism uncovered in this work.

\section*{Acknowledgments} 
This work (P.S. and C.W.) has been supported by the Cluster of Excellence “Advanced Imaging of Matter” of the Deutsche Forschungsgemeinschaft (DFG)-EXC 2056, Project ID No. 390715994. \\[15pt]

\appendix

\section*{APPENDIX}

\section{Topological Characterization of the Continuous System via Zak Phase}
\label{app:ZakPhase}

The Zak Phase ~\cite{PhysRevLett.62.2747} is the one-dimensional analogue of the Berry phase ~\cite{berry1984phase}, which serves as a well-established topological invariant that distinguishes between topologically trivial and non-trivial band structures. 
It is a geometric phase acquired by a particle's wavefunction as it traverses the Brillouin zone in a one-dimensional periodic system and reads:
\begin{equation}
\gamma_{Zak}=i\int_{BZ}\bra{u_k}\ket{\frac{\partial}{\partial k}u_k}dk
\end{equation}
where $\ket{u_k}$ is the periodic part of the Bloch wavefunction and the integration is performed over the first Brillouin zone (BZ). In systems with inversion or chiral symmetry, the Zak phase take quantized values ($0$ or $\pi$). Importantly, the topological significance of the Zak phase is encoded in its value modulo $2\pi$, or equivalently, in the differences between Zak phases computed for different bands or system configurations.

In our case, we are interested in classifying the optical superlattice system as topologically trivial or non-trivial, with respect to the combination of its phase factor $\phi$ and the number of cells $M$. To do so, for each phase factor $\phi=0$ and $\pi/2$, we consider the corresponding Bloch Hamiltonian. Specifically, we construct the Hamiltonian in real space for one unit cell ($M=1$) with periodic boundary conditions and then perform a Fourier transform to get the $k$-dependent Bloch Hamiltonian ($\mathcal{H}(k, \phi)$). For each quasi-momentum $k$, the $\mathcal{H}(k, \phi)$ is diagonalized using finite-difference methods to obtain the lowest energy eigenstate. We compute numerically the Zak phase, as a discretized Berry phase, by summing the phases of overlaps between adjacent Bloch eigenstates along the Brillouin zone. The accumulated phase from these overlaps yields the Zak phase module $2\pi$, providing a topological invariant of the first band. Finally we compute the difference:
\begin{equation}
\Delta\gamma_{Zak} =\gamma_{Zak}(\pi/2)-\gamma_{Zak}(0)
\end{equation}
which is the desired difference of Zak phases for the classification of the two optical superlattice configurations. 
\begin{figure}[!ht]
\begin{center}
\begin{overpic}[width=0.48\textwidth]{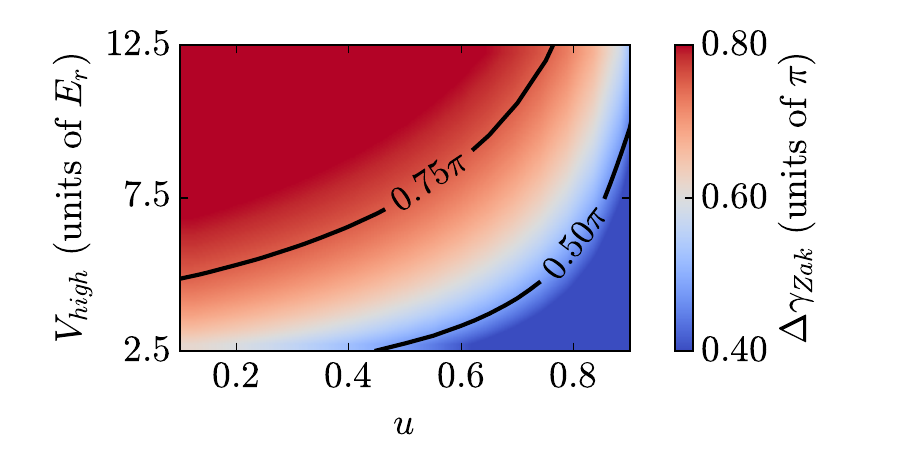}
\end{overpic}
\caption{Difference of the Zak phase $\Delta \gamma_{Zak}=\gamma_{Zak}(\pi/2)-\gamma_{Zak}(0)$ between the two configurations of the optical superlattice system corresponding to $\phi = \pi/2$ and $\phi = 0$ versus the height of the higher barrier $V_{high}$ and the ratio of barrier heights $u$. The region above $0.5\pi$ corresponds to parameters where the system exhibits two distinguishable phases, while values above $0.75\pi$ indicate that the topological and trivial phases are well-separated. Below $0.5\pi$, the distinction between phases is lost.}
\label{fig:ZakPhaseDiff}
\end{center}
\end{figure}
In Fig.~\ref{fig:ZakPhaseDiff} we present $\Delta\gamma_{Zak}$ as a function of $V_{high}$ and $u$. As we can see, for fixed $V_{high}$, increasing $u$ decreases $\Delta\gamma_{Zak}$, while at fixed $u$, increasing $V_{high}$ increases $\Delta\gamma_{Zak}$. Numerically, $\Delta\gamma_{Zak}$ is found to range from $0.45\pi$ to $0.85\pi$. These values are affected both by numerical precision limits and by the $V_{high}$ values we consider in our study. Using the Cauchy argument principle, it is analytically predicted  that the topological phase transition occurs at the critical point where $\Delta\gamma_{Zak}=\pi/2$. Based on this, we consider the optical superlattice system to exhibit topologically distinct phases when $\Delta\gamma_{Zak}>\pi/2$, 
corresponding to the top-left corner of Fig.~\ref{fig:ZakPhaseDiff}. For most of the parameter space, the system remains within this well-defined topological regime. However, as we move toward the bottom-right corner, it gradually enters a region where the two topological phases become indistinguishable. In finite systems with open boundaries, these two topological phases are characterized by the configuration of the barrier heights at the system edges, specifically, whether the system starts and ends with a high or a low barrier. This arises from the combination of phase $\phi$ and number of cells $M$, as shown in Table~\ref{Table:TopoClas}. 

\section{Definition of the Optimal Extension Length ($\vb*{d}_{opt}$)}\label{app:Optimal_d}

The importance of the linear extension of the potential in restoring the TESs in an optical superlattice system with HWBC has been studied in detail in ~\cite{Katsaris2024}. Here we employ a simple and straightforward method to determine the optimal extension length for maximal restoration of the TESs energies. We define the $\delta$ measure as:
\begin{equation}
\delta = (E_{M+2}-E_{M+1})-(E_{M}-E_{M-1})
\end{equation}
which quantifies the difference between the energy gaps separating the two edge states (upper and lower) from the adjacent bulk bands.
The optimal extension length $d_{opt}$ is defined as the value of $d$ that minimizes $\delta$. To find $d_{opt}$, we solve the eigenvalue problem of the optical superlattice system for varying extension lengths $d$, while keeping the starting-point fixed at $x_0=L$. 
\begin{figure}[!ht]
\begin{center}
\begin{overpic}[width=0.48\textwidth]{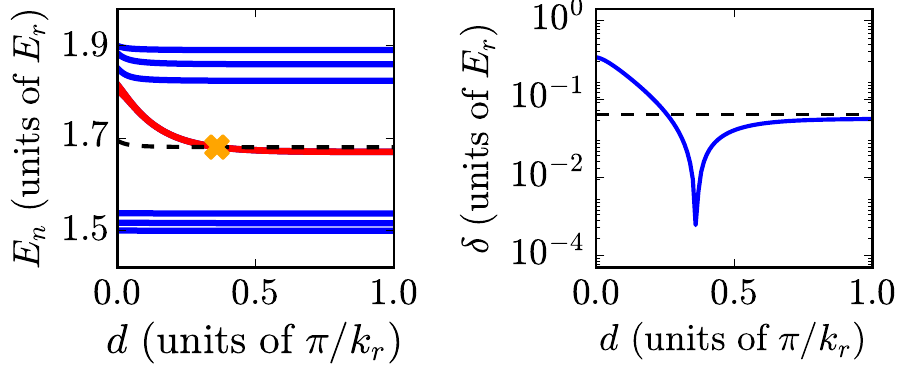}
\put(37,20) {\textbf{(a)}}
\put(89,20) {\textbf{(b)}}
\end{overpic}
\\[10pt]
\begin{overpic}[width=0.48\textwidth]{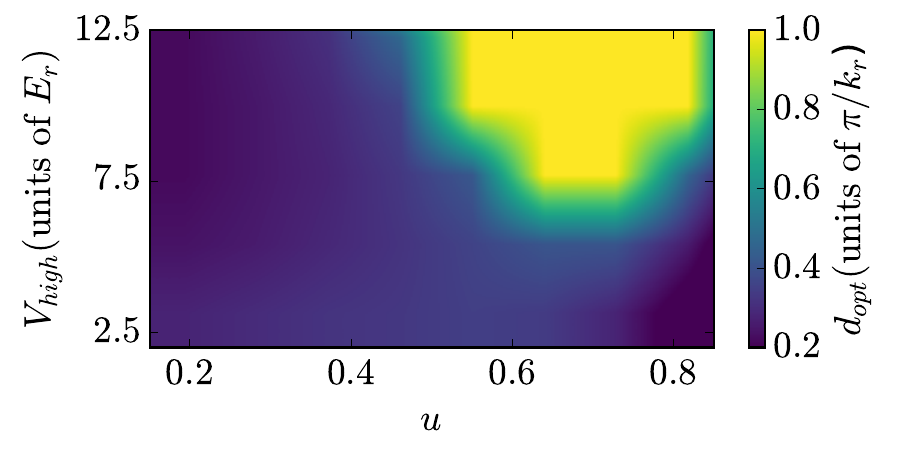}
\put(92,8) {\textbf{(c)}}
\end{overpic}
\\[10pt]
\begin{overpic}[width=0.48\textwidth]{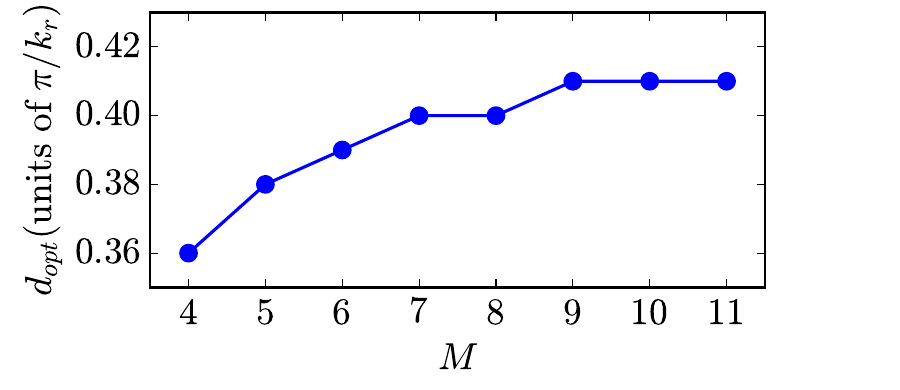}
\put(77,15) {\textbf{(d)}}
\end{overpic}
\caption{(a) Energy spectrum versus the extension length $d$ of the extended optical superlattice. The dashed line indicates the midpoint of the band gap. The cross scatter marker denotes the $d_{opt}$ value for the system for that specific case. (b) The $\delta$ measure we use to determine the optimal extension length ($d_{opt}$). The dashed line indicates the value $E_r/20$. (c) $d_{opt}$ versus $V_{high}$ and ratio $u$. (d) $d_{opt}$ versus the number of cells $M$. Parameters of the extended optical superlattice are (a,b,d) $V_{high}=5.0E_r$, $u = 0.6$, (a-c) $M=4$ and (d) $M\in[4,11]$. For odd values of $M$ we have $\phi = 0$ and for even $M$ $\phi = \pi/2$, enabling it to support TESs. The extension starts at $x_0 = L$ for all cases.}
\label{fig:d_opt_results}
\end{center}
\end{figure}
This procedure is best illustrated in Fig~\ref{fig:d_opt_results}(a) and (b), where we present the energy spectrum as a function of $d$, along with the corresponding $\delta$ measure, for an optical superlattice system with $V_{high}=5.0E_r$, $u=0.6$ and $M=4$. Evidently, although $d_{\text{opt}}$ in most cases is well-defined as the value that minimizes $\delta$, we observe that beyond a certain $d$, the value of $\delta$ remains relatively small (approximately $E_r/20$). This implies that, with respect to an experimental realization, the exact value of $d$ is irrelevant, as long as it exceeds a certain threshold. Following the same procedure for various combinations of $V_{high}$ and $u$, while keeping $M = 4$  fixed, yields the color plot shown in Fig.~\ref{fig:d_opt_results}(c). As we can see, for values of $u<0.5$, increasing $V_{high}$ results in a nearly constant $d_{opt}$. 
Also, there is a region where, for sufficiently large $V_{high}$ and $u$, the definition of $d_{opt}$ as the value that minimizes $\delta$ no longer hold, since although $\delta$ decreases, it never reaches a minimum but instead approaches an asymptotic value. These cases can be explored following the methods presented in \cite{Katsaris2024}. In Fig.~\ref{fig:d_opt_results}(d), we present the dependence of $d_{opt}$ on the system size. As $M$ increases, $d_{opt}$ generally increases as well. In Fig.~\ref{fig:d_opt_results}(d), certain values of $d_{opt}$ appear identical due to the finite grid resolution and the chosen scale of the $y$-axis. Finally, in this work we fix the starting-point of the extension at $x_0 = L$ for convenience. We note for clarity that the same analysis can be applied to any starting-point, provided the corresponding optimal extension length is used in each case.
\begin{figure*}[!ht]
\begin{center}
\begin{overpic}[width=0.98\textwidth]{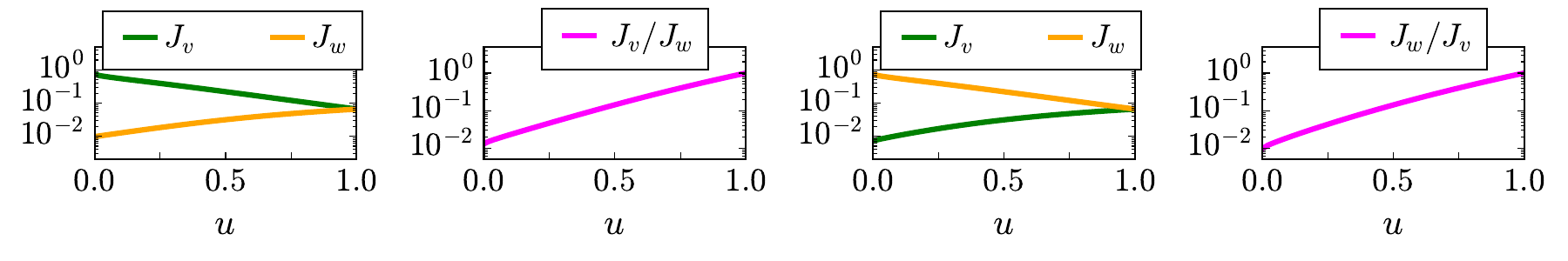}
\put(3.5,14.5) {\textbf{(a)}}
\put(28,14.5) {\textbf{(b)}}
\put(53,14.5) {\textbf{(c)}}
\put(77.5,14.5) {\textbf{(d)}}
\end{overpic}
\caption{(a,c) The hopping amplitudes $J_{1,3}=J_v$ and $J_2=J_w$ versus the ratio of the heights of the barriers $u = V_{low}/V_{high}$ for the minimal extended optical superlattice system. (b,d) The ratio of weak-to-strong hopping amplitudes versus $u$. The minimal extended optical superlattice system has $V_{high}=5.0 E_r$. The extension starts at $x_0=L$ and has the optimal length $d_{opt}=0.28 \pi/k_r$. The phase factor is (a,b) $\phi = 0$ and (c,d) $\phi = \pi/2$, corresponding to the topologically trivial and non-trivial phase, respectively.}
\label{fig:M2_TB}
\end{center}
\end{figure*}
\begin{figure}[!ht]
\begin{center}
\begin{overpic}[width=0.48\textwidth]{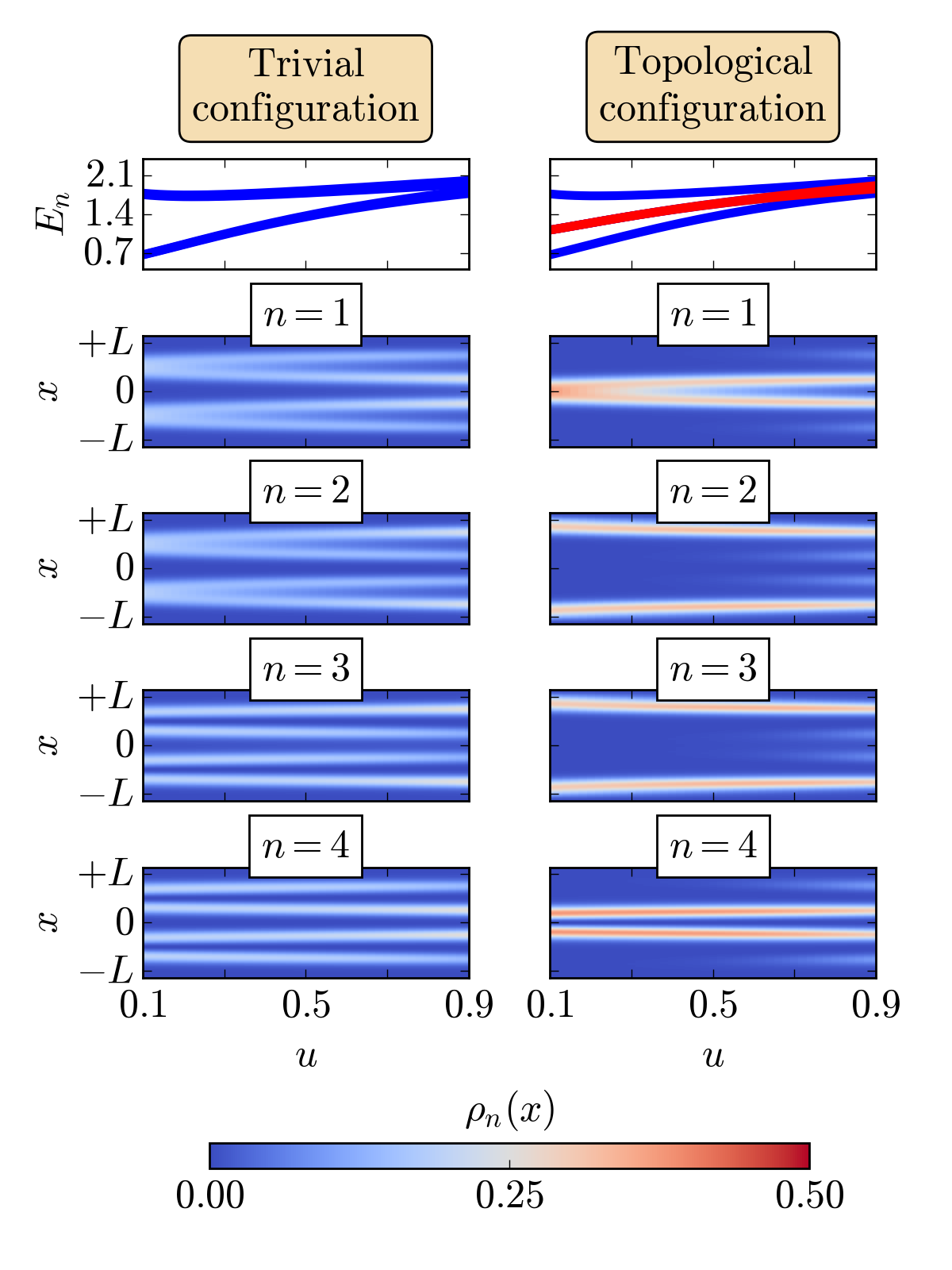}
\put(7,92.5) {\textbf{(a)}}
\put(39,92.5) {\textbf{(b)}}
\end{overpic}
\caption{(a,b) The spectrum and the spatial profile of the densities of the first four eigenstates of the minimal extended optical superlattice system versus $u$, in (a) the trivial and (b) the topological configuration.
The minimal extended optical superlattice system has $V_{high}=5.0 E_r$. The extension starts at $x_0=L$ and has the optimal length $d_{opt}=0.28 \pi/k_r$.}
\label{fig:M2_Spectra_States_vs_u}
\end{center}
\end{figure}
\begin{figure}[!ht]
\begin{center}
\begin{overpic}[width=0.48\textwidth]{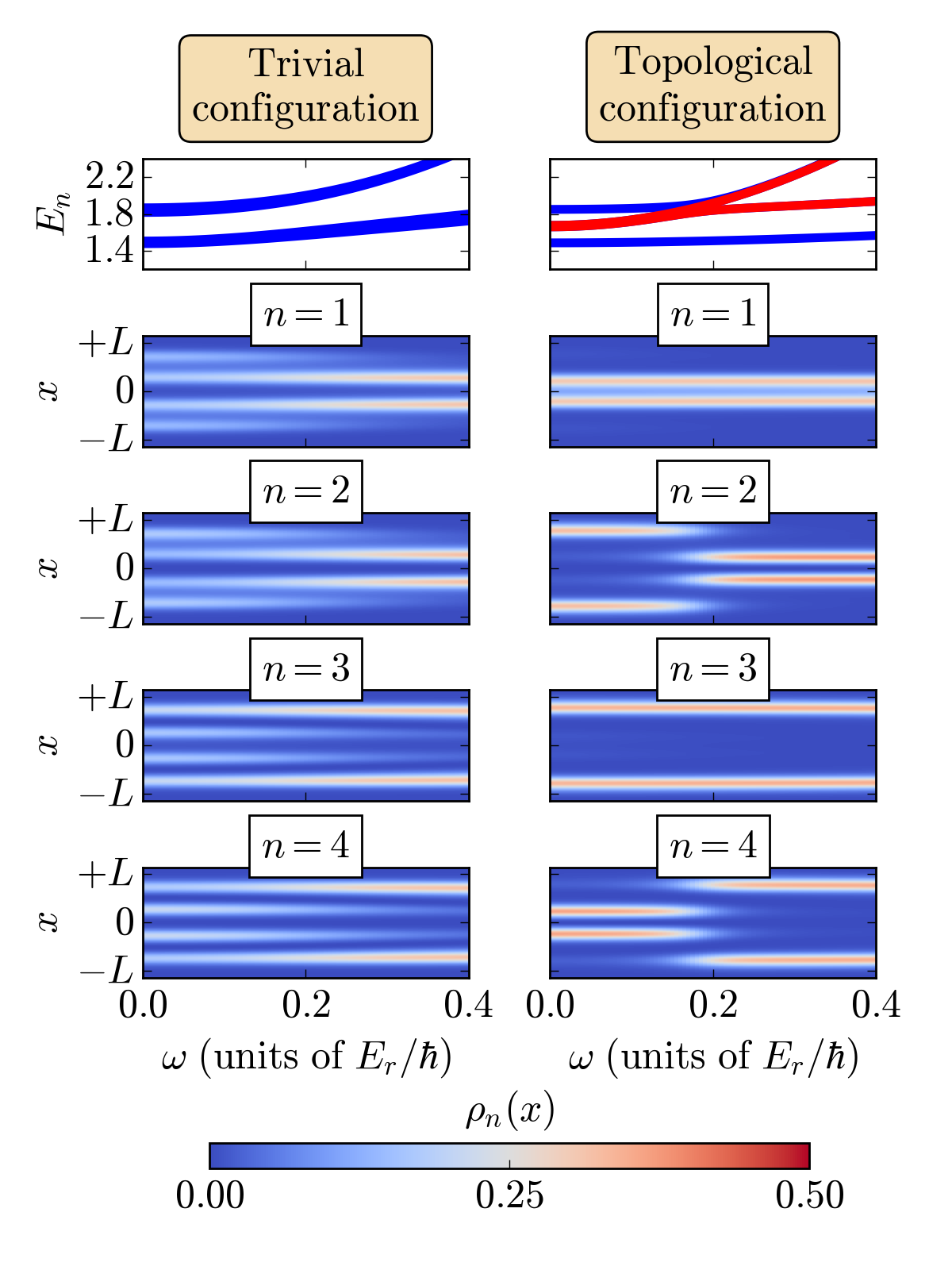}
\put(7,92.5) {\textbf{(a)}}
\put(39,92.5) {\textbf{(b)}}
\end{overpic}
\caption{The spectrum and the spatial profile of the densities of the first four eigenstates of the system versus the trapping frequency $\omega$, in (a) the trivial and (b) the topological configuration.
The minimal extended optical superlattice system has $V_{high}=5.0 E_r$, $u=0.6$. The extension starts at $x_0=L$ and has the optimal length $d_{opt}=0.28 \pi/k_r$.}
\label{fig:Spectra_States_vs_omega}
\end{center}
\end{figure}
\section{Minimal System Exhibiting Nontrivial Topology (The case of $\vb*{M=2}$)}\label{app:MinimalSystem}

In this work, we consider only systems with an even number of sites, since $\mathcal{N}=2M$. 
Based on this, and noting that the case of $M=1$ corresponds effectively to a two-level system, we see that the minimal (i.e., smallest) system that can display either topological or trivial phases is the one with $M=2$.
That is, with respect to $\phi$, the three barriers have a spatial profile of either low-high-low ($\phi = 0$) or high-low-high ($\phi = \pi/2$), corresponding to a TB system with strong-weak-strong or weak-strong-weak hopping amplitudes, respectively.
Due to the mirror symmetry of the system, we have two distinct hopping amplitudes. We set the first and last hopping amplitudes as $J_v$ and the center one as $J_w$. 
In Fig.~\ref{fig:M2_TB}, we see that for fixed $V_{high}$ and varying $u$, the two hopping amplitudes start with a weak-to-strong ratio of approximately $10^{-2}$, and as $u$ increases, we reach to the case of $u=1$ where the ratio is $1$. This means that the smaller the $u$ the deeper we are in either the trivial or non-trivial topological phase. This behavior is strongly highlighted in the spectral behavior and spatial profiles of the eigenstates, as shown in Fig.~\ref{fig:M2_Spectra_States_vs_u}, for the minimal extended optical superlattice system with $V_{high}=5.0E_r$. As expected, for the trivial configuration the first four states form a band consisting of two states in each sub-band, while the spatial profiles of all four states generally extend across the entire optical superlattice. On the other hand, in the topological configuration, the band structure consists of one state in each sub-band and two states residing in the middle of the sub-band gap.  The spatial profiles of the two 
states in the middle of the gap show that they are the minimal TESs of the system, localized at the two edge sites. 
Interestingly, the other two states occupy only the two middle sites and thus represent the minimal bulk states.
Moving on to Fig.~\ref{fig:Spectra_States_vs_omega}, we see that once the harmonic trap is applied, the system preserves its behavior up to a finite value of trapping frequency. Then, when the harmonic trap becomes dominant, the system transitions to the phase where the first two states form a pair and the next two states form another pair, with each pair localized on symmetric sites. 

\bibliographystyle{apsrev4-2}
\bibliography{ref_drops}

@article{Gaunt_Hadzibabic2013Box,
  title = {Bose-Einstein Condensation of Atoms in a Uniform Potential},
  author = {Gaunt, Alexander L. and Schmidutz, Tobias F. and Gotlibovych, Igor and Smith, Robert P. and Hadzibabic, Zoran},
  journal = {Phys. Rev. Lett.},
  volume = {110},
  issue = {20},
  pages = {200406},
  numpages = {5},
  year = {2013},
  month = {May},
  publisher = {American Physical Society},
  doi = {10.1103/PhysRevLett.110.200406},
  url = {https://link.aps.org/doi/10.1103/PhysRevLett.110.200406}
}

@article{Navon_Hadzibabic2021Box,
  author    = {Nir Navon and Robert P. Smith and Zoran Hadzibabic},
  title     = {Quantum gases in optical boxes},
  journal   = {Nature Phys.},
  year      = {2021},
  volume    = {17},
  number    = {12},
  pages     = {1334--1341},
  doi       = {10.1038/s41567-021-01403-z},
  url       = {https://doi.org/10.1038/s41567-021-01403-z},
  issn      = {1745-2481},
}

@article{Rigol2004HOL,
  title = {Confinement control by optical lattices},
  author = {Rigol, Marcos and Muramatsu, Alejandro},
  journal = {Phys. Rev. A},
  volume = {70},
  issue = {4},
  pages = {043627},
  numpages = {12},
  year = {2004},
  month = {Oct},
  publisher = {American Physical Society},
  doi = {10.1103/PhysRevA.70.043627},
  url = {https://link.aps.org/doi/10.1103/PhysRevA.70.043627}
}

@article{Hooley2004HOL,
  title = {Single-Atom Density of States of an Optical Lattice},
  author = {Hooley, Chris and Quintanilla, Jorge},
  journal = {Phys. Rev. Lett.},
  volume = {93},
  issue = {8},
  pages = {080404},
  numpages = {4},
  year = {2004},
  month = {Aug},
  publisher = {American Physical Society},
  doi = {10.1103/PhysRevLett.93.080404},
  url = {https://link.aps.org/doi/10.1103/PhysRevLett.93.080404}
}

@article{Ruuska2004HOL,
doi = {10.1088/1367-2630/6/1/059},
url = {https://dx.doi.org/10.1088/1367-2630/6/1/059},
year = {2004},
month = {jun},
publisher = {},
volume = {6},
number = {1},
pages = {59},
author = {Ruuska, V and Törmä, P},
title = {Quantum transport of non-interacting Fermi gas in an optical lattice combined with harmonic trapping},
journal = {New J. Phys.}
}

@article{Ott2004HOL,
  title = {Radio Frequency Selective Addressing of Localized Atoms in a Periodic Potential},
  author = {Ott, H. and de Mirandes, E. and Ferlaino, F. and Roati, G. and T\"urck, V. and Modugno, G. and Inguscio, M.},
  journal = {Phys. Rev. Lett.},
  volume = {93},
  issue = {12},
  pages = {120407},
  numpages = {4},
  year = {2004},
  month = {Sep},
  publisher = {American Physical Society},
  doi = {10.1103/PhysRevLett.93.120407},
  url = {https://link.aps.org/doi/10.1103/PhysRevLett.93.120407}
}

@article{Pezze2004HOL,
  title = {Insulating Behavior of a Trapped Ideal Fermi Gas},
  author = {Pezz\`e, L. and Pitaevskii, L. and Smerzi, A. and Stringari, S. and Modugno, G. and de Mirandes, E. and Ferlaino, F. and Ott, H. and Roati, G. and Inguscio, M.},
  journal = {Phys. Rev. Lett.},
  volume = {93},
  issue = {12},
  pages = {120401},
  numpages = {4},
  year = {2004},
  month = {Sep},
  publisher = {American Physical Society},
  doi = {10.1103/PhysRevLett.93.120401},
  url = {https://link.aps.org/doi/10.1103/PhysRevLett.93.120401}
}

@article{Rey2005HOL,
  title = {Ultracold atoms confined in an optical lattice plus parabolic potential: A closed-form approach},
  author = {Rey, Ana Maria and Pupillo, Guido and Clark, Charles W. and Williams, Carl J.},
  journal = {Phys. Rev. A},
  volume = {72},
  issue = {3},
  pages = {033616},
  numpages = {17},
  year = {2005},
  month = {Sep},
  publisher = {American Physical Society},
  doi = {10.1103/PhysRevA.72.033616},
  url = {https://link.aps.org/doi/10.1103/PhysRevA.72.033616}
}

@article{Batrouni2002HOL,
  title = {Mott Domains of Bosons Confined on Optical Lattices},
  author = {Batrouni, G. G. and Rousseau, V. and Scalettar, R. T. and Rigol, M. and Muramatsu, A. and Denteneer, P. J. H. and Troyer, M.},
  journal = {Phys. Rev. Lett.},
  volume = {89},
  issue = {11},
  pages = {117203},
  numpages = {4},
  year = {2002},
  month = {Aug},
  publisher = {American Physical Society},
  doi = {10.1103/PhysRevLett.89.117203},
  url = {https://link.aps.org/doi/10.1103/PhysRevLett.89.117203}
}

@article{Ali2025HOL,
  title = {Wave packet dynamics in parabolic optical lattices: From Bloch oscillations to long-range dynamical tunneling},
  author = {Ali, Usman and Holthaus, Martin and Meier, Torsten},
  journal = {Phys. Rev. Res.},
  volume = {7},
  issue = {1},
  pages = {013141},
  numpages = {9},
  year = {2025},
  month = {Feb},
  publisher = {American Physical Society},
  doi = {10.1103/PhysRevResearch.7.013141},
  url = {https://link.aps.org/doi/10.1103/PhysRevResearch.7.013141}
}

@article{Schneider2012BoxHubbard,
  author    = {Schneider, Ulrich and Hackermüller, Lucia and Ronzheimer, Jens Philipp and Will, Sebastian and Braun, Simon and Best, Thorsten and Bloch, Immanuel and Demler,Eugene  and Mandt, Stephan and Rasch, David and Rosch, Achim},
  title     = {Fermionic transport and out-of-equilibrium dynamics in a homogeneous Hubbard model with ultracold atoms},
  journal   = {Nature Phys.},
  year      = {2012},
  volume    = {8},
  number    = {3},
  pages     = {213--218},
  doi       = {10.1038/nphys2205},
  url       = {https://doi.org/10.1038/nphys2205},
  issn      = {1745-2481}
}

@article{Katsaris2024,
  title = {Restoring the topological edge states in a finite optical superlattice},
  author = {Katsaris, A. and Englezos, I. A. and Weitenberg, C. and Diakonos, F. K. and Schmelcher, P.},
  journal = {Phys. Rev. A},
  volume = {110},
  issue = {1},
  pages = {013316},
  numpages = {10},
  year = {2024},
  month = {Jul},
  publisher = {American Physical Society},
  doi = {10.1103/PhysRevA.110.013316},
  url = {https://link.aps.org/doi/10.1103/PhysRevA.110.013316}
}

@article{tweezerArrays2024,
  title = {Hubbard parameters for programmable tweezer arrays},
  author = {Wei, Hao-Tian and Ibarra-Garcia-Padilla, Eduardo and Wall, Michael L. and Hazzard, Kaden R. A.},
  journal = {Phys. Rev. A},
  volume = {109},
  issue = {1},
  pages = {013318},
  numpages = {13},
  year = {2024},
  month = {Jan},
  publisher = {American Physical Society},
  doi = {10.1103/PhysRevA.109.013318},
  url = {https://link.aps.org/doi/10.1103/PhysRevA.109.013318}
}

@article{WannierReview,
  title = {Maximally localized Wannier functions: Theory and applications},
  author = {Marzari, Nicola and Mostofi, Arash A. and Yates, Jonathan R. and Souza, Ivo and Vanderbilt, David},
  journal = {Rev. Mod. Phys.},
  volume = {84},
  issue = {4},
  pages = {1419--1475},
  numpages = {0},
  year = {2012},
  month = {Oct},
  publisher = {American Physical Society},
  doi = {10.1103/RevModPhys.84.1419},
  url = {https://link.aps.org/doi/10.1103/RevModPhys.84.1419}
}

@article{WannierStates,
  title = {Maximally localized generalized Wannier functions for composite energy bands},
  author = {Marzari, Nicola and Vanderbilt, David},
  journal = {Phys. Rev. B},
  volume = {56},
  issue = {20},
  pages = {12847--12865},
  numpages = {0},
  year = {1997},
  month = {Nov},
  publisher = {American Physical Society},
  doi = {10.1103/PhysRevB.56.12847},
  url = {https://link.aps.org/doi/10.1103/PhysRevB.56.12847}
}

@article{BlochNature2012,
  title = {Quantum simulations with ultracold quantum gases},
  author = {Bloch, Immanuel and Dalibard, Jean and Nascimbène, Sylvain},
  journal = {Nature Phys.},
  volume = {8},
  issue = {4},
  pages = {267--276},
  numpages = {10},
  year = {2012},
  month = {April},
  doi = {10.1038/nphys2259},
  url = {https://doi.org/10.1038/nphys2259}
}

@article{Chin2010RevModPhysFeshbach,
  title = {Feshbach resonances in ultracold gases},
  author = {Chin, Cheng and Grimm, Rudolf and Julienne, Paul and Tiesinga, Eite},
  journal = {Rev. Mod. Phys.},
  volume = {82},
  issue = {2},
  pages = {1225--1286},
  numpages = {0},
  year = {2010},
  month = {Apr},
  publisher = {American Physical Society},
  doi = {10.1103/RevModPhys.82.1225},
  url = {https://link.aps.org/doi/10.1103/RevModPhys.82.1225}
}

@article{BECexpAnderson,
author = {M. H. Anderson  and J. R. Ensher  and M. R. Matthews  and C. E. Wieman  and E. A. Cornell },
title = {Observation of Bose-Einstein Condensation in a Dilute Atomic Vapor},
journal = {Science},
volume = {269},
number = {5221},
pages = {198-201},
year = {1995},
doi = {10.1126/science.269.5221.198},
URL = {https://www.science.org/doi/abs/10.1126/science.269.5221.198}
}

@article{BECexpBradley,
  title = {Evidence of Bose-Einstein Condensation in an Atomic Gas with Attractive Interactions},
  author = {Bradley, C. C. and Sackett, C. A. and Tollett, J. J. and Hulet, R. G.},
  journal = {Phys. Rev. Lett.},
  volume = {75},
  issue = {9},
  pages = {1687--1690},
  numpages = {0},
  year = {1995},
  month = {Aug},
  publisher = {American Physical Society},
  doi = {10.1103/PhysRevLett.75.1687},
  url = {https://link.aps.org/doi/10.1103/PhysRevLett.75.1687}
}

@article{BECexpDavis,
  title = {Bose-Einstein Condensation in a Gas of Sodium Atoms},
  author = {Davis, K. B. and Mewes, M. -O. and Andrews, M. R. and van Druten, N. J. and Durfee, D. S. and Kurn, D. M. and Ketterle, W.},
  journal = {Phys. Rev. Lett.},
  volume = {75},
  issue = {22},
  pages = {3969--3973},
  numpages = {0},
  year = {1995},
  month = {Nov},
  publisher = {American Physical Society},
  doi = {10.1103/PhysRevLett.75.3969},
  url = {https://link.aps.org/doi/10.1103/PhysRevLett.75.3969}
}

@article{BlochLatticeRev,
  title = {Many-body physics with ultracold gases},
  author = {Bloch, Immanuel and Dalibard, Jean and Zwerger, Wilhelm},
  journal = {Rev. Mod. Phys.},
  volume = {80},
  issue = {3},
  pages = {885--964},
  numpages = {0},
  year = {2008},
  month = {Jul},
  publisher = {American Physical Society},
  doi = {10.1103/RevModPhys.80.885},
  url = {https://link.aps.org/doi/10.1103/RevModPhys.80.885}
}

@article{FeshbachExpCourteille,
  title = {Observation of a Feshbach Resonance in Cold Atom Scattering},
  author = {Courteille, Ph. and Freeland, R. S. and Heinzen, D. J. and van Abeelen, F. A. and Verhaar, B. J.},
  journal = {Phys. Rev. Lett.},
  volume = {81},
  issue = {1},
  pages = {69--72},
  numpages = {0},
  year = {1998},
  month = {Jul},
  publisher = {American Physical Society},
  doi = {10.1103/PhysRevLett.81.69},
  url = {https://link.aps.org/doi/10.1103/PhysRevLett.81.69}
}

@article{FeshbachExpInouye,
  title = {Observation of Feshbach resonances in a Bose–Einstein condensate},
  author = {Inouye, S. and Andrews, M. R. and Stenger, J. and Miesner, H.-J. and  Stamper-Kurn, D. M. and Ketterle, W. },
  journal = {Nature},
  volume = {392},
  pages = {151--154},
  numpages = {4},
  year = {1998},
  month = {Mar},
  doi = {10.1038/32354},
  url = {https://doi.org/10.1038/32354}
}

@article{GreinerExplattice,
  title = {Quantum phase transition from a superfluid to a Mott insulator in a gas of ultracold atoms},
  author = {Greiner, M. and Mandel, O. and Esslinger, T. and Hänsch, T. W. and  Bloch, I. },
  journal = {Nature},
  volume = {415},
  pages = {39--44},
  numpages = {5},
  year = {2002},
  month = {Jan},
  doi = {10.1038/415039a},
  url = {https://doi.org/10.1038/415039a}
}

@article{JakschBH,
  title = {Cold Bosonic Atoms in Optical Lattices},
  author = {Jaksch, D. and Bruder, C. and Cirac, J. I. and Gardiner, C. W. and Zoller, P.},
  journal = {Phys. Rev. Lett.},
  volume = {81},
  issue = {15},
  pages = {3108--3111},
  numpages = {0},
  year = {1998},
  month = {Oct},
  publisher = {American Physical Society},
  doi = {10.1103/PhysRevLett.81.3108},
  url = {https://link.aps.org/doi/10.1103/PhysRevLett.81.3108}
}

@article{DalibardReview,
  title = {Topological bands for ultracold atoms},
  author = {Cooper, N. R. and Dalibard, J. and Spielman, I. B.},
  journal = {Rev. Mod. Phys.},
  volume = {91},
  issue = {1},
  pages = {015005},
  numpages = {55},
  year = {2019},
  month = {Mar},
  publisher = {American Physical Society},
  doi = {10.1103/RevModPhys.91.015005},
  url = {https://link.aps.org/doi/10.1103/RevModPhys.91.015005}
}

@article{Ryu_2010,
doi = {10.1088/1367-2630/12/6/065010},
url = {https://dx.doi.org/10.1088/1367-2630/12/6/065010},
year = {2010},
month = {jun},
publisher = {},
volume = {12},
number = {6},
pages = {065010},
author = {Shinsei Ryu and Andreas P Schnyder and Akira Furusaki and Andreas W W Ludwig},
title = {Topological insulators and superconductors: tenfold way and dimensional hierarchy},
journal = {New J. Phys.}
}

@article{PhysRevResearch.2.033475,
  title = {Fast and robust quantum state transfer in a topological Su-Schrieffer-Heeger chain with next-to-nearest-neighbor interactions},
  author = {D'Angelis, Felippo M. and Pinheiro, Felipe A. and Gu\'ery-Odelin, David and Longhi, Stefano and Impens, Franifmmode},
  journal = {Phys. Rev. Res.},
  volume = {2},
  issue = {3},
  pages = {033475},
  numpages = {11},
  year = {2020},
  month = {Sep},
  publisher = {American Physical Society},
  doi = {10.1103/PhysRevResearch.2.033475},
  url = {https://link.aps.org/doi/10.1103/PhysRevResearch.2.033475}
}

@article{PhysRevA.103.052409,
  title = {Fast and robust quantum state transfer via a topological chain},
  author = {Palaiodimopoulos, N. E. and Brouzos, I. and Diakonos, F. K. and Theocharis, G.},
  journal = {Phys. Rev. A},
  volume = {103},
  issue = {5},
  pages = {052409},
  numpages = {9},
  year = {2021},
  month = {May},
  publisher = {American Physical Society},
  doi = {10.1103/PhysRevA.103.052409},
  url = {https://link.aps.org/doi/10.1103/PhysRevA.103.052409}
}

@article{PhysRevA.106.022419,
  title = {Fast and robust quantum state transfer assisted by zero-energy interface states in a splicing Su-Schrieffer-Heeger chain},
  author = {Huang, Lijun and Tan, Zhi and Zhong, Honghua and Zhu, Bo},
  journal = {Phys. Rev. A},
  volume = {106},
  issue = {2},
  pages = {022419},
  numpages = {9},
  year = {2022},
  month = {Aug},
  publisher = {American Physical Society},
  doi = {10.1103/PhysRevA.106.022419},
  url = {https://link.aps.org/doi/10.1103/PhysRevA.106.022419}
}

@article{PhysRevB.106.245109,
  title = {Electronic and topological properties of extended two-dimensional Su-Schrieffer-Heeger models and realization of flat edge bands},
  author = {Ma, Hongyu and Zhang, Ze and Fu, Pei-Hao and Wu, Jiansheng and Yu, Xiang-Long},
  journal = {Phys. Rev. B},
  volume = {106},
  issue = {24},
  pages = {245109},
  numpages = {10},
  year = {2022},
  month = {Dec},
  publisher = {American Physical Society},
  doi = {10.1103/PhysRevB.106.245109},
  url = {https://link.aps.org/doi/10.1103/PhysRevB.106.245109}
}

@article{PhysRevB.107.054105,
  title = {Exploring interacting topological insulator in the extended Su-Schrieffer-Heeger model},
  author = {Zhou, Xiaofan and Pan, Jian-Song and Jia, Suotang},
  journal = {Phys. Rev. B},
  volume = {107},
  issue = {5},
  pages = {054105},
  numpages = {8},
  year = {2023},
  month = {Feb},
  publisher = {American Physical Society},
  doi = {10.1103/PhysRevB.107.054105},
  url = {https://link.aps.org/doi/10.1103/PhysRevB.107.054105}
}

@article{Kanungo_2022,
   title={Realizing topological edge states with Rydberg-atom synthetic dimensions},
   volume={13},
   pages={972},
   ISSN={2041-1723},
   url={http://dx.doi.org/10.1038/s41467-022-28550-y},
   number={1},
   journal={Nature Comm.},
   publisher={Springer Science and Business Media LLC},
   author={Kanungo, S. K. and Whalen, J. D. and Lu, Y. and Yuan, M. and Dasgupta, S. and Dunning, F. B. and Hazzard, K. R. A. and Killian, T. C.},
   year={2022},
   month=feb }

@article{weitenberg2021tailoring,
  title={Tailoring quantum gases by Floquet engineering},
  author={Weitenberg, Christof and Simonet, Juliette},
  journal={Nature Phys.},
  volume={17},
  number={12},
  pages={1342--1348},
  year={2021},
  publisher={Nature Publishing Group UK London}
}

@article{PhysRevLett.108.133001,
  title = {Quantum Simulation of an Extra Dimension},
  author = {Boada, O. and Celi, A. and Latorre, J. I. and Lewenstein, M.},
  journal = {Phys. Rev. Lett.},
  volume = {108},
  issue = {13},
  pages = {133001},
  numpages = {6},
  year = {2012},
  month = {Mar},
  publisher = {American Physical Society},
  doi = {10.1103/PhysRevLett.108.133001},
  url = {https://link.aps.org/doi/10.1103/PhysRevLett.108.133001}
}

@article{Mancini_2015,
   title={Observation of chiral edge states with neutral fermions in synthetic Hall ribbons},
   volume={349},
   ISSN={1095-9203},
   url={http://dx.doi.org/10.1126/science.aaa8736},
   DOI={10.1126/science.aaa8736},
   number={6255},
   journal={Science},
   publisher={American Association for the Advancement of Science (AAAS)},
   author={Mancini, M. and Pagano, G. and Cappellini, G. and Livi, L. and Rider, M. and Catani, J. and Sias, C. and Zoller, P. and Inguscio, M. and Dalmonte, M. and Fallani, L.},
   year={2015},
   month=sep, pages={1510}}

@article{Stuhl_2015,
   title={Visualizing edge states with an atomic Bose gas in the quantum Hall regime},
   volume={349},
   ISSN={1095-9203},
   url={http://dx.doi.org/10.1126/science.aaa8515},
   DOI={10.1126/science.aaa8515},
   number={6255},
   journal={Science},
   publisher={American Association for the Advancement of Science (AAAS)},
   author={Stuhl, B. K. and Lu, H.-I. and Aycock, L. M. and Genkina, D. and Spielman, I. B.},
   year={2015},
   month=sep, pages={1514} }

@article{
doi:10.1126/science.aav9105,
author = {Sylvain de Léséleuc  and Vincent Lienhard  and Pascal Scholl  and Daniel Barredo  and Sebastian Weber  and Nicolai Lang  and Hans Peter Büchler  and Thierry Lahaye  and Antoine Browaeys },
title = {Observation of a symmetry-protected topological phase of interacting bosons with Rydberg atoms},
journal = {Science},
volume = {365},
number = {6455},
pages = {775-780},
year = {2019},
doi = {10.1126/science.aav9105},
URL = {https://www.science.org/doi/abs/10.1126/science.aav9105}}

@article{PhysRevLett.110.260405,
  title = {Topological Edge States in the One-Dimensional Superlattice Bose-Hubbard Model},
  author = {Grusdt, Fabian and H\"oning, Michael and Fleischhauer, Michael},
  journal = {Phys. Rev. Lett.},
  volume = {110},
  issue = {26},
  pages = {260405},
  numpages = {5},
  year = {2013},
  month = {Jun},
  publisher = {American Physical Society},
  doi = {10.1103/PhysRevLett.110.260405},
  url = {https://link.aps.org/doi/10.1103/PhysRevLett.110.260405}
}

@article{PhysRevA.99.053614,
  title = {Quantum phases and topological properties of interacting fermions in one-dimensional superlattices},
  author = {Stenzel, L. and Hayward, A. L. C. and Hubig, C. and Schollw\"ock, U. and Heidrich-Meisner, F.},
  journal = {Phys. Rev. A},
  volume = {99},
  issue = {5},
  pages = {053614},
  numpages = {17},
  year = {2019},
  month = {May},
  publisher = {American Physical Society},
  doi = {10.1103/PhysRevA.99.053614},
  url = {https://link.aps.org/doi/10.1103/PhysRevA.99.053614}
}

@article{Nakajima_2016,
   title={Topological Thouless pumping of ultracold fermions},
   volume={12},
   ISSN={1745-2481},
   url={http://dx.doi.org/10.1038/nphys3622},
   DOI={10.1038/nphys3622},
   number={4},
   journal={Nature Phys.},
   publisher={Springer Science and Business Media LLC},
   author={Nakajima, Shuta and Tomita, Takafumi and Taie, Shintaro and Ichinose, Tomohiro and Ozawa, Hideki and Wang, Lei and Troyer, Matthias and Takahashi, Yoshiro},
   year={2016},
   month=jan, pages={296} }

@article{Lohse_2015,
   title={A Thouless quantum pump with ultracold bosonic atoms in an optical superlattice},
   volume={12},
   ISSN={1745-2481},
   url={http://dx.doi.org/10.1038/nphys3584},
   DOI={10.1038/nphys3584},
   number={4},
   journal={Nature Phys.},
   publisher={Springer Science and Business Media LLC},
   author={Lohse, M. and Schweizer, C. and Zilberberg, O. and Aidelsburger, M. and Bloch, I.},
   year={2016},
   month=dec, pages={350} }

@article{Anderlini_2007,
   title={Controlled exchange interaction between pairs of neutral atoms in an optical lattice},
   volume={448},
   ISSN={1476-4687},
   url={http://dx.doi.org/10.1038/nature06011},
   DOI={10.1038/nature06011},
   number={7152},
   journal={Nature},
   publisher={Springer Science and Business Media LLC},
   author={Anderlini, Marco and Lee, Patricia J. and Brown, Benjamin L. and Sebby-Strabley, Jennifer and Phillips, William D. and Porto, J. V.},
   year={2007},
   month=jul, pages={452} }

@article{F_lling_2007,
   title={Direct observation of second-order atom tunnelling},
   volume={448},
   ISSN={1476-4687},
   url={http://dx.doi.org/10.1038/nature06112},
   DOI={10.1038/nature06112},
   number={7157},
   journal={Nature},
   publisher={Springer Science and Business Media LLC},
   author={Fölling, S. and Trotzky, S. and Cheinet, P. and Feld, M. and Saers, R. and Widera, A. and Müller, T. and Bloch, I.},
   year={2007},
   month=aug, pages={1029} }

@article{bakr2009quantum,
  title={A quantum gas microscope for detecting single atoms in a Hubbard-regime optical lattice},
  author={Bakr, Waseem S and Gillen, Jonathon I and Peng, Amy and F{\"o}lling, Simon and Greiner, Markus},
  journal={Nature},
  volume={462},
  number={7269},
  pages={74--77},
  year={2009},
  publisher={Nature Publishing Group UK London}
}

@article{PhysRevB.106.205111,
  title = {Topological properties of subsystem-symmetry-protected edge states in an extended quasi-one-dimensional dimerized lattice},
  author = {Jangjan, Milad and Hosseini, Mir Vahid},
  journal = {Phys. Rev. B},
  volume = {106},
  issue = {20},
  pages = {205111},
  numpages = {9},
  year = {2022},
  month = {Nov},
  publisher = {American Physical Society},
  doi = {10.1103/PhysRevB.106.205111},
  url = {https://link.aps.org/doi/10.1103/PhysRevB.106.205111}
}

@article{Aubry1980,
author = {Aubry, Serge and André, Gilles},
year = {1980},
month = {01},
pages = {133},
title = {Analyticity breaking and Anderson localization in incommensurate lattices},
volume = {3},
journal = {Proceedings, VIII International Colloquium on Group-Theoretical Methods in Physics}
}

@article{Lahini2009,
  title = {Observation of a Localization Transition in Quasiperiodic Photonic Lattices},
  author = {Lahini, Y. and Pugatch, R. and Pozzi, F. and Sorel, M. and Morandotti, R. and Davidson, N. and Silberberg, Y.},
  journal = {Phys. Rev. Lett.},
  volume = {103},
  issue = {1},
  pages = {013901},
  numpages = {4},
  year = {2009},
  month = {Jun},
  publisher = {American Physical Society},
  doi = {10.1103/PhysRevLett.103.013901},
  url = {https://link.aps.org/doi/10.1103/PhysRevLett.103.013901}
}

@article{Roati2008,
   title={Anderson localization of a non-interacting Bose–Einstein condensate},
   volume={453},
   ISSN={1476-4687},
   url={http://dx.doi.org/10.1038/nature07071},
   DOI={10.1038/nature07071},
   number={7197},
   journal={Nature},
   publisher={Springer Science and Business Media LLC},
   author={Roati, Giacomo and D’Errico, Chiara and Fallani, Leonardo and Fattori, Marco and Fort, Chiara and Zaccanti, Matteo and Modugno, Giovanni and Modugno, Michele and Inguscio, Massimo},
   year={2008},
   month=jun, pages={895} }

@article{Anderson1958,
  title = {Absence of Diffusion in Certain Random Lattices},
  author = {Anderson, P. W.},
  journal = {Phys. Rev.},
  volume = {109},
  issue = {5},
  pages = {1492--1505},
  numpages = {0},
  year = {1958},
  month = {Mar},
  publisher = {American Physical Society},
  doi = {10.1103/PhysRev.109.1492},
  url = {https://link.aps.org/doi/10.1103/PhysRev.109.1492}
}

@article{Lahini2008,
  title = {Anderson Localization and Nonlinearity in One-Dimensional Disordered Photonic Lattices},
  author = {Lahini, Yoav and Avidan, Assaf and Pozzi, Francesca and Sorel, Marc and Morandotti, Roberto and Christodoulides, Demetrios N. and Silberberg, Yaron},
  journal = {Phys. Rev. Lett.},
  volume = {100},
  issue = {1},
  pages = {013906},
  numpages = {4},
  year = {2008},
  month = {Jan},
  publisher = {American Physical Society},
  doi = {10.1103/PhysRevLett.100.013906},
  url = {https://link.aps.org/doi/10.1103/PhysRevLett.100.013906}
}

@article{berry1984phase,
  author       = {Berry, Michael V.},
  title        = {Quantal phase factors accompanying adiabatic changes},
  journal      = {Proceedings of the Royal Society of London. A. Mathematical and Physical Sciences},
  volume       = {392},
  number       = {1802},
  pages        = {45--57},
  year         = {1984},
  publisher    = {The Royal Society},
  doi          = {10.1098/rspa.1984.0023},
  url          = {https://doi.org/10.1098/rspa.1984.0023}
}

@article{PhysRevLett.62.2747,
  title = {Berry's phase for energy bands in solids},
  author = {Zak, J.},
  journal = {Phys. Rev. Lett.},
  volume = {62},
  issue = {23},
  pages = {2747--2750},
  numpages = {0},
  year = {1989},
  month = {Jun},
  publisher = {American Physical Society},
  doi = {10.1103/PhysRevLett.62.2747},
  url = {https://link.aps.org/doi/10.1103/PhysRevLett.62.2747}
}

@article{L_onard_2023,
   title={Realization of a fractional quantum Hall state with ultracold atoms},
   volume={619},
   ISSN={1476-4687},
   url={http://dx.doi.org/10.1038/s41586-023-06122-4},
   DOI={10.1038/s41586-023-06122-4},
   number={7970},
   journal={Nature},
   publisher={Springer Science and Business Media LLC},
   author={Léonard, Julian and Kim, Sooshin and Kwan, Joyce and Segura, Perrin and Grusdt, Fabian and Repellin, Cécile and Goldman, Nathan and Greiner, Markus},
   year={2023},
   month=jun, pages={495–499} }

@article{Gross_2021,
   title={Quantum gas microscopy for single atom and spin detection},
   volume={17},
   ISSN={1745-2481},
   url={http://dx.doi.org/10.1038/s41567-021-01370-5},
   DOI={10.1038/s41567-021-01370-5},
   number={12},
   journal={Nature Phys.},
   publisher={Springer Science and Business Media LLC},
   author={Gross, Christian and Bakr, Waseem S.},
   year={2021},
   month=nov, pages={1316–1323} }

@article{Su_2025,
   title={Fast single atom imaging for optical lattice arrays},
   volume={16},
   ISSN={2041-1723},
   url={http://dx.doi.org/10.1038/s41467-025-56305-y},
   DOI={10.1038/s41467-025-56305-y},
   number={1},
   journal={Nature Comm.},
   publisher={Springer Science and Business Media LLC},
   author={Su, Lin and Douglas, Alexander and Szurek, Michal and Hébert, Anne H. and Krahn, Aaron and Groth, Robin and Phelps, Gregory A. and Marković, Ognjen and Greiner, Markus},
   year={2025},
   month=jan }

@article{Chalopin_2020,
   title={Probing chiral edge dynamics and bulk topology of a synthetic Hall system},
   volume={16},
   ISSN={1745-2481},
   url={http://dx.doi.org/10.1038/s41567-020-0942-5},
   DOI={10.1038/s41567-020-0942-5},
   number={10},
   journal={Nature Phys.},
   publisher={Springer Science and Business Media LLC},
   author={Chalopin, Thomas and Satoor, Tanish and Evrard, Alexandre and Makhalov, Vasiliy and Dalibard, Jean and Lopes, Raphael and Nascimbene, Sylvain},
   year={2020},
   month=jun, pages={1017–1021} }

@article{Braun_2024,
   title={Real-space detection and manipulation of topological edge modes with ultracold atoms},
   volume={20},
   ISSN={1745-2481},
   url={http://dx.doi.org/10.1038/s41567-024-02506-z},
   DOI={10.1038/s41567-024-02506-z},
   number={8},
   journal={Nature Phys.},
   publisher={Springer Science and Business Media LLC},
   author={Braun, Christoph and Saint-Jalm, Raphaël and Hesse, Alexander and Arceri, Johannes and Bloch, Immanuel and Aidelsburger, Monika},
   year={2024},
   month=jun, pages={1306–1312} }

@article{Yao_2024,
   title={Observation of chiral edge transport in a rapidly rotating quantum gas},
   volume={20},
   ISSN={1745-2481},
   url={http://dx.doi.org/10.1038/s41567-024-02617-7},
   DOI={10.1038/s41567-024-02617-7},
   number={11},
   journal={Nature Phys.},
   publisher={Springer Science and Business Media LLC},
   author={Yao, Ruixiao and Chi, Sungjae and Mukherjee, Biswaroop and Shaffer, Airlia and Zwierlein, Martin and Fletcher, Richard J.},
   year={2024},
   month=sep, pages={1726–1731} }

@article{PhysRev.117.432,
  title = {Wave Functions and Effective Hamiltonian for Bloch Electrons in an Electric Field},
  author = {Wannier, Gregory H.},
  journal = {Phys. Rev.},
  volume = {117},
  issue = {2},
  pages = {432--439},
  numpages = {0},
  year = {1960},
  month = {Jan},
  publisher = {American Physical Society},
  doi = {10.1103/PhysRev.117.432},
  url = {https://link.aps.org/doi/10.1103/PhysRev.117.432}
}

@article{PhysRevB.36.7353,
  title = {Existence of Wannier-Stark localization},
  author = {Emin, David and Hart, C. F.},
  journal = {Phys. Rev. B},
  volume = {36},
  issue = {14},
  pages = {7353--7359},
  numpages = {0},
  year = {1987},
  month = {Nov},
  publisher = {American Physical Society},
  doi = {10.1103/PhysRevB.36.7353},
  url = {https://link.aps.org/doi/10.1103/PhysRevB.36.7353}
}

@article{PhysRevLett.61.1639,
  title = {Observation of the Wannier-Stark Quantization in a Semiconductor Superlattice},
  author = {Voisin, P. and Bleuse, J. and Bouche, C. and Gaillard, S. and Alibert, C. and Regreny, A.},
  journal = {Phys. Rev. Lett.},
  volume = {61},
  issue = {14},
  pages = {1639--1642},
  numpages = {0},
  year = {1988},
  month = {Oct},
  publisher = {American Physical Society},
  doi = {10.1103/PhysRevLett.61.1639},
  url = {https://link.aps.org/doi/10.1103/PhysRevLett.61.1639}
}

\end{document}